\documentclass[times,twocolumn,final,authoryear]{elsarticle}

\usepackage{jasr}
\usepackage{framed,multirow}
\usepackage{amssymb}
\usepackage{latexsym}
\usepackage[switch]{lineno}  
\usepackage{url}
\usepackage{xcolor}
\usepackage[citebordercolor=white]{hyperref}
\usepackage{graphicx}
\usepackage{hyperref}
\usepackage{hypcap}
\usepackage{wrapfig,booktabs}
\usepackage[export]{adjustbox}
\usepackage{needspace}
\usepackage{bibentry}
\journal{Advances in Space Research}
\graphicspath{{figures/}}

\definecolor{newcolor}{rgb}{.8,.349,.1}
\definecolor{ao(english)}{rgb}{0.0, 0.5, 0.0} 

\newcommand{\BE}{\begin{equation}}
\newcommand{\EE}{\end{equation}}
\newcommand{\BA}{\begin{align}}
\newcommand{\EA}{\end{align}}

\newcommand{\degree}{\ensuremath{^\circ}}

\begin{document}
\verso{Hosteaux \textit{et al}}

\begin{frontmatter}

\title{Analysis of Voyager 1 and Voyager 2 in situ CME observations}

\author[1]{Skralan \snm{Hosteaux}}
\author[2]{Luciano \snm{Rodiguez}}
\author[1,3]{Stefaan \snm{Poedts}\corref{cor1}} 
\cortext[cor1]{Corresponding author:
  Tel.: +32-16-327023;
  fax: +32-16-327998;}%
\address[1]{CmPA/Department of Mathematics, KU Leuven, Celestijnenlaan 200 B, 3001 Leuven, Belgium}
\address[2]{Solar-Terrestrial Centre of Excellence – SIDC, Royal Observatory of Belgium; Avenue Circulaire 3, 1180 Brussels, Belgium}
\address[3]{Institute of Physics, University of Maria Curie-Sk{\l}odowska, Pl. M. Curie-Sk{\l}odowskiej 5, 20-031 Lublin, Poland}
\received{31 August 2021}
\finalform{2021}
\accepted{2021}
\availableonline{2021}
\communicated{}

\begin{abstract}
This paper studies ICMEs detected by both Voyager spacecraft during propagation from 1 to 10~AU, with observations from 1977 to 1980. ICMEs are detected by using several signatures in the in-situ data, the primary one being the low measured to expected proton temperature ratio. We found 21 events common to both spacecraft and study their internal structure in terms of plasma and magnetic field properties. We find that ICMEs are expanding as they propagate outwards, with decreasing density and magnetic field intensities, in agreement with previous studies. We first carry out a statistical study and then a detailed analysis of each case. Furthermore, we analyse one case in which a shock can be clearly detected by both spacecraft. The methods described here can be interesting for other studies combining data sets from heliospheric missions. Furthermore, they highlight the importance of exploiting useful data from past missions.

\end{abstract}


\end{frontmatter}

\section{Introduction}
Coronal Mass ejections (CMEs) are one of the most influential transient events in the solar system. They comprise huge amounts of plasma and magnetic fields ejected from the Sun, with velocities typically in a range of 400 to 1000 km/s but can also go faster than 2000 km/s \citep{Hundhausen1994}. It takes about three days for a CME to reach the Earth. Upon arrival, its interaction with the Earth’s magnetosphere can produce geomagnetic storms. When CMEs are detected in situ by a spacecraft, they are termed Interplanetary CMEs (ICMEs). If the speed of the ICME exceeds the speed of the fast MHD wave in the solar wind frame, a shock will develop in front of the eruption. This forms a discontinuity
in plasma properties such as density, pressure and velocity. Behind the shock, a dense sheath region will form. The thickness of the sheath region is also called the shock standoff distance. The compression of plasma inside the sheath region makes it a very
suitable location for magnetic reconnection \citep{Kilpua2017}. The sheath
region of the ICME is normally followed by a flux rope, which is a large closed field
structure of increased magnetic field strength, smooth rotation of magnetic field
direction and below average temperature \citep{Burlaga1981}. {Not all observed ICMEs display this
structure, as not all ICMEs are fast enough to develop a shock. Moreover, only approximately a third \citep{Gosling1990, Rodriguez2004} to a half \citep{Cane1997} of all observed ICMEs show signatures of a magnetic cloud. This, however, might be due to the fact that most observations are made by one spacecraft on a single track through the ICME and thus might monitor only part of it.}

CMEs also have an influence in the far reaches of the solar system. \cite{Wang2004} identified and characterized ICMEs observed by Voyager 2 in the heliosphere between 1 and 30 AU. They found that the average radial width of ICMEs increases with distance up to $\sim$15$\;$AU, and it remains constant after that.  \cite{Leitner2007} studied magnetic clouds detected mostly by Helios 1 and 2 between 0.3 and 1 AU \citep[also][]{Bothmer1998} , and complemented those with some events observed by the Voyager spacecraft and others. \citet{Gulisano2012} did a study on the expansion of magnetic clouds using in-situ Ulysses data from July 1992 to November 2001, finding events between approximately 1.5 and 5.5~AU. {In this paper, we present an exhaustive comparative study of ICMEs that were observed by both the Voyager 1 and Voyager 2 spacecraft from 1 to 10 AU, with observations from 1977 to 1980. This is then a multi-spacecraft study, as the same events are observed by both spacecraft.} 

A case study performed by~\citet{Burlaga1982} identified five ICMEs that were observed by both Voyagers between 2-4~AU. In fact, three of those cases were also found as common events in this work. This also means that two events in that study were actually not identified here, which illustrates the subjectivity of identifying ICMEs by eye, an effect which is amplified by increasing radial distance from the Sun, as the events become more diluted, and difficult to differentiate from the background solar wind.

Besides ICMEs, we also studied the shocks that preceded them. Unfortunately, many of our events contained data gaps where the shock was most likely located. That, in addition to the fact that not all ICMEs are preceded by a shock, severely lowers our statistics, resulting in only one suitable shock front. This case was analysed using the Minimum Variance Analysis method \citep[MVA, ][]{Sonnerup1967}, that allows for the determination of the shock normal and which will be explained in Section \ref{shock}.

In Section \ref{data}, we explain how the data was analysed, together with our events selection. Then, in Section \ref{AllICMEs}, we describe our results in terms of general properties of the ICMEs detected and with respect to distance from the Sun. In Section \ref{shock}, we examine shock properties. Finally, in Section \ref{Conclusions}, we present our conclusions.

\section{Data and event selection} \label{data}

The ICMEs analysed here were observed by Voyager 1 (VOY1) and Voyager 2 (VOY2) between November 1977 and December 1980, between 1 and 10 AU.

The data used to identify the ICME events was obtained from the NASA National Space Science Data Center and the Space Physics Data Facility. To set up the list of events for this study, we mainly used the following ICME signatures \citep{Zurbuchen2006}:

\begin{itemize}
\item Low ratio between measured and expected proton temperature $\mathrm{T_p/T_{exp}<0.5}$ \citep{Gosling1973,Richardson1995}
\item Smoothly rotating B-field \citep{Klein1982}
\item B-field variance decrease \citep{Pudovkin1979, Klein1982}
\item Low density $\leq$1~$\mathrm{cm^{-3}}$ \citep{Richardson2000}
\item Linearly decreasing bulk velocity \citep{Klein1982,Russell2003}
\end{itemize}

The correlation between solar wind bulk velocity and solar wind proton temperature has long been established \citep{Neugebauer1966, Hundhausen1970}. ICMEs can be regarded as thermally isolated expanding regions of plasma, due to thermal conduction being inefficient across magnetic field lines. Thus, if an ICME crosses a spacecraft, the measured $\mathrm{T_P}$ is lower than it would be for typical solar wind measurements. Therefore, a low measured to expected temperature ratio is an indication of a passing ICME. The method used for detecting regions of unusually low solar wind proton temperatures consists {in determining the ratio of the observed proton temperature with the expected proton temperature. The latter is derived from solar wind velocity measurements} \citep{Richardson1993,Richardson1995} by using the formula found by \citet{Lopez1986} and \citet{Lopez1987}:

\begin{equation}
    \mathrm{T_{exp}}[\mathrm{10^3K}]=\left\{
                \begin{array}{ll}
                  \mathrm{(0.031v_{sw}-5.1)^2/R^{\beta} \hspace{0.35cm} v<500~km/s}\\
                  \mathrm{(0.51v_{sw}-142)/R^{\beta} \hspace{0.6cm}v\geq500~km/s}\\
                \end{array}
              \right.
              \label{eqTexp}
\end{equation}

where the heliocentric distance $R$ is in $AU$ and $v_{sw}$ is in $km/s$. Since our data set does not reach beyond 10~AU, $\beta$ is set to 0.7 according to the conclusions of \citet{Gazis1994}. This relationship was also used by \citet{Wang2004} to identify and perform a statistical analysis for ICMEs between 1 and 30~AU using VOY2 data, albeit with $\beta=0.6$ because of their larger radial distance. They used abnormally low proton temperatures as the primary identification signature of ICMEs and compared with other plasma and magnetic field data to verify these identifications. {At 1 AU, the elevated magnetic field inside ICMEs is normally used as a signature for identifying them. Nevertheless, with increasing distance from the Sun, the magnetic field magnitude in ICMEs decreases and is no longer a clear signature for their identification. This justifies the use of temperature here as a primary identifying signature.}  {We use then the abnormally low proton temperatures as the primary condition for the ICME detection and determination of boundaries. The presence of more signatures serves to increase confidence in the identification. This process is done by eye, ICME identification “is still something of an art” \citep{Gosling1997}, in particular when dealing with a sparse dataset containing gaps and other complications. Furthermore, some signatures may be present only in parts of the ICME, making the determination of boundaries complicated. Usually a low $\mathrm{T_p/T_{exp}}$, complemented with other signatures, are used for classification as an ICME. To decide if the same ICME was observed by both spacecraft, two conditions have to be fulfilled. {First, we use the speed of the ICME (should be similar in both spacecraft) to calculate the travel time between the two locations and require that the estimated travel time should be consistent with the arrival time of the ICME at the second spacecraft.} Second, we examine that the plasma and magnetic field characteristics of the ICMEs are similar in both spacecraft. This can be checked in the images of each event provided in the Appendix. Normally, ICMEs are prominent in the data, and the Voyagers do not see a large amount of them in the periods concerned, so misidentifications are possible but not likely.}

We analysed the full VOY1 and VOY2 dataset between November 1977 and December 1980. Those events classified as an ICME in both spacecraft were added to our list of common ICMEs. The boundaries of the ICMEs were found using the low proton temperature condition, $\mathrm{T_p/T_{exp}<0.5}$. In this manner, a total of 21 ICMEs that were measured by both VOY1 and VOY2 were found. Table \ref{table:CMEs} presents the ICMEs that were adequately measured by both Voyager 1 and Voyager 2. For simplicity each event in the table is represented by a letter, which is then also used in the following sections. 

\begin{center}
\begin{table*}[htb!]
\begin{minipage}{0.99\textwidth}
\begin{center}
\begin{tabular}{l|lll|l|lll}
  & year & doy VOY1 & doy VOY2 &  & year & doy VOY1 & doy VOY2 \\  \cline{1-8}
A & 77   & 265.46   & 265.58 & L & 79   & 49.83    & 48.5  \\
B & 77   & 350.33   & 350.21 & M & 79   & 107.88   & 105.96 \\
C & 78   & 29.83    & 29.42  & N & 79   & 134.08   & 132.29  \\
D & 78   & 39.71    & 39.75  & O & 79   & 146.04   & 142.83  \\
E & 78   & 52.17    & 52.08  & P & 79   & 167.08   & 166   \\
F & 78   & 153.75   & 153.29 & Q & 80   & 35.25    & 32.88 \\
G & 78   & 158.08   & 157.42 & R & 80   & 134.67   & 128.83  \\
H & 78   & 267.92   & 266.33 & S & 80   & 157.79   & 153.79  \\
I & 78   & 321.88   & 320.54 & T & 80   & 181.04   & 175.71   \\
J & 78   & 344.17   & 342.92 & U & 80   & 241.33   & 233.92  \\
K & 79   & 4.96     & 3.88   & & & & \\
\end{tabular}
\end{center}
\end{minipage}
  \caption{The 21 events that were identified in the data of both Voyager 1 and Voyager 2. The first column gives the letter by which the event will be referenced in the text, the latter two give the decimal day of year of the front of the ICME for both spacecraft.}\label{table:CMEs}
\end{table*}
\end{center}

\begin{figure*}[!htb]
\centering
  \centering
  \includegraphics[trim=1cm 1.5cm 1.5cm 2cm, clip=true,width=0.495\linewidth]{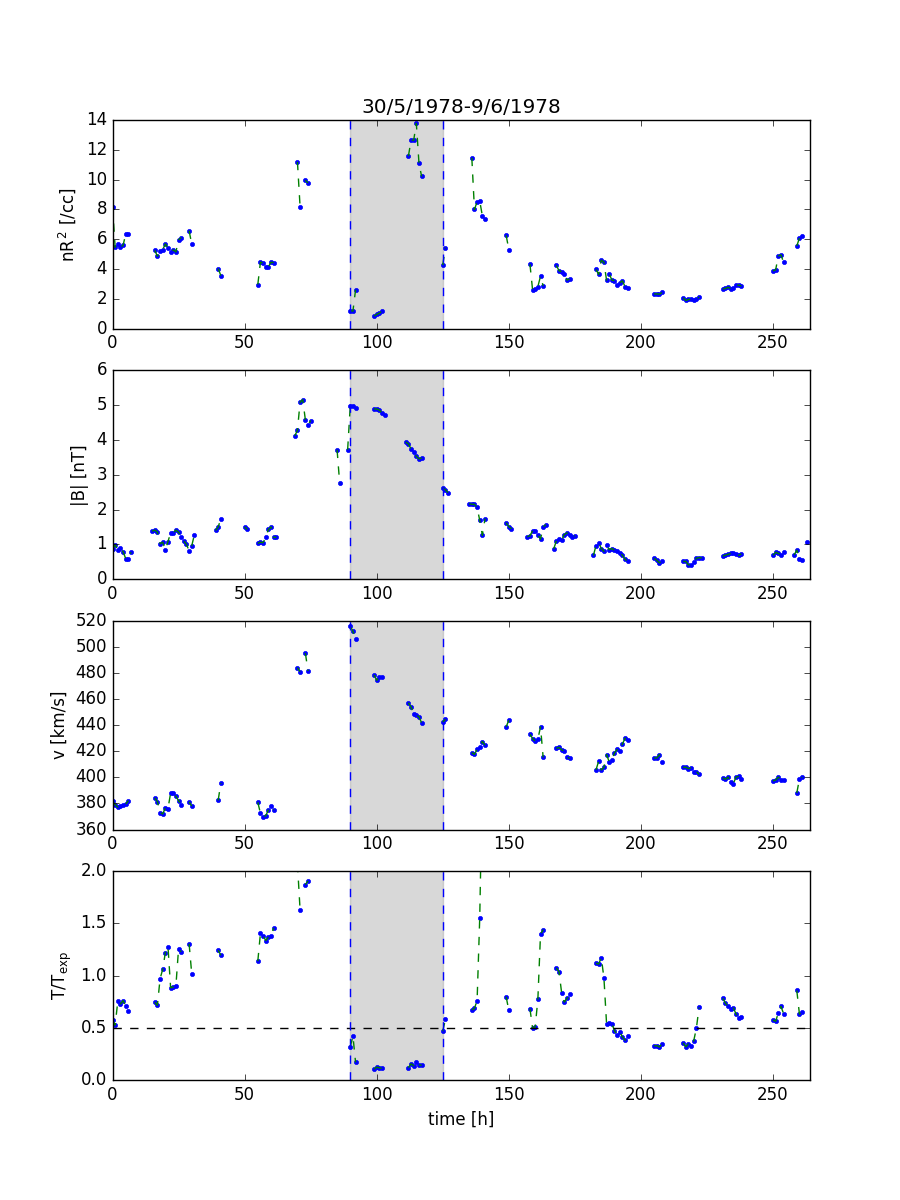} 
  \includegraphics[trim=1cm 1.5cm 1.5cm 2cm, clip=true,width=0.495\linewidth]{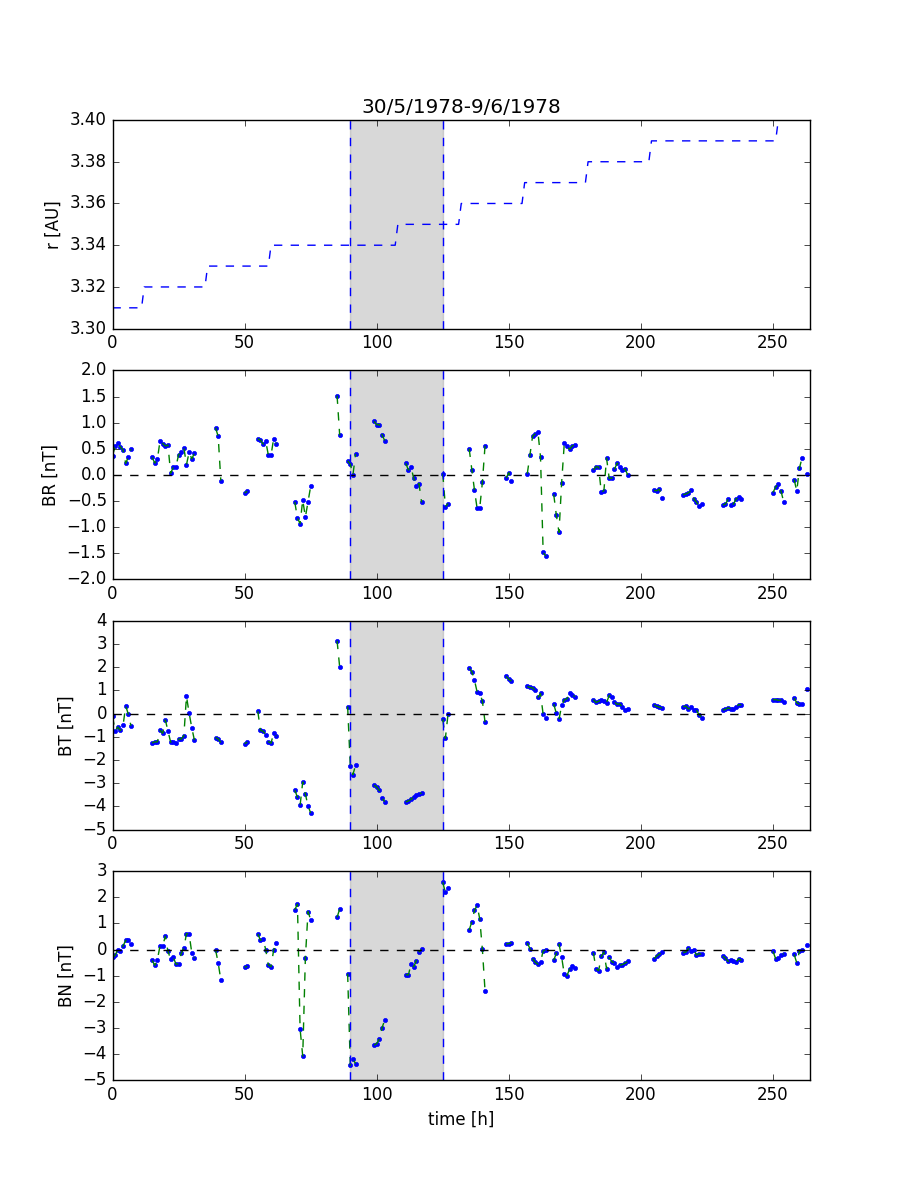}
  \includegraphics[trim=1cm 1.5cm 1.5cm 2cm, clip=true,width=0.495\linewidth]{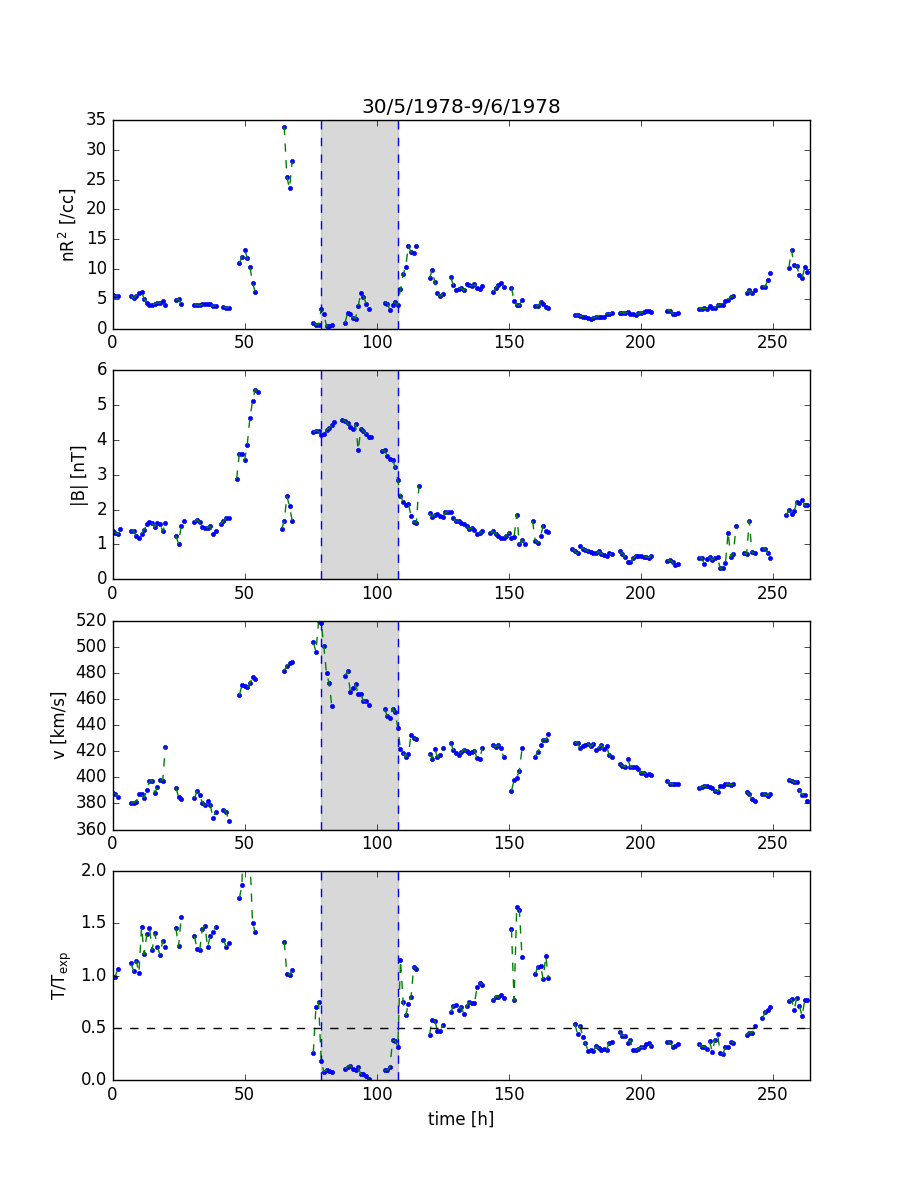}
  \includegraphics[trim=1cm 1.5cm 1.5cm 2cm, clip=true,width=0.495\linewidth]{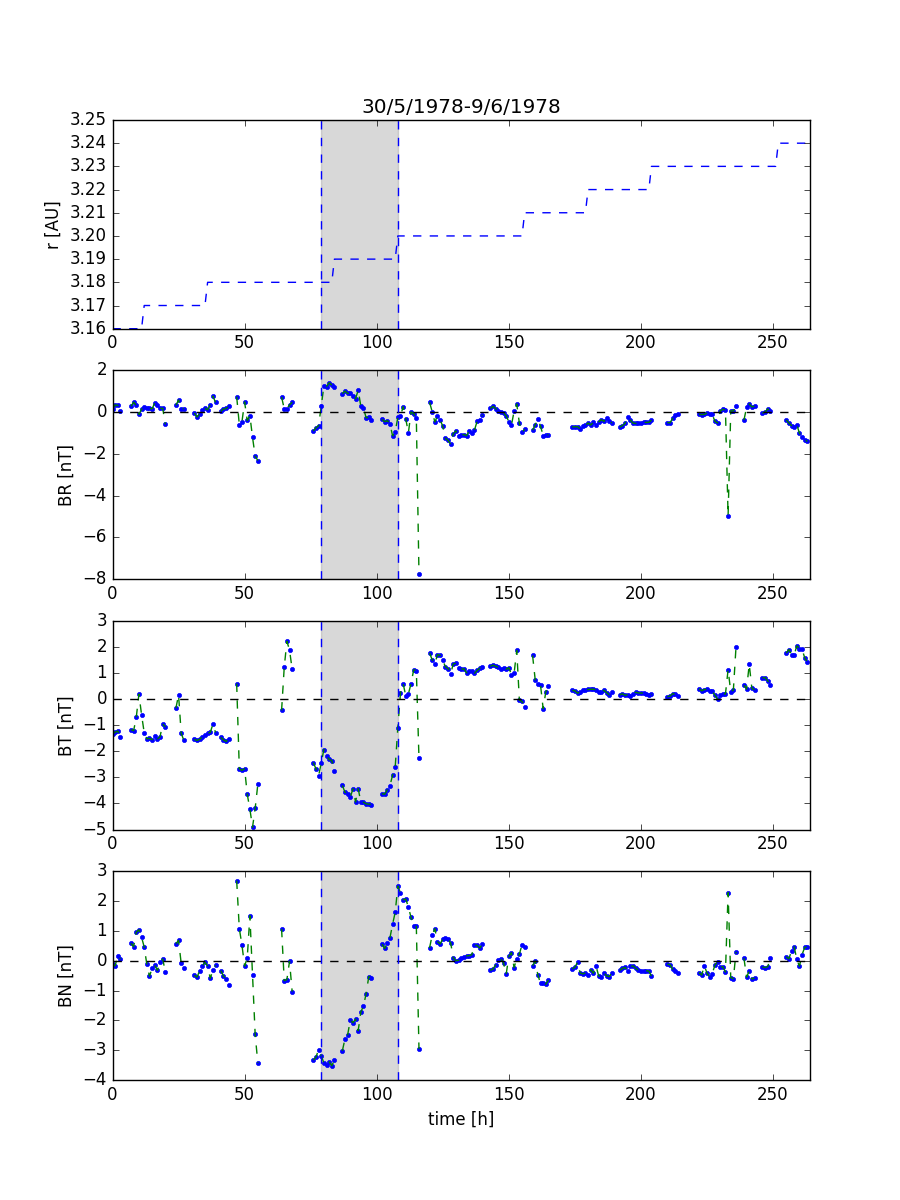} 
\caption{Voyager 1 (upper half) and Voyager 2 (lower half) time series for Event F, observed on 6 June 1978. The ICME region is marked in grey. The parameters plotted are (from top left to bottom right): density, magnetic field magnitude, speed, temperature ratio, radial distance from the Sun, the three components of the magnetic field vector.  }\label{fig:fig1_eventF}
\end{figure*}

{An example ICME (Event F) is show in Figure \ref{fig:fig1_eventF}.
VOY1 measured event F on 3 June 1978 at 18:00 UT while VOY2 measured the event on the same day at 06:00 UT, at a radial distance of 3.34 and 3.18~AU, respectively. A summary of its properties can be seen in Table \ref{tab:exampleeventF}. The ICME boundaries are obtained from the low proton temperature condition, $\mathrm{T_p/T_{exp}<0.5}$. Figure~\ref{fig:fig1_eventF} shows that both Voyagers observe an enhanced magnetic field, with a rotation typical of a magnetic cloud, seen in the gradual variation of the magnetic field components. These signatures are similar in both spacecraft, with the ICME width increasing by 0.06~AU as it travels between them.}

\Needspace{100pt}


\begin{table}
\begin{center}
\begin{footnotesize}
\begin{tabular}{c|c|c}
 & VOY1 & VOY 2 \\ \cline{1-3}
r [AU] & 3.34 & 3.18 \\
$\mathrm{N_{avg} [cm^{-3}]}$ & 0.51 & 0.28 \\
$\mathrm{B_{avg} [nT]}$ & 4.16 & 4.11 \\
$\mathrm{V_{avg} [km/s]}$ & 468.3 & 464.6  \\
$\mathrm{V_{beg} [km/s]}$ & 515.9 & 518.7 \\
width [AU]& 0.39 & 0.32 \\
\end{tabular}
\caption{Table of properties for event F, observed on 03/06/1978. The properties displayed are average density, average magnetic field, average speed, leading edge speed and radial width of the ICME.}\label{tab:exampleeventF}
\vspace{-20pt}
\end{footnotesize}
\end{center}
\end{table}

{A similar description of all other events listed in Table~\ref{table:CMEs} can be found in \ref{appendix_event_description}. The in-situ data plots of the events discussed in this section and in \ref{appendix_event_description} is shown in \ref{appendix_time_series}. The magnetic field components are shown in the Radial Tangential Normal (RTN) Heliographic coordinate system, where R is directed along the Sun-to-spacecraft line, T is parallel to the Solar Equatorial plane and perpendicular to R, and N is the normal component that completes the right-handed triad.}

\begin{figure*}[!htb]
   \centering
   \includegraphics[width=\hsize]{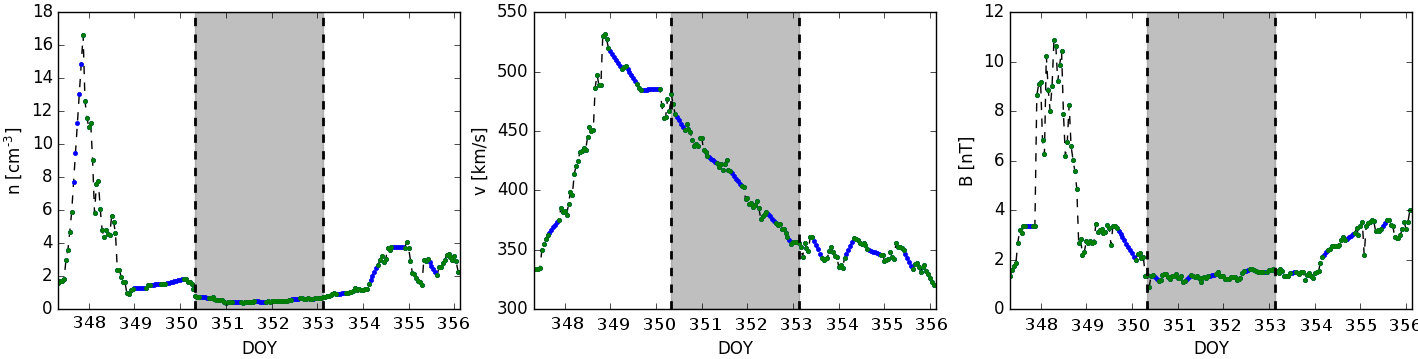}
      \caption{Hourly time series of an ICME detected by VOY1 on 17/12/1977. Green dots represent data points, while blue dots represent data gaps that have been closed using linear interpolation. The ICME itself is confined in the greyed out region. {The increase in V, n and B seen before the ICME is due to a compression region generated by the high speed solar wind pushing into the slow wind in front.}
              }
         \label{fig:17-12-1977_interpol}
   \end{figure*}

Time series from both VOY1 and VOY2 suffer from severe data gaps. If a gap is present in a relevant part of the data but it does not prevent a satisfactory identification of the ICME and its boundaries, the gap is closed by linearly interpolating between its edges. An example ICME from 1977, where this process is applied, can be seen in Figure \ref{fig:17-12-1977_interpol}. This ICME is classified as event B further in this paper.

\section{Variation of ICME properties with radial distance from the Sun} \label{AllICMEs}

\begin{figure}[!h]
   \centering
   \includegraphics[width=0.8\hsize]{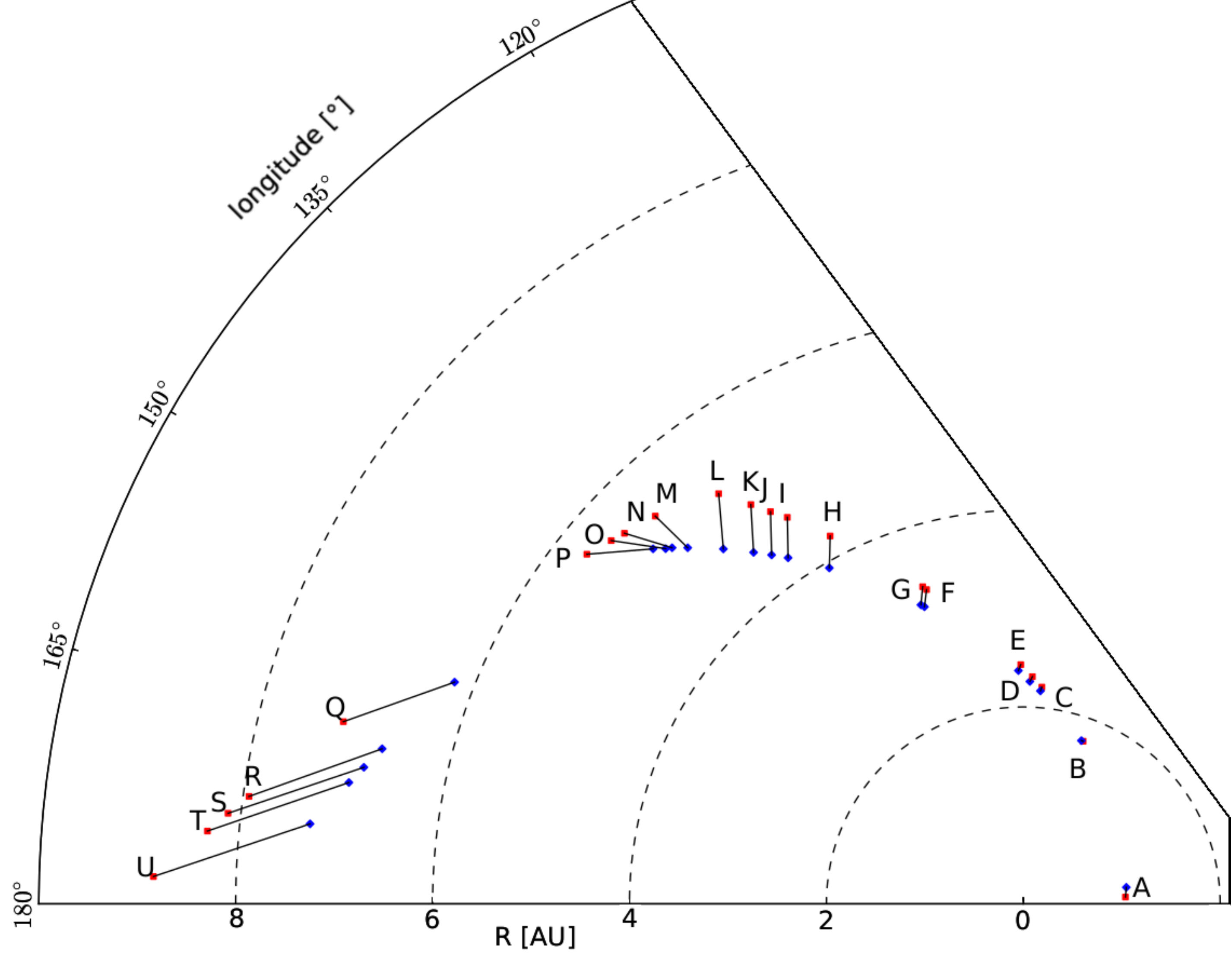}
      \caption{Locations of Voyager 1 (red) and Voyager 2 (blue) at the moment of the ICME measurements at that spacecraft in Heliographic Inertial Coordinates.
              }
         \label{fig:coordinates}
   \end{figure}

{{In this section, we study how the differences in properties of ICMEs detected by two Voyager spacecraft evolve with distance from the Sun. Typically previous studies published in the literature have focused on how the properties of different ICMEs evolve with distance. In some cases we describe how a particular ICME evolves between the two spacecraft.}} The plasma and magnetic field data sets were investigated by eye and using Equation \ref{eqTexp}. Figure \ref{fig:coordinates} shows the locations of the Voyager spacecraft when the ICMEs were detected {(both spacecraft were at the same latitudes for the period under study, between approximately -5$^{\circ}$ and 5$^{\circ}$ in heliographic latitude). Normally an ICME would be seen first by Voyager 2 (closer to the Sun) and then by Voyager 1. Nonetheless, up to event F, the order of detection by the two Voyagers can vary. This is due to the spacecraft being close to each other, producing two effects: one spacecraft overtaking the other, and ambiguities in the identification of the ICME boundaries. } Event E is such a case, as shown in Figure~\ref{fig:eventE}. {One final effect is related to intrinsic longitudinal distortions in the ICMEs. The longitudinal separation is very small close to the Sun, but this effect may increase with radial distance (see Figure~\ref{fig:coordinates}).}

\begin{figure*}[!h]
   \centering
   \includegraphics[width=0.495\hsize]{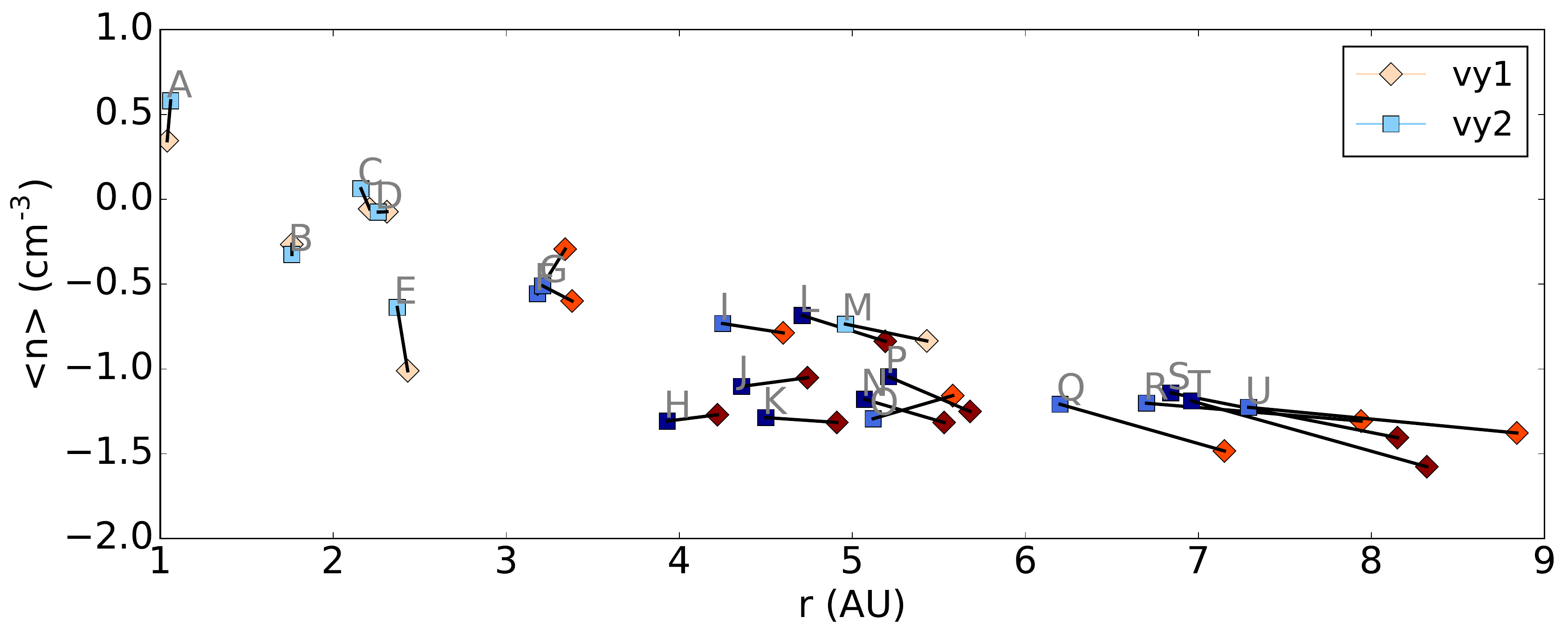}
   \includegraphics[width=0.495\hsize]{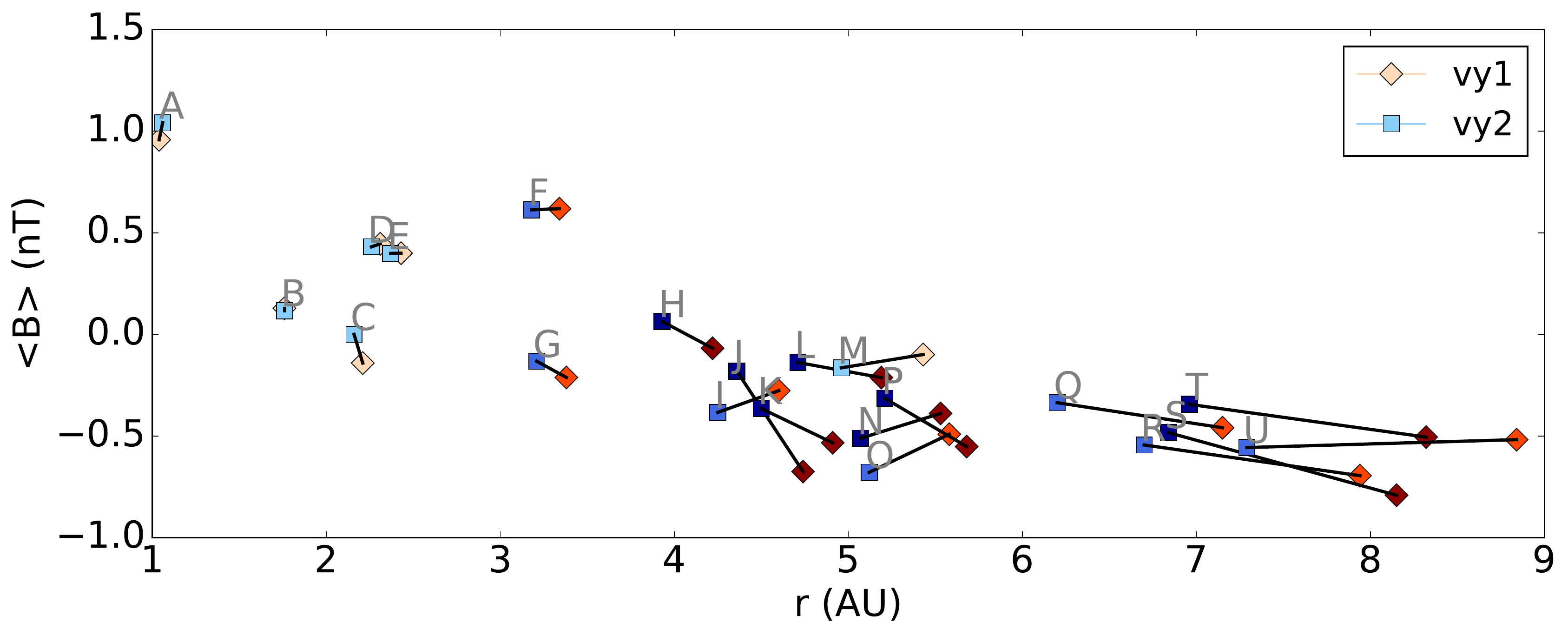}
   \includegraphics[width=0.495\hsize]{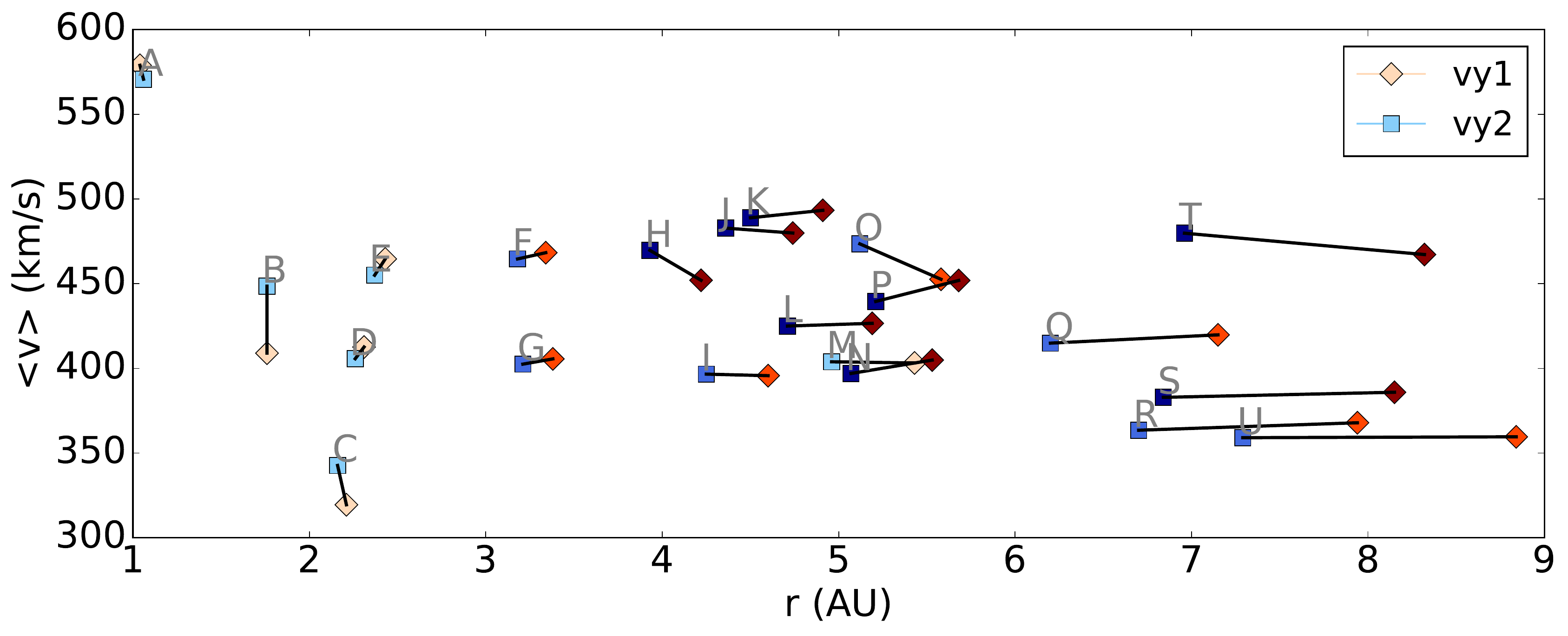}
   \includegraphics[width=0.495\hsize]{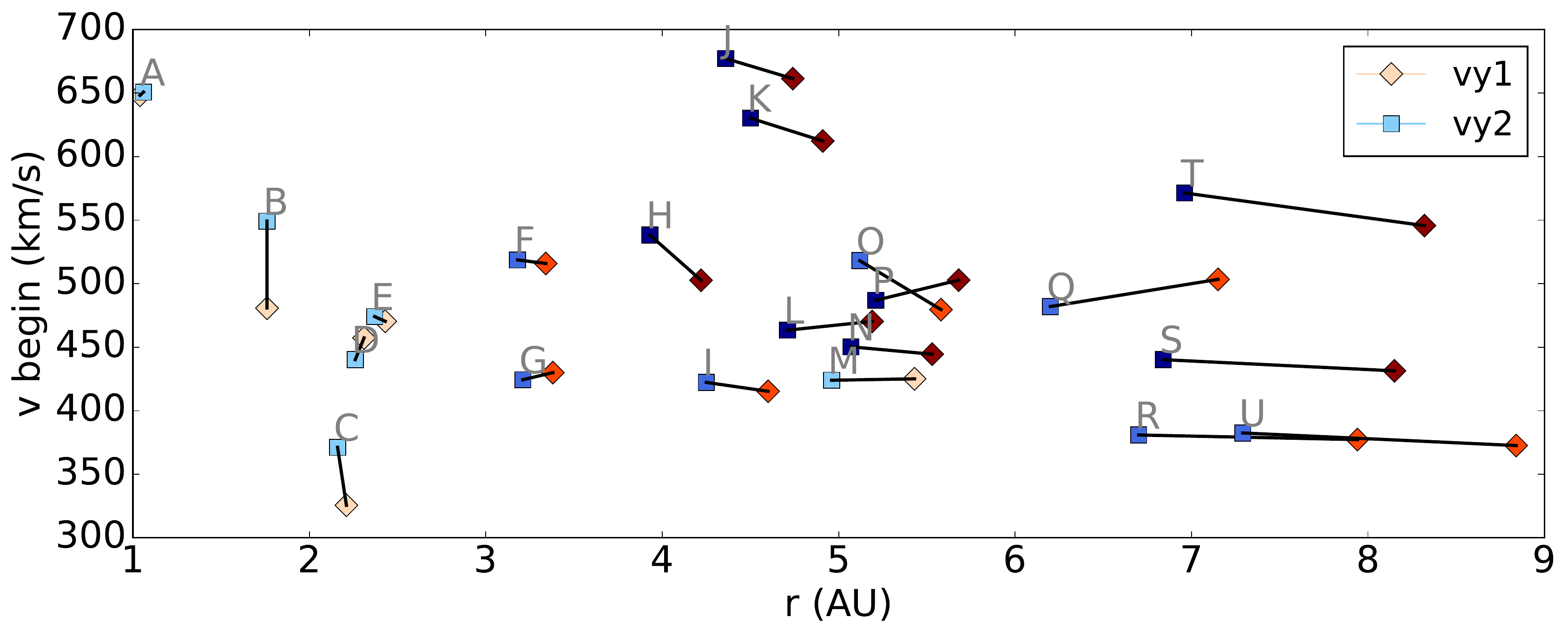}
   \includegraphics[width=0.495\hsize]{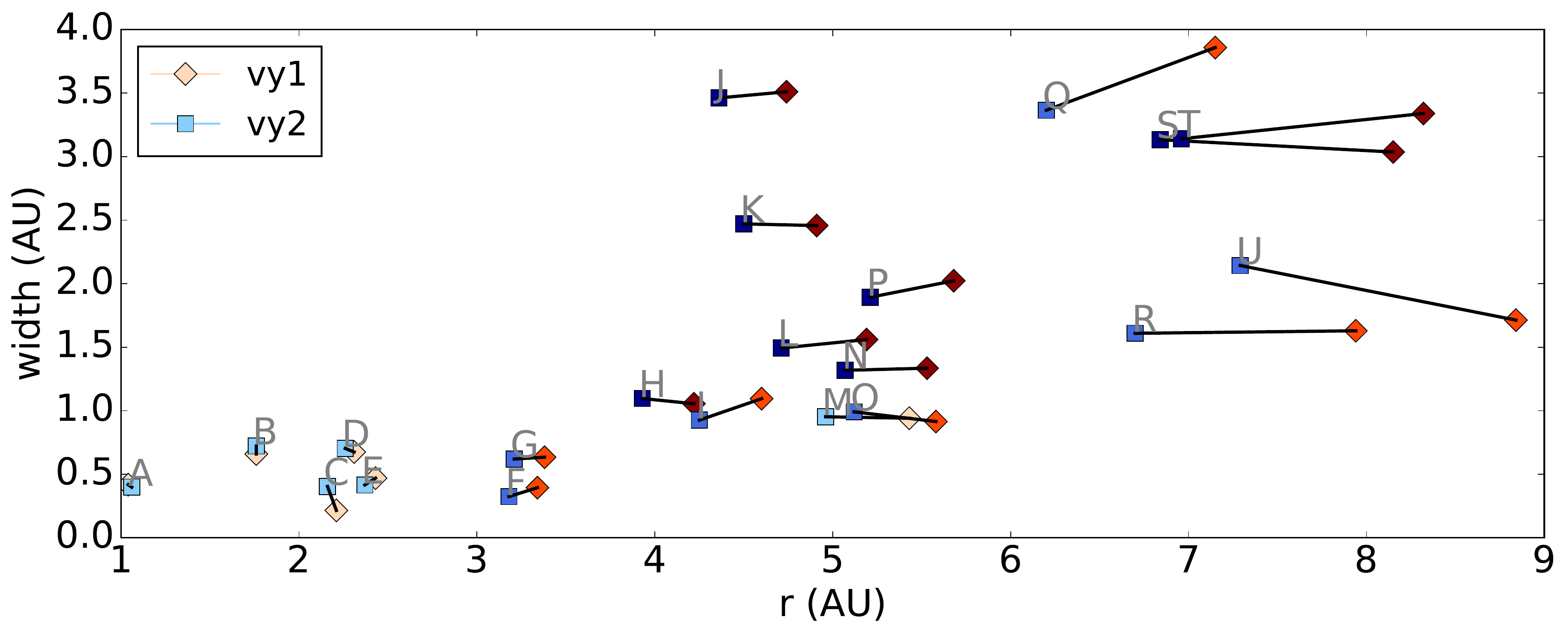}
      \caption{Top row: Logarithm of the average ICME number density and average ICME magnetic field strength. Middle row: Average ICME velocity and ICME front velocity. Bottom row: ICME width. Orange diamonds depict VOY1 events while blue squares depict VOY2 events. The darker the colour of the event, the more confidence there is that the observation is reliable, due to the presence of more ICME signatures or the lack of data gaps. Same events detected by both spacecraft are connected by a black line.
              }
         \label{fig:voy_comp}
   \end{figure*}

The upper left panel of Figure \ref{fig:voy_comp} shows the logarithm of the average number density over the total length of the ICME. Each event has a black line connecting its measurement by Voyager 1 to its measurement by Voyager 2. The first event, measured close to 1~AU, has an average number density of approximately 3~$\mathrm{cm^{-3}}$  \citep[comparable with the {typical} values found by][for magnetic clouds at 1 AU]{Rodriguez2016}. {For the events occurring after,} there is a clear downward trend until the last event where Voyager 1 was close to 9 AU and the average number density was approximately 0.03~$\mathrm{cm^{-3}}$. These values are comparable with previously found numbers \citep{Leitner2007, Gulisano2012}. ICMEs are normally expanding as they travel out from the Sun. {This is caused mostly by the decrease of external pressure \citep{Demoulin2009}, as the solar wind surrounding the ICME becomes less dense as the distance from the Sun increases. A second factor is the internal expansion of the ICME, due to the internal pressure acting on the structure. This expansion is observed for the ICMEs studied here. We see in the upper left panel of Figure \ref{fig:voy_comp} that the ICMEs detected further away from the Sun tend to have lower density than those observed closer to the Sun. The same effect can be seen when a ICME is observed by both Voyager spacecraft, given that the separation between the two spacecraft is significant (normally found in those cases beyond 5 AU).} It is also possible that the ICMEs become compressed due to interaction with other structures in the solar wind, which would increase their average density. {This can happen if the ICME is being compressed by another ICME or by a high speed solar wind stream \citep[this was found to be a common situation by][]{Rodriguez2016}. Another possibility is that one satellite is detecting high density fine structures while the other does not. The differences could also be caused by ambiguities in the determination of ICME borders}. 

The upper right panel of Figure~\ref{fig:voy_comp} shows the average magnetic field strength. {There is a clear trend of magnetic field magnitude inside ICMEs to decrease when radial distance from the Sun increases. } The figure shows the same trend as the number density, created by expansion, with event A having approximately 10~nT and event U having approximately 0.3~nT \citep[again comparable to values found in][]{Leitner2007,Rodriguez2016,Gulisano2012}. Apart from expansion, magnetic clouds can experience erosion during their propagation, stripping the ICME from magnetic field lines and causing a lower average magnetic field \citep{Dasso2006,Ruffenach2015}. \citet{Hosteaux2021} showed that, in case of an inverse ICME {(i.e.\ a CME with the same magnetic polarity as the background coronal magnetic field)}, the opposite occurs and magnetic field lines are added to the ICME, increasing the average magnetic field strength. The few events that do not show a decreasing $\langle B \rangle$ might be caused by magnetic reconnection adding magnetic field lines in a similar process as the inverse ICMEs in {the \citet{Hosteaux2021} study}. {As a matter of fact, \citet{Hosteaux2021} showed that, for inverse CMEs, continuous magnetic reconnection with the surrounding magnetic field at the rear of the separatrix (determining the CME cloud) adds more magnetic flux to the magnetic cloud.  Another possibility could be similar reasons as given in the previous paragraph for the average density not decreasing.}

The left panel of the middle row of Figure \ref{fig:voy_comp} shows the average plasma speed over the magnetic cloud. In contrast to the number density and the magnetic field strength, the average plasma speed does not show a downward trend from 1 to 10 AU, as expected. {ICMEs experience aerodynamic drag forces acting upon them \citep[e.g.][]{Cargill2004}. These forces are proportional to the velocity difference between the ICME and the solar wind. At the distances these events are observed, the effects of the drag forces on the ICMEs have in most cases already resulted in the ICME taking approximately the same velocity as the solar wind they are propagating in. Thus, little acceleration or deceleration of the ICMEs occurs.} {For most events the speeds} fall in a range of [350,500]~km/s and stay approximately constant. The right panel of the middle row of Figure \ref{fig:voy_comp} shows the ICME leading edge speed (which is close to the highest speed of the event in most cases). The range is much larger for this panel, with front speeds being as low as \textasciitilde350~km/s and going up to almost 700~km/s. {This is because for this case we are not using an average speed taken over many hours, as it was the case for the average speed {shown in the left panel of the middle row of Figure \ref{fig:voy_comp}}, so the variation is larger here. The leading edge speed is composed by the bulk velocity of the CME plus the expansion speed of the CME.}

The panel in the lower row of Figure \ref{fig:voy_comp} shows the ICME width. The width of the ICME is calculated by multiplying the average plasma speed with its time duration:

\begin{equation} \label{eq:v_avg}
\mathrm{width} = \langle v \rangle \Delta t .
\end{equation}
{The general trend is to observe an increase in ICME width with respect to radial distance, due to expansion. The deviations in this trend are most likely being caused by different spacecraft trajectories within the ICMEs. An ICME encountered in its flank by the spacecraft will turn into a different calculated width as a central encounter, even for the same ICME.} The widths go up to 3.5-4~AU for the events observed furthest from the Sun. These results are comparable with those found by \cite{Bothmer1998}, \cite{Leitner2007} and \cite{Gulisano2012}, and show again the expanding nature of ICMEs as they travel outwards in the solar system. The ICME width stays approximately constant between the two Voyager spacecraft for most events or increases slightly, meaning that the ICME expansion can still occur at distances larger than 5 AU.

\section{Shock properties} \label{shock}

In this section, an analysis on the properties of the shocks that are captured by both VOY1 and VOY2 is presented. All MHD shocks satisfy the Rankine-Hugoniot (RH) equations, which describe the relations between the up (denoted by subscript u) and downstream (denoted by subscript d) shock conditions. One of these equations represents the conservation of mass across the shock, which can be described as:

\begin{equation}
\label{eq:RH_masscons}
\rho_u \vec{v'}_{nu} = \rho_d \vec{v'}_{nd}\hspace{0.3cm},
\end{equation}

where the subscript n denotes components in the direction of the shock normal. The {variables with a prime symbol} represents variables in the reference frame of the shock. The other RH condition that is used in the following analysis is:

\begin{equation}
\label{eq:RH_mag}
    B_{nu} = B_{nd},
\end{equation}

meaning that the normal component of the magnetic field stays constant over the shock.

 Calculations are performed using the up-and downstream plasma and magnetic field quantities. To reduce the effect of scatter in data, a period {of approximately 10 minutes} before and after the shock is selected. The average of each variable is then taken over this interval and used in the subsequent analysis. 
 Using the Minimum Variance Analysis (MVA) \citep{Sonnerup1967} we are able to find the direction of the normal to the shock by identifying the direction of the magnetic field measurements with minimum variance. This is done by finding the eigenvalues of the magnetic variance matrix, defined as:

\begin{equation}
M_{i,j} = \left( \langle B_i B_j \rangle - \langle B_i \rangle \langle B_j \rangle \right) ,
\end{equation}

where $B_i$ and $B_j$ are the components of a single magnetic field measurement. This matrix has three eigenvectors $\vec{x_i}$ and three corresponding eigenvalues $\lambda_i$. The direction of minimum variance is then given by the eigenvector $\vec{x_3}$ corresponding to the smallest eigenvalue $\lambda_3$, which is also an estimator for the vector normal component to the shock due to the RH condition given in Equation~\ref{eq:RH_mag}. $\lambda_3$ represents the variance of the magnetic field component along the normal. The other two eigenvectors, $\vec{x_1}$ and $\vec{x}_{2}$, signify the direction of maximum and intermediate variance, respectively, and are tangential to the shock. There are two conditions that determine that the MVA results are adequate \citep{Siscoe1972, Lepping1980}:

\begin{equation}
    \frac{\lambda_2}{\lambda_3} > 2 ,
\end{equation}
\begin{equation}
    \angle(\vec{B}_u,\vec{B}_n) > 30\degree
\end{equation}

The ratio of the smallest and the intermediate eigenvalue acts as a reliability factor for the result, the higher the ratio the more trustworthy the result is. The second condition represents that directional changes of $\vec{B}$ should be larger than 30\degree. The boundaries of the up-and downstream regions in the magnetic field time series are selected as to maximise the ratio of eigenvalues. From the set of ICMEs shown in table \ref{table:CMEs}, not all events showed a fully formed shock and only the ICMEs that showed no data gap at the location of the shock were included in the shock analysis. 

 Using the RH condition in Equation~\ref{eq:RH_masscons} and inserting the shock normal found using MVA to get the up- and downstream plasma velocity in the reference frame of the shock, it is possible to obtain an expression for the shock speed $v_{sh}$ \citep{Kivelson1995}:

\begin{equation}
    \rho_u(\vec{v}_u \cdot \vec{n} - v_{sh}) = \rho_d(\vec{v}_d \cdot \vec{n} - v_{sh}) \hspace{0.3cm},
\end{equation}

\begin{equation}
    v_{sh} = \frac{\rho_u \vec{v}_u - \rho_d \vec{v}_d}{\rho_u-\rho_d} \cdot \vec{n} \hspace{0.3cm}.
\end{equation}

A well known representation of the strength of an ICME-driven shock is the Alfv\'en Mach number, given by:

\begin{equation}
    M_a = (v_{sh}-v_{nu})/v_{Au},
\end{equation}

where $v_{Au}$ is the upstream Alfv\'en speed $v_{Au}=|B_u|/\sqrt{\mu_0 \rho_u}$. The Alfv\'en Mach number indicates the intensity of the shock. At 1~AU, the majority of ICME driven shocks have a Mach number close to two, though values $>$ 8 can also be reached \citep{Makela2011}.

Another important shock characteristic is the density compression ratio $\rho_d/\rho_u$. ICME shocks can be a location where Solar Energetic Particles (SEP) are being generated over a large heliolongitudinal range \citep{Cliver1995}, and the intensity spectrum depends on the density compression ratio according to the diffusive shock acceleration theory \citep{Desai2016}.

Knowing the position of the shock, it is also possible to determine the ICME's stand-off distance. This is calculated by taking the time difference of the crossing of the shock and the crossing of the front of the ICME with the spacecraft and multiplying this with the speed of the front of the ICME:

\begin{equation}
    \Delta = (t_{front} - t_{shock}) \times |v_{front}|
\end{equation}

{Only one event (S) was found in the data set with sufficient data coverage both upstream  and downstream of the shock to determine shock parameters at both Voyager locations.}

\Needspace{100pt}

\begin{table}
\begin{center}
\begin{footnotesize}
\begin{tabular}{c|c|c}
 & VOY1 & VOY 2 \\ \cline{1-3}
r [AU] & 8.06 & 6.69 \\
N & 0.99 0.04 0.14 & 0.93 -0.02 0.36 \\
$\rho_d/\rho_u$ & 2.29 & 4.11 \\
$\mathrm{v_{sh}}$ [km/s] & 457$\pm$17 & 517$\pm$15 \\
$\mathrm{M_A}$ & 2.58$\pm$0.37 & 2.31$\pm$0.21  \\
$\Delta [\mathrm{R_{\odot}}]$ & 517.54 & 441.47
\end{tabular}
\caption{Table of shock properties for event S}\label{tab:shockS}
\end{footnotesize}
\end{center}
\end{table}

The shock driven by the ICME in event S was detected by VOY1 at 03:13:40 UT on 29 May 1980 and by VOY2 at 17:14:45 UT on 24 May 1980. This means that the time separation between the two detections is 4 days 12 hours and 58 minutes, with their radial distance separation being approximately 1.37~AU. Over this stretch, the stand-off distance has increased by 17\%. Table \ref{tab:shockS} shows that the shock normal lies close to the direction both spacecraft are propagating in (N in the table). Though the upstream Alfv\'en speed is much higher for VOY2 than for VOY1 (77.27~km/s and 43.56~km/s, respectively), the shock speed is much higher as well while the upstream solar wind speed is comparable, resulting in {$\mathrm{M_A}$ numbers with a slight increase, but this increase is within error bounds of each other. The errors on the shock speed and Mach number were determined by taking the error on the minimum variance direction as 10$^{\circ}$ \citep{Burlaga1982}}. This data would claim that the shock is getting weaker going from VOY2 to VOY1 as it slows down and the compression ratio becomes lower. However, the result that $\Delta$ increases while $M_A$ {stays constant or increases slightly is surprising (although the values are within error bars)}, since they are supposed to be anti-correlated. {Taking the relation between the standoff distance to radius of curvature and the Mach number from \citet{Russell2002}, a strong shock has a shorter standoff distance than a weak shock for a CME of similar size}. A possible reason for the behaviour observed here is that $\Delta$ is expressed in R$_\odot$ and not as a function of the size of the ICME.

   \begin{figure*}[!h]
   \centering
   \includegraphics[width=0.495\hsize]{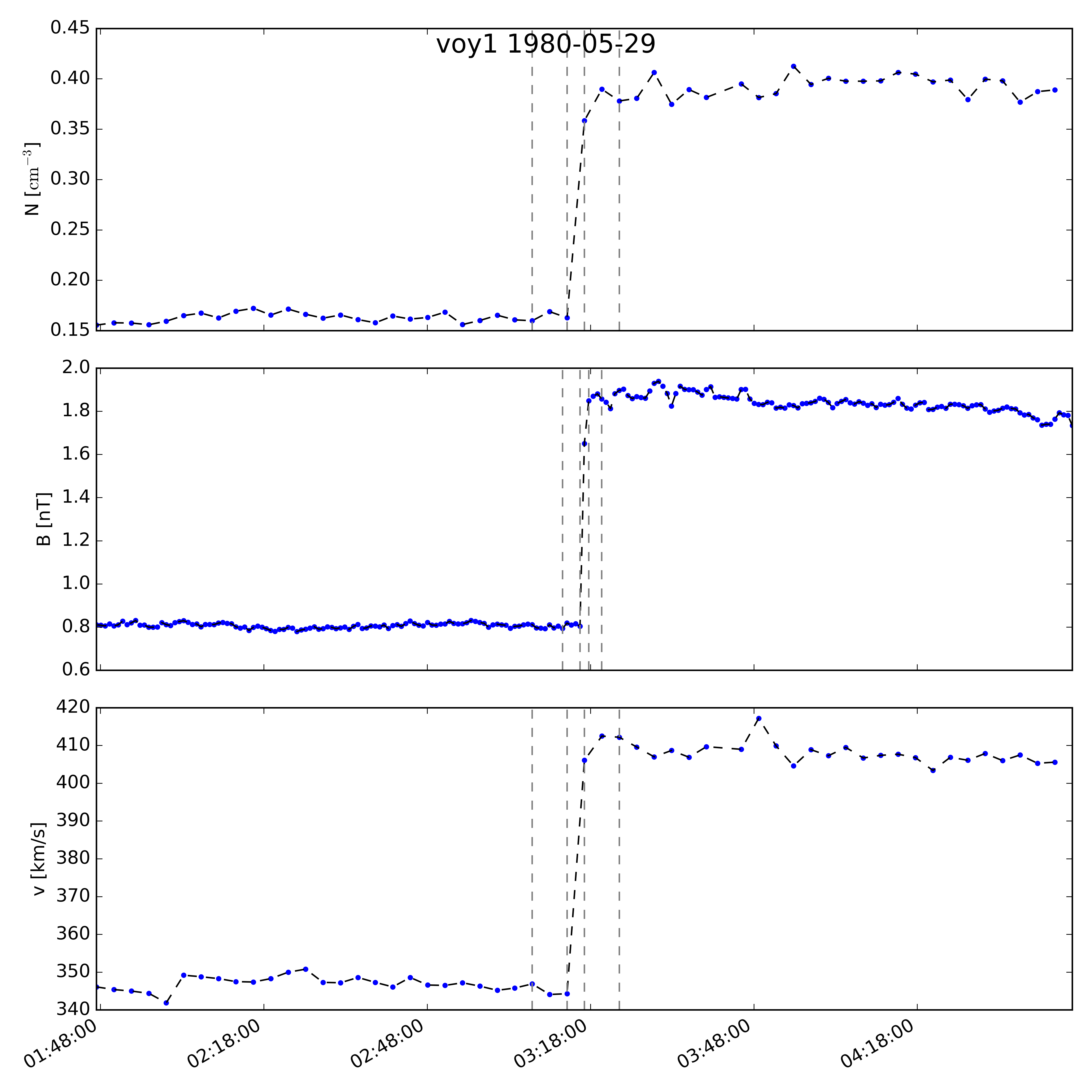}
   \includegraphics[width=0.495\hsize]{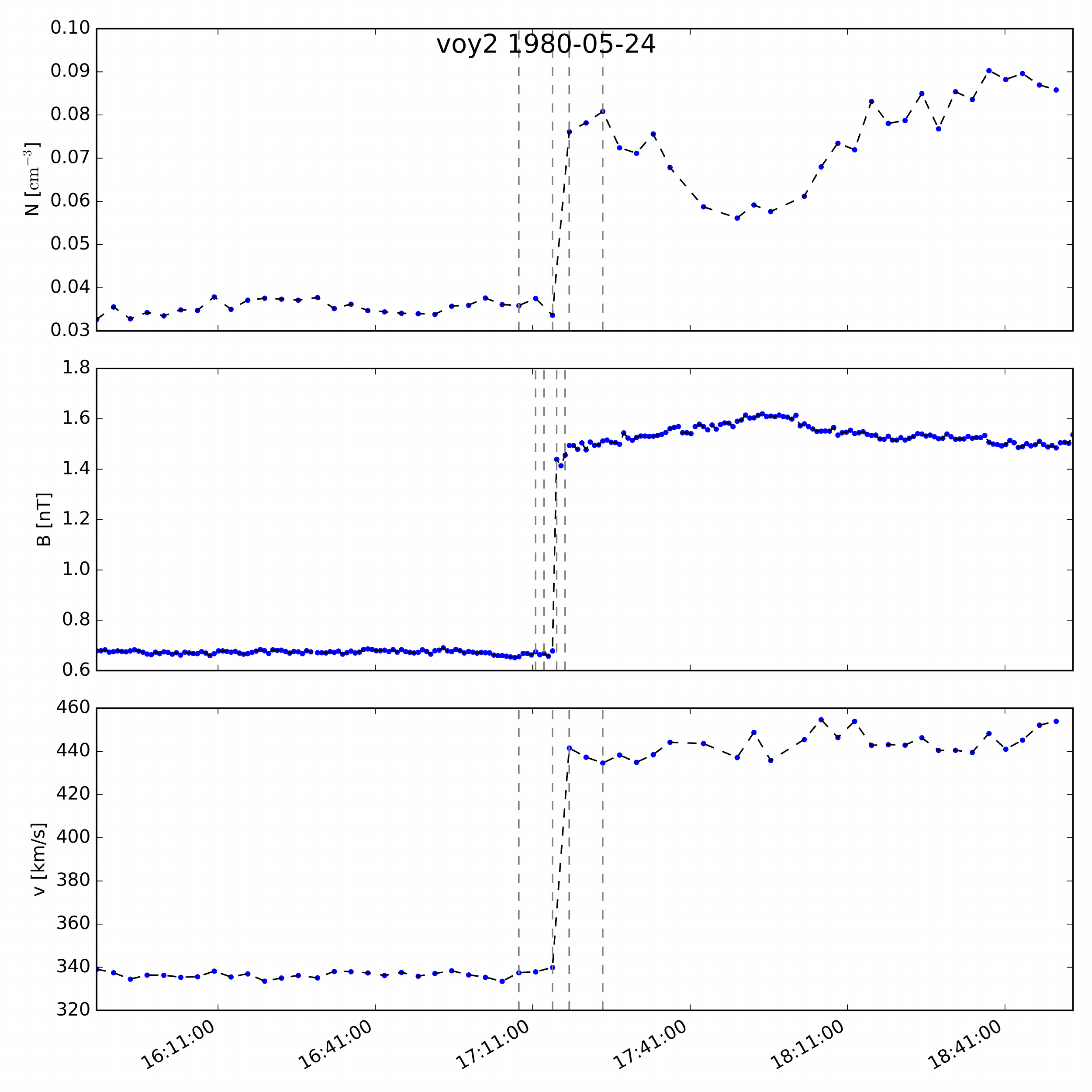}
      \caption{High resolution VOY1 and VOY2 time series of the shock that precedes event~S.
              }
         \label{fig:shockS}
   \end{figure*}

\begin{figure}[!h]
   \centering
   \includegraphics[width=\hsize]{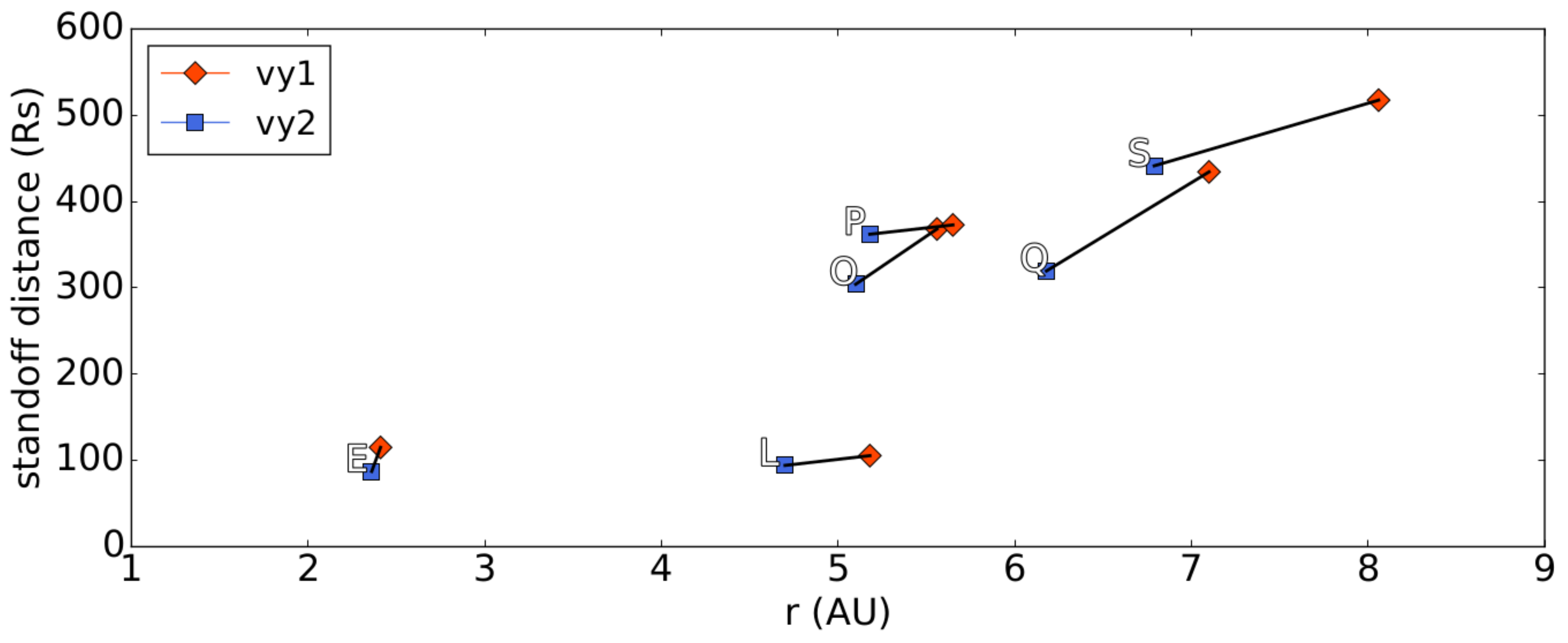}
   \caption{Plot of the standoff distance for all events for which a shock front was approximately located. Common events are connected by a black line. Each shock is represented by the same letter as the events in table~\ref{table:CMEs}.}\label{fig:standoff_voy}
   \end{figure}

While the aforementioned shock was the only one deemed satisfactory for a determination of the normal via MVA, {five} events with possible shock fronts were found among the common events dataset. These events either had a small gap at the location of the shock or were not developed enough for accurate MVA. However, since their location is approximately known, the stand-off distance of these events can be estimated. Figure~\ref{fig:standoff_voy} shows the standoff distance for each of these events. Event E, which has the closest to Earth recorded shock at \textasciitilde2.4~AU, has a stand-off distance of approximately 115~R$_\odot$. The other events all have a stand-off distance in the range 300-550~R$_\odot$, except for event L which has a stand-off distance similar to that of event E. {Moreover, Figure 6 shows that for each event the stand-off distance is larger for the Voyager that detected the event at larger heliospheric distance. This could be the result of the shocks becoming weaker during the ICME propagation.} Another possible reason is that the shock speed is higher than the ICME speed. This might be due to changes in the upstream conditions, as the shock needs to move to adapt to new values of the upstream $\mathrm{M_A}$, or the ICME might still be expanding, as $\Delta$ depends on both $\mathrm{M_A}$ and the size of the magnetic cloud. If the magnetic cloud gets bigger, $\Delta$ increases as well and the shock moves away from the magnetic cloud. {\citet{Farris1994} determined the standoff distance of a bow shock which is influenced by the size/shape of the body, magnetic field orientation, magnetosonic Mach number, and plasma beta. These authors noticed that it is most appropriate to compare the standoff distance of the bow shock to the radius of curvature of the obstacle, as opposed to the distance from the focus of the object to the nose. 
We would like to point out that the non-radial separation was not taken into account. Nevertheless, the time frame of the observations (1977 to 1980) was chosen so that the non-radial separation between the Voyagers is minimal.} 

\section{Conclusions} \label{Conclusions}

This work discusses the observations of ICMEs that were detected by both the Voyager 1 and Voyager 2 spacecraft between November 1977 and December 1980, { comprising distances between 1 and 10 AU. Using a number of different observational characteristics, primarily the low ratio of measured to expected proton temperature ($\mathrm{T/T_{exp}}$), 21 common ICMEs were listed. This is to our knowledge the largest {multi-spacecraft} list of ICMEs studied by the two Voyagers.  \citet{Burlaga1982} identified five ICMEs that were observed also by these spacecraft (plus Helios and IMP 8), those events were observed at distances up to 4 AU form the Sun. \cite{Wang2004} provides an extensive list of ICMEs observed by Voyager 2 up to 30 AU. Other studies used other datasets such as those provided by Ulysses up to 5 AU \citep{Gulisano2012} or Helios within 1 AU form the Sun \citep{Leitner2007}}

{The ICMEs studied here were expanding (even the events seen furthest from the Sun,{ in agreement with previous studies}), the average number density and the average magnetic field strength was observed to decrease with respect to radial distance from the Sun. The radial width of the ICMEs was also clearly increasing for ICMEs observed further away. The average ICME speed and the ICME leading edge speed stay approximately constant, meaning that processes that infer acceleration (magnetic reconnection or drag forces) do not influence the ICME significantly at these distances. These results are in line with previous studies \citep{Burlaga1982,Wang2004,Leitner2007, Gulisano2012}}. {They also provide a continuation to similar multi-spacecraft results observed within 1 AU \citep[e.g.][]{Good2019}.}

In some events, the density time series show a rise of the measured density immediately following the magnetic cloud. Such cases include events J, K, and L. \citet{Rodriguez2016} studied the internal structure of the magnetic clouds of 63 events that were observed by the ACE spacecraft at 1~AU. One of their results is that a sudden density increase at the trailing region of a magnetic cloud is most likely due to compression as a result of a high-speed stream, originating from a coronal hole, following and overtaking the CME. In some cases, the density increase in the trailing region of the magnetic cloud exhibits distinct peaks, such as event Q. The aforementioned study suggests that this is most likely caused by a combination of an overtaking high-speed stream and other processes that lead to density enhancements, such as expansion or intrinsic magnetic cloud processes. The presence of a density enhancements in the trailing regions of an ICME may lead to increased geoeffectiveness of the eruption { \citep{Reeves1998,Kilpua2012,fenrich1998}}. {In the present study, we can confirm that these effects found previously mostly in events observed at 1 AU, can also be detected at larger distances from the Sun.}

If the ICMEs were found to be driving a shock, the shock was investigated using MVA. Unfortunately, due to data gaps and many shocks not being fully developed, only one shock was suitable. MVA showed that the shock velocity decreased significantly over the distance between the two spacecraft, but due to the Alfv\'en speed decreasing as well, the Alfv\'en mach number increased slightly, {though staying within error bounds of both spacecraft}. Six other events were identified to have shocks as well, but were not suited for MVA. For these events it was found that standoff distances reach values over 400~R$\odot$. {The stand-off distances clearly increase with distance from the Sun}.

The events considered in this study were limited to a radial distance of $<$10 AU. The Voyagers went farther than that, their data can be used to study ICMEs at even larger distances. Nevertheless, the spacecraft angular separation becomes larger with increasing distance, lowering the chances of finding common events. At large distances from the Sun, most of the ICME signatures that were used in this work (such as low proton temperature and magnetic field rotation) may become undetectable due to expansion and loss of internal pressure of the ICME. It becomes more difficult to differentiate the ICME from the background solar wind. Nonetheless, \citet{Paulerana2001} were able to track an ICME from 5~AU to 58~AU, with Ulysses and Voyager 2 data, using alpha particle enhancements. Studying how ICMEs evolve at large distances is important to increase our knowledge of ICME processes. This is relevant with regard to their impact at 1~AU, but also because they can create significant activity at large distances. \citet{Dunn2016} reported auroral enhancements at Jupiter (at approximately 9.5~AU). In this respect, the combination of these rare observations with advanced 3D MHD heliospheric models, such as EUHFORIA \citep{Pomoell2018} is of great relevance for heliospheric sciences. EUHFORIA currently reaches up to 2 AU, but there are plans to extend its coverage and be used as an advanced tool for simulations at large distances from the Sun.

{The methods presented in this work can result of interest when applied to events detected by multiple spacecraft at different radial distances. The datasets used here are not straightforward to analyse, as they contain data gaps and other caveats, but we believe that they still present interesting opportunities for future studies of the inner and outer heliosphere.}

\section*{Acknowledgments}
This project (EUHFORIA 2.0) has received funding from the European Union’s Horizon 2020 research and innovation programme under grant agreement No 870405.
These results were also obtained in the framework of the projects
C14/19/089  (C1 project Internal Funds KU Leuven), G.0D07.19N  (FWO-Vlaanderen), SIDC Data Exploitation (ESA Prodex-12), and Belspo projects BR/165/A2/CCSOM and B2/191/P1/SWiM.
The simulations were carried out at the VSC – Flemish Supercomputer Centre, funded by the Hercules foundation and the Flemish Government – Department EWI.
The authors are very grateful for the invaluable contributions of Dr.\ Emmanuel Chan\'e.\\

\appendix

\section{Individual event properties extended list} \label{appendix_event_description}

In this section, all the events listed in Table~\ref{table:CMEs} are individually presented, with exception of event F which can be found in Section~\ref{data}.

\subsubsection*{\textbf{EVENT A}}

\begin{figure*}[tb]
\centering
  \centering
  \includegraphics[trim=1cm 1.5cm 1.5cm 2cm, clip=true,width=0.495\linewidth]{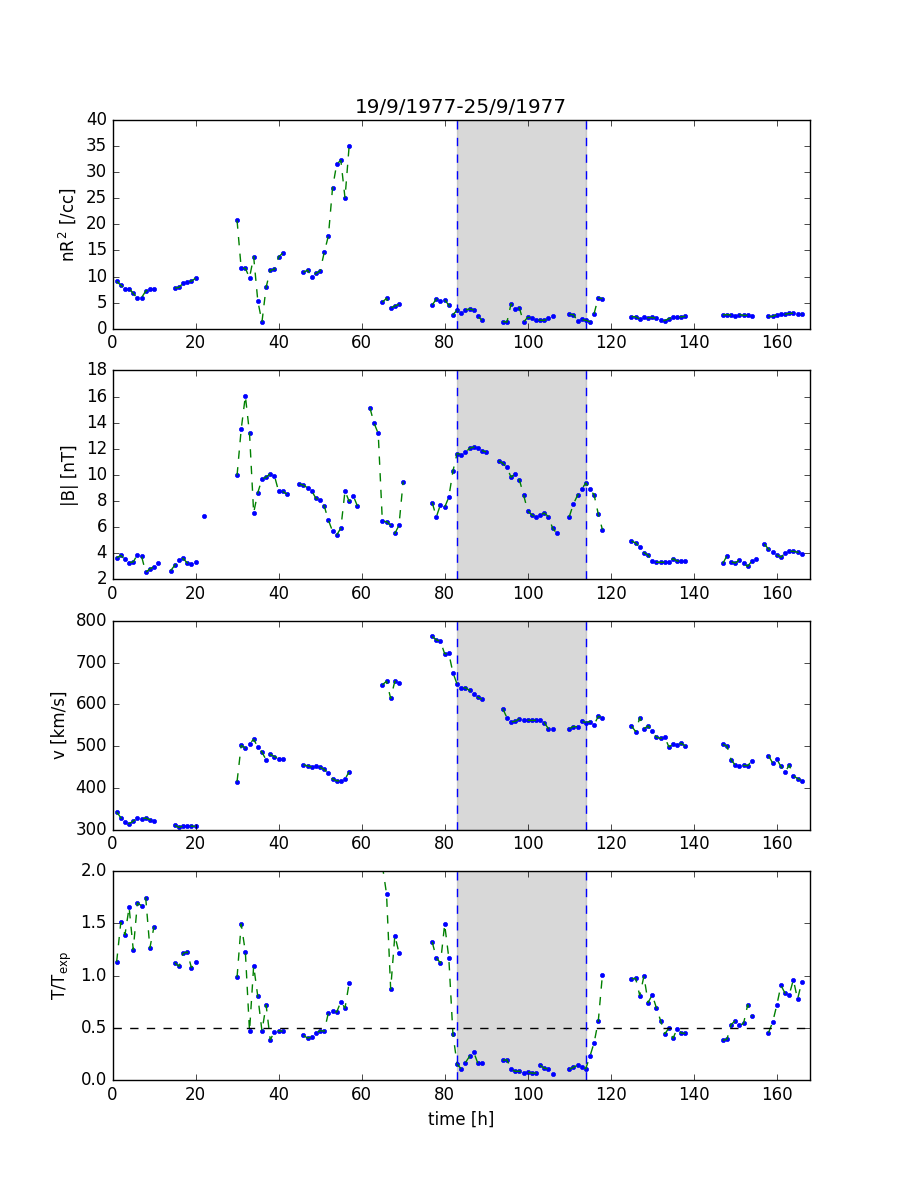} 
  \includegraphics[trim=1cm 1.5cm 1.5cm 2cm, clip=true,width=0.495\linewidth]{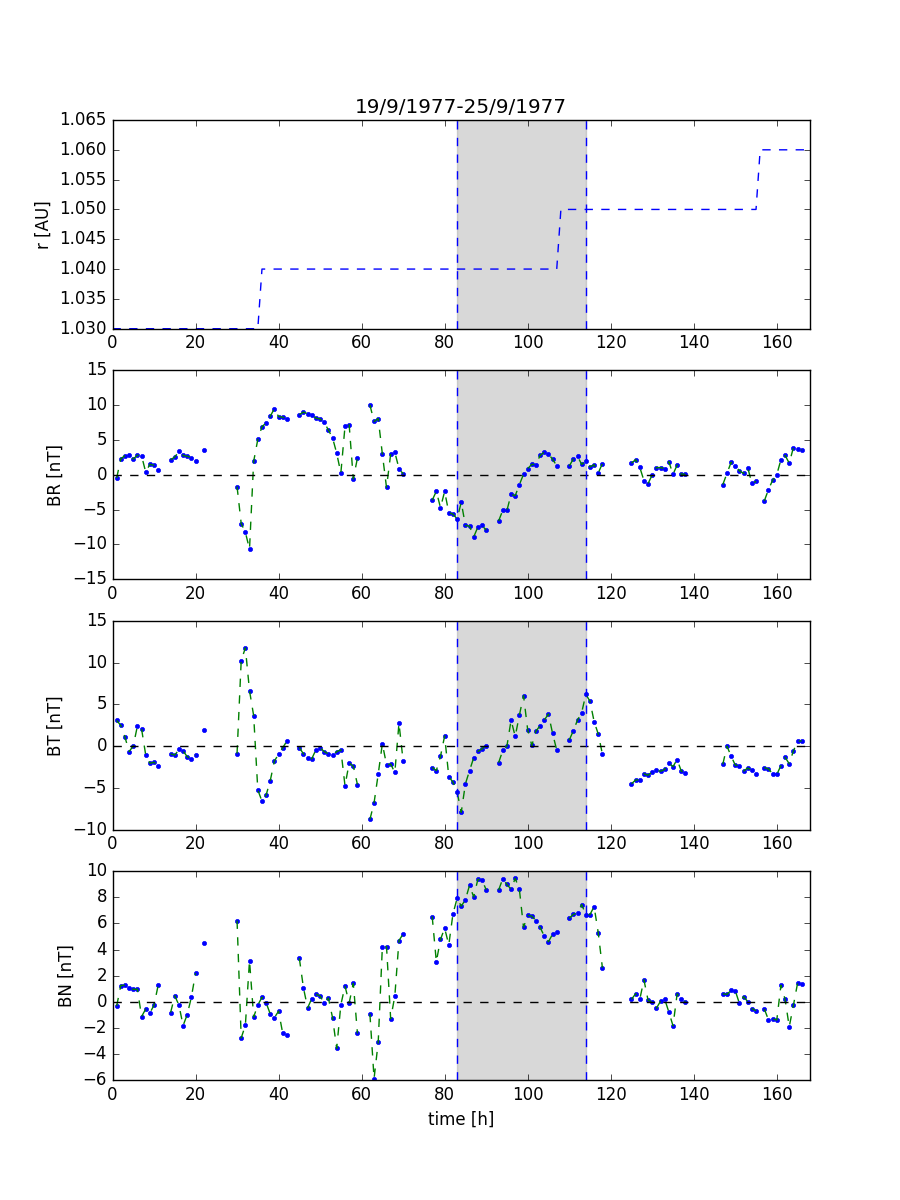} 
   \includegraphics[trim=1cm 1.5cm 1.5cm 2cm, clip=true,width=0.495\linewidth]{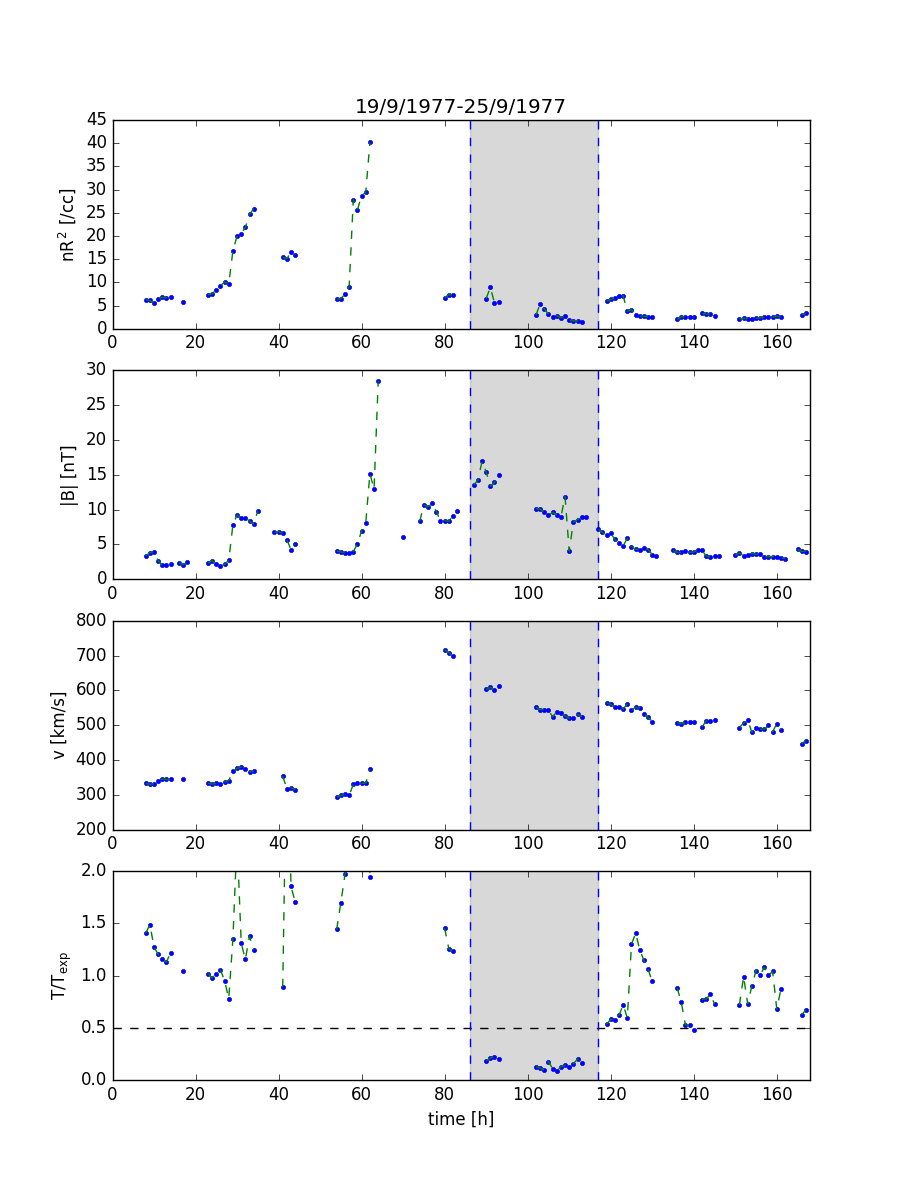}
  \includegraphics[trim=1cm 1.5cm 1.5cm 2cm, clip=true,width=0.495\linewidth]{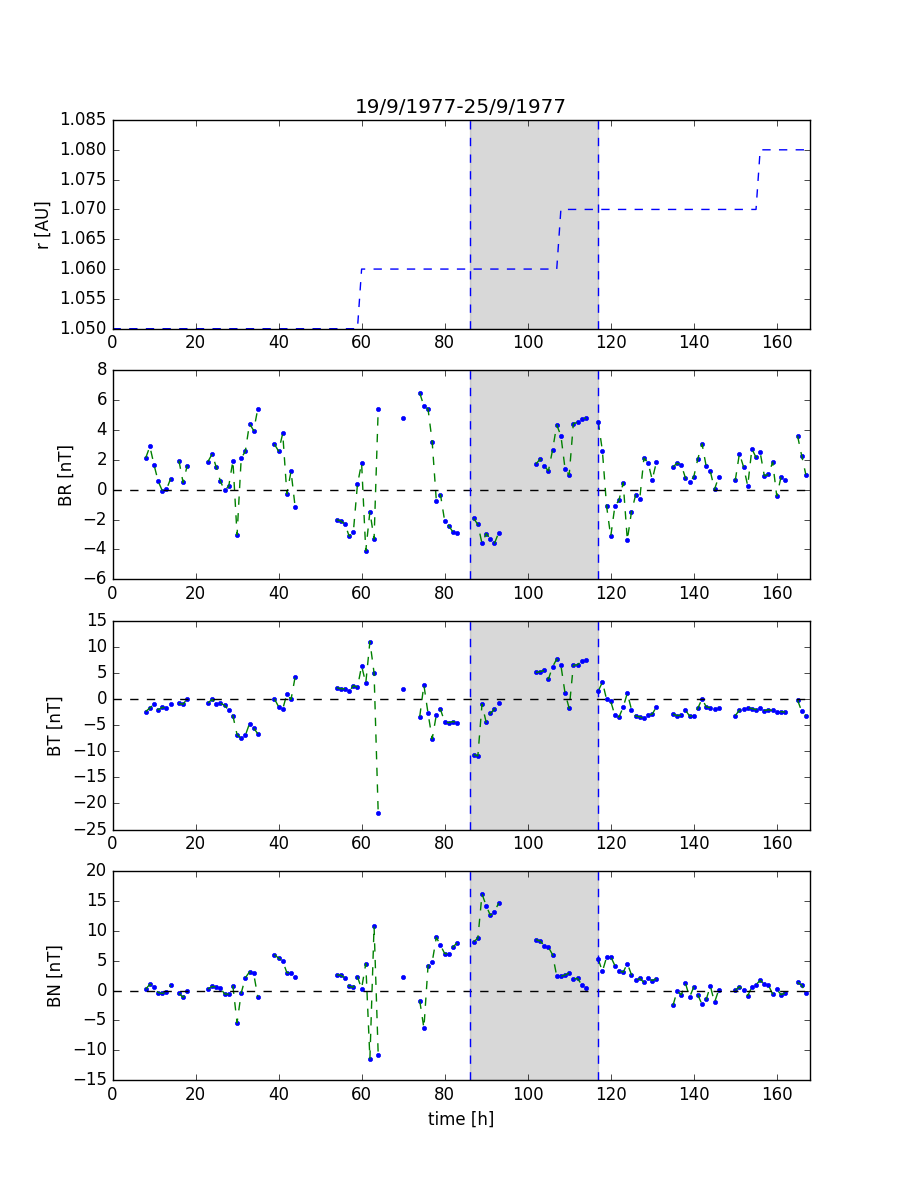}
\caption{Voyager 1 (upper row) and 2 (lower row) time series for Event A.}\label{fig:eventA}
\end{figure*}

Event A was detected by VOY1 on 23/09/1977 at 11:00 UT. with a duration of 31 hours, and by VOY2 two hours later on the same day. As Figure \ref{fig:eventA} shows, VOY2 is approximately 0.02~AU farther than VOY1. A summary of its properties can be seen in Table \ref{tab:eventA}.

\Needspace{100pt}
\begin{wraptable}{r}{0.52\linewidth}
\begin{center}
\begin{footnotesize}
\vspace{-20pt}
\begin{tabular}{c|c|c}
 & VOY1 & VOY 2 \\ \cline{1-3}
r [AU] & 1.04 & 1.06 \\
$\mathrm{N_{avg} [cm^{-3}]}$ & 2.21 & 3.8 \\
$\mathrm{B_{avg} [nT]}$ & 9.07 & 11.01 \\
$\mathrm{V_{avg} [km/s]}$ & 578.9 & 570.7  \\
$\mathrm{V_{beg} [km/s]}$ & 648.3 & 650.8 \\
width [AU]& 0.42 & 0.40 \\
\end{tabular}
\caption{Table of properties from event A}\label{tab:eventA}
\vspace{-20pt}
\end{footnotesize}
\end{center}
\end{wraptable}
 The ICME boundaries are relatively well defined in the VOY1 time series, with a clear established region of low $\mathrm{T/T_{exp}}$, a low average number density, a somewhat linearly decreasing velocity profile and a rotating magnetic field. The VOY2 time series shows the same ICME characteristics, but the data gaps make it impossible to accurately define the ICME boundaries. Figure \ref{fig:voy_comp} shows that the average magnetic field strength, the average velocity, the front velocity ($\mathrm{V_{beg}}$) and the width measured by both spacecraft are very similar, as expected since the spacecraft are still close to each other. However, there is a relatively large discrepancy in average measured number density, as it can be seen in the upper left panel of Figure \ref{fig:voy_comp}. The bottom panel of the left row of the VOY2 time series in Figure \ref{fig:eventA} shows that there are data gaps at the ICME boundaries, and that a large jump in number density is present at the second (rightmost) ICME boundary. The data gaps are closed by connecting the closest data points by linear interpolation. Thus, the number density that is considered might be higher than the true value. The magnetic field components show a rotating field, although the variability does not decrease significantly.

\subsubsection*{\textbf{EVENT B}}

\begin{figure*}[tb]
\centering
  \centering
  \includegraphics[trim=1cm 1.5cm 1.5cm 2cm, clip=true,width=0.495\linewidth]{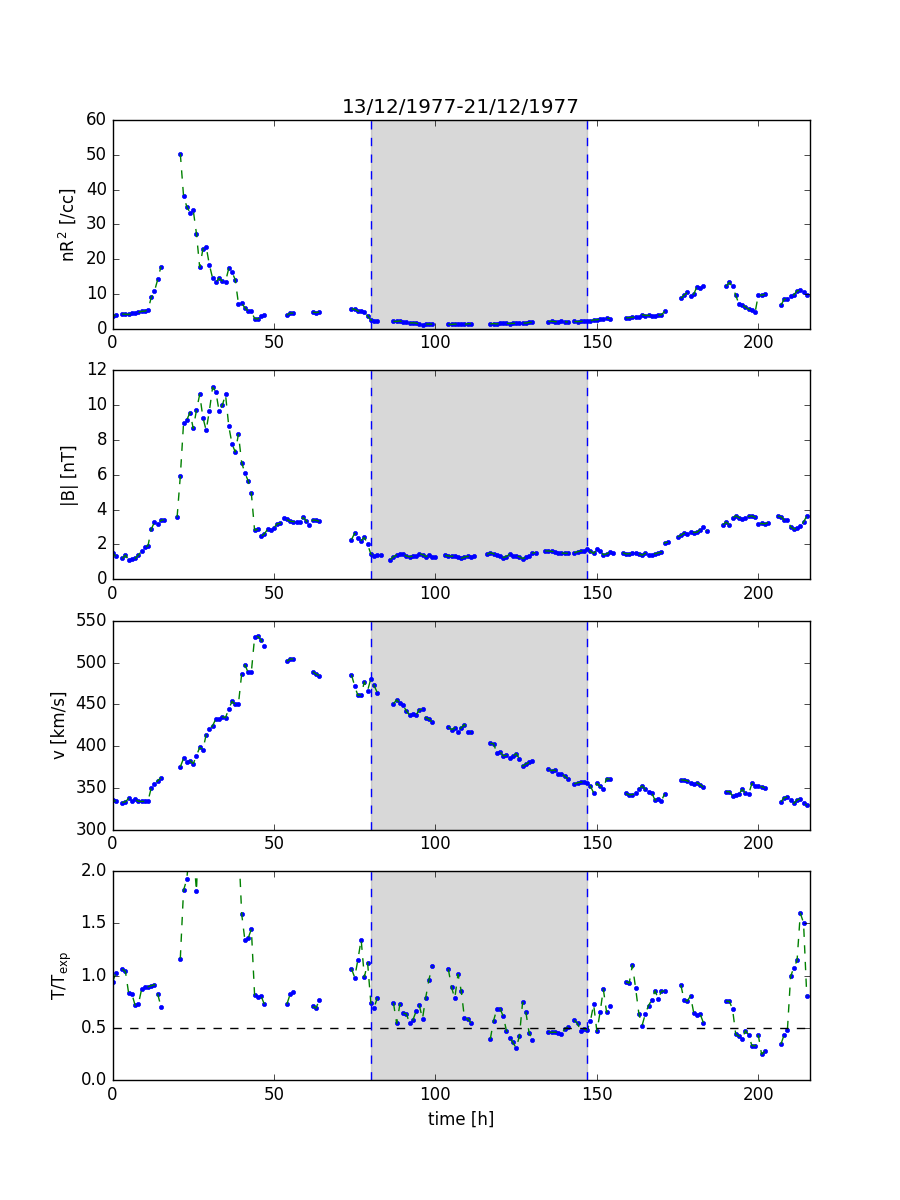} 
  \includegraphics[trim=1cm 1.5cm 1.5cm 2cm, clip=true,width=0.495\linewidth]{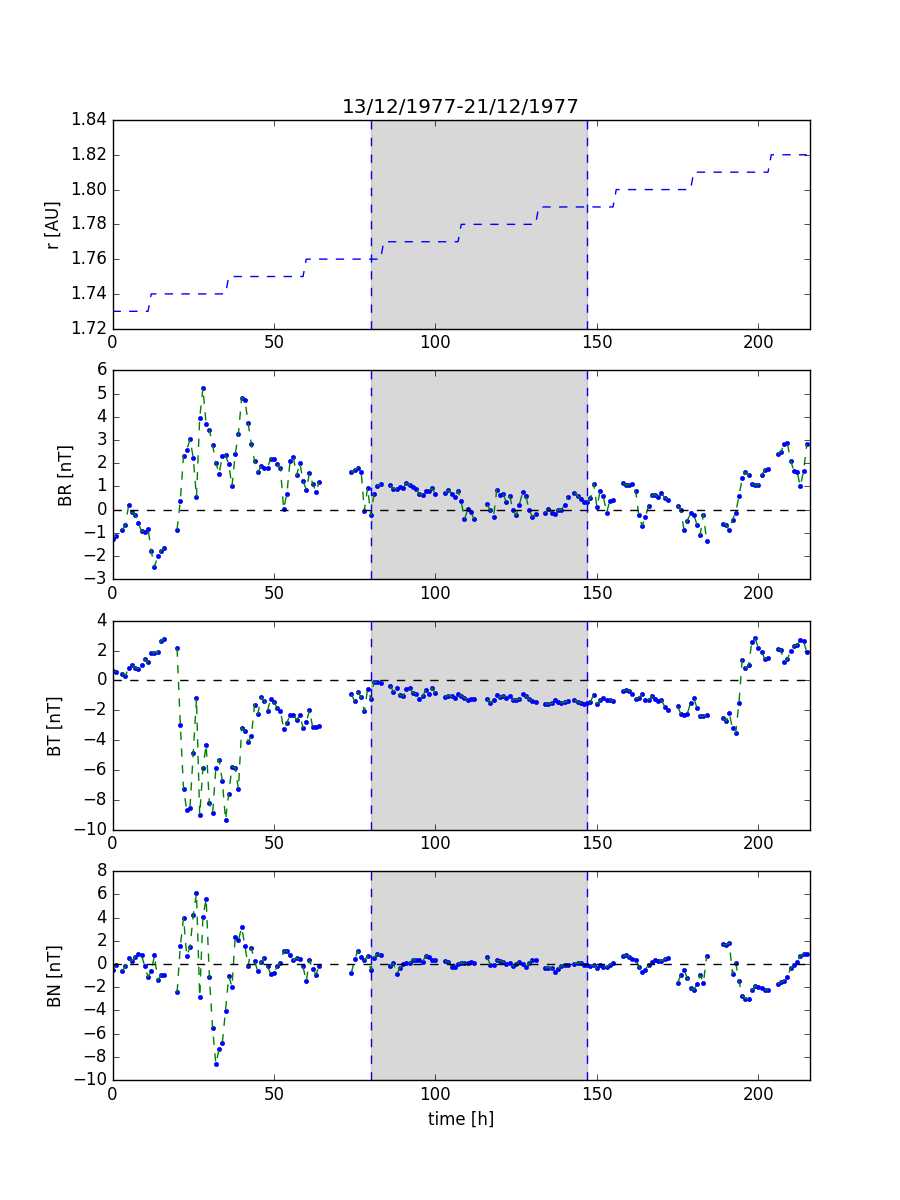}
  \includegraphics[trim=1cm 1.5cm 1.5cm 2cm, clip=true,width=0.495\linewidth]{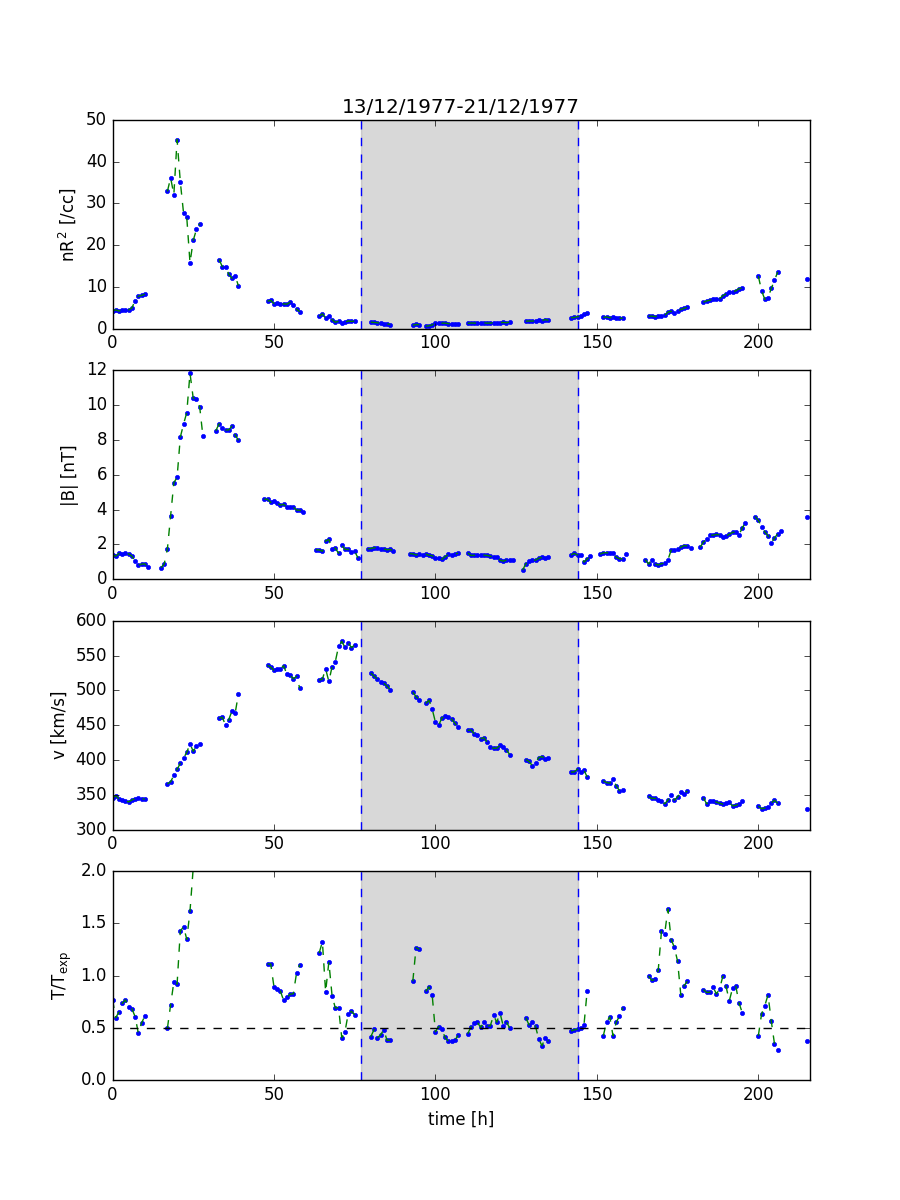}
  \includegraphics[trim=1cm 1.5cm 1.5cm 2cm, clip=true,width=0.495\linewidth]{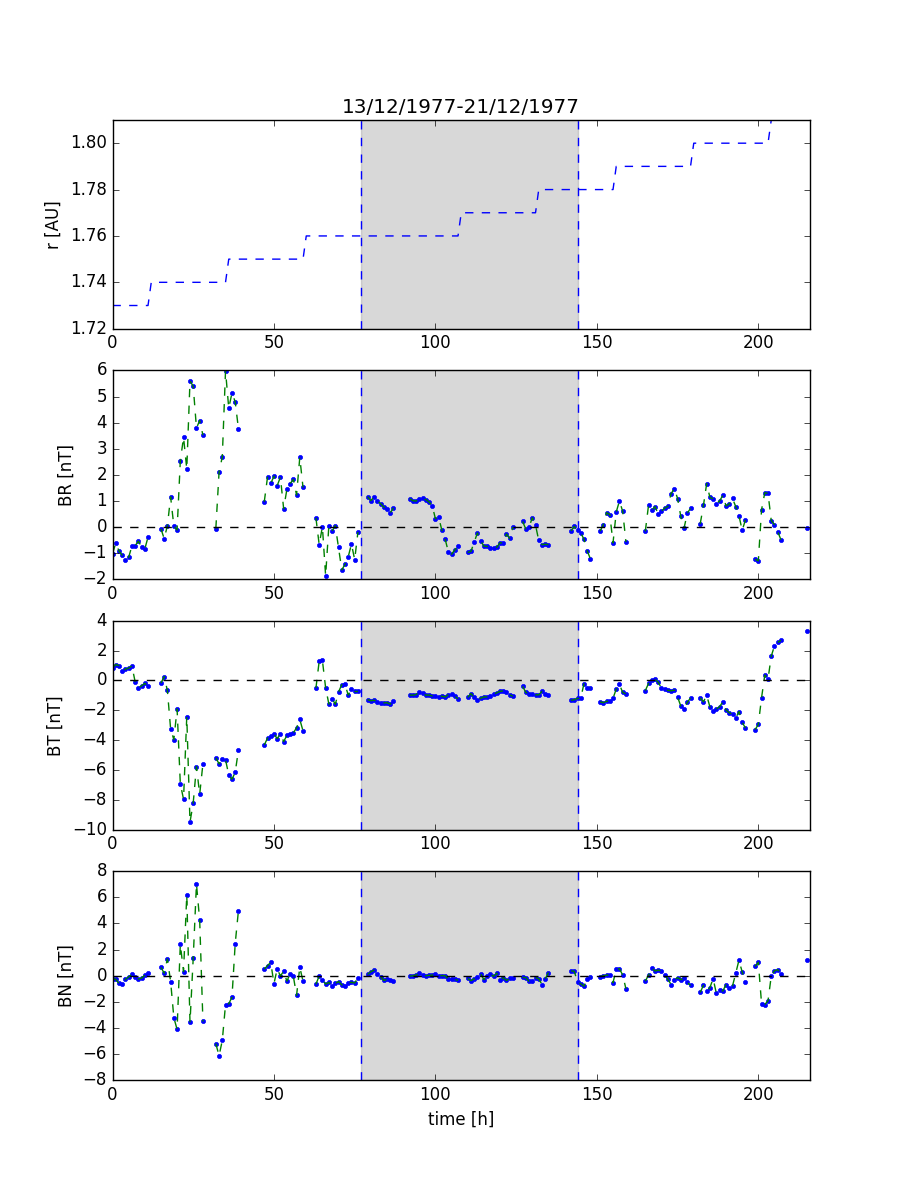} 
\caption{Voyager 1 (upper row) and 2 (lower row) time series for Event B.}\label{fig:eventB}
\end{figure*}

Event B was detected by both voyagers at approximately the same distance of 1.76~AU on 17/12/1977 at 07:00 UT for VOY1 and 05:00 UT for VOY2. A summary of its properties can be seen in Table \ref{tab:eventB}.
\Needspace{100pt}
\begin{wraptable}{r}{0.52\linewidth}
\begin{center}
\begin{footnotesize}
\vspace{-20pt}
\begin{tabular}{c|c|c}
 & VOY1 & VOY 2 \\ \cline{1-3}
r [AU] & 1.76 & 1.76 \\
$\mathrm{N_{avg} [cm^{-3}]}$ & 0.54 & 0.47 \\
$\mathrm{B_{avg} [nT]}$ & 1.35 & 1.31 \\
$\mathrm{V_{avg} [km/s]}$ & 409 & 448.5  \\
$\mathrm{V_{beg} [km/s]}$ & 480.7 & 549.2 \\
width [AU]& 0.66 & 0.72 \\
\end{tabular}
\caption{Table of properties for event B}\label{tab:eventB}
\vspace{-20pt}
\end{footnotesize}
\end{center}
\end{wraptable}
Like event A, most measured quantities are similar due to the spacecraft being very close together. However, as can be seen in Figure~\ref{fig:voy_comp}, the velocity measured by VOY2 is higher than that of VOY1. Comparing the plasma speed time series in Figure~\ref{fig:eventB}, the front part (just before the ICME boundaries) is very similar, with the speed swiftly rising to approximately 550~km/s. For VOY1, the plasma speed decreases quasi linearly following this peak, while for VOY2 a second, higher velocity peak develops. This might be due to VOY2 hitting a different part of the ICME, one that contains a high velocity region. \citet{Hosteaux2018} shows that ICMEs consist of fine structures which show different observational characteristics. As a result of Equation~\ref{eq:v_avg}, the higher average speed also results in a higher ICME width of 0.06~AU, as can be seen in table \ref{tab:eventB}.

\subsubsection*{\textbf{EVENT C}}

\begin{figure*}[!htb]
\centering
  \centering
  \includegraphics[trim=1cm 1.5cm 1.5cm 2cm, clip=true,width=0.495\linewidth]{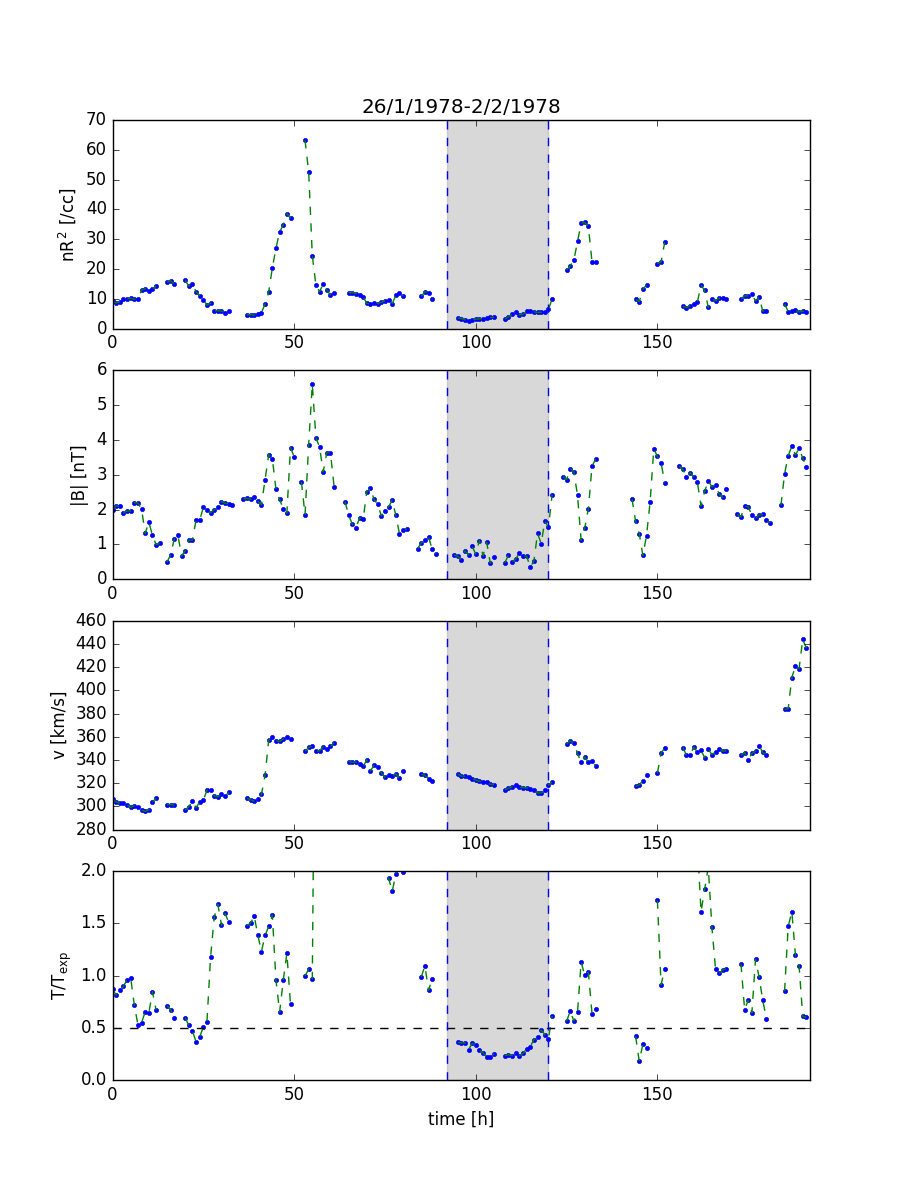} 
  \includegraphics[trim=1cm 1.5cm 1.5cm 2cm, clip=true,width=0.495\linewidth]{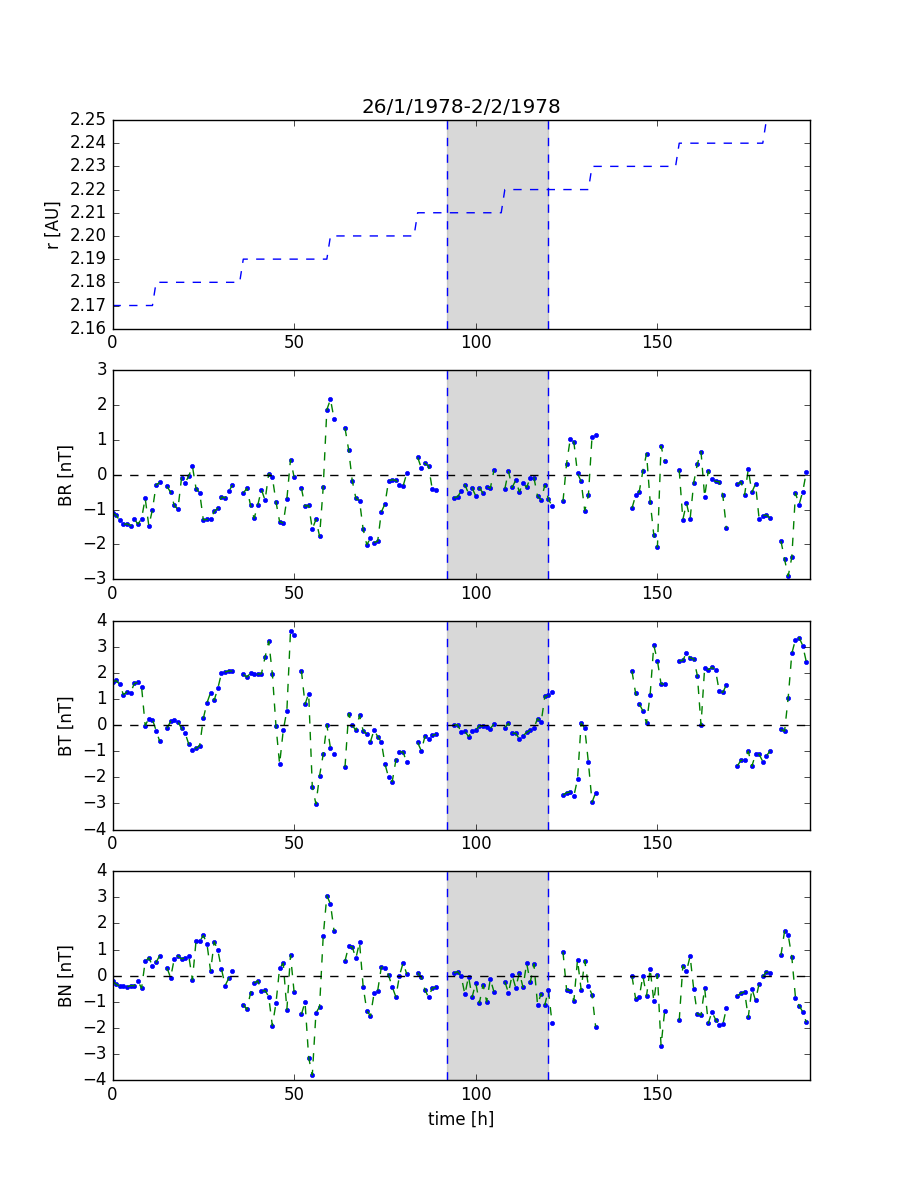}
  \includegraphics[trim=1cm 1.5cm 1.5cm 2cm, clip=true,width=0.495\linewidth]{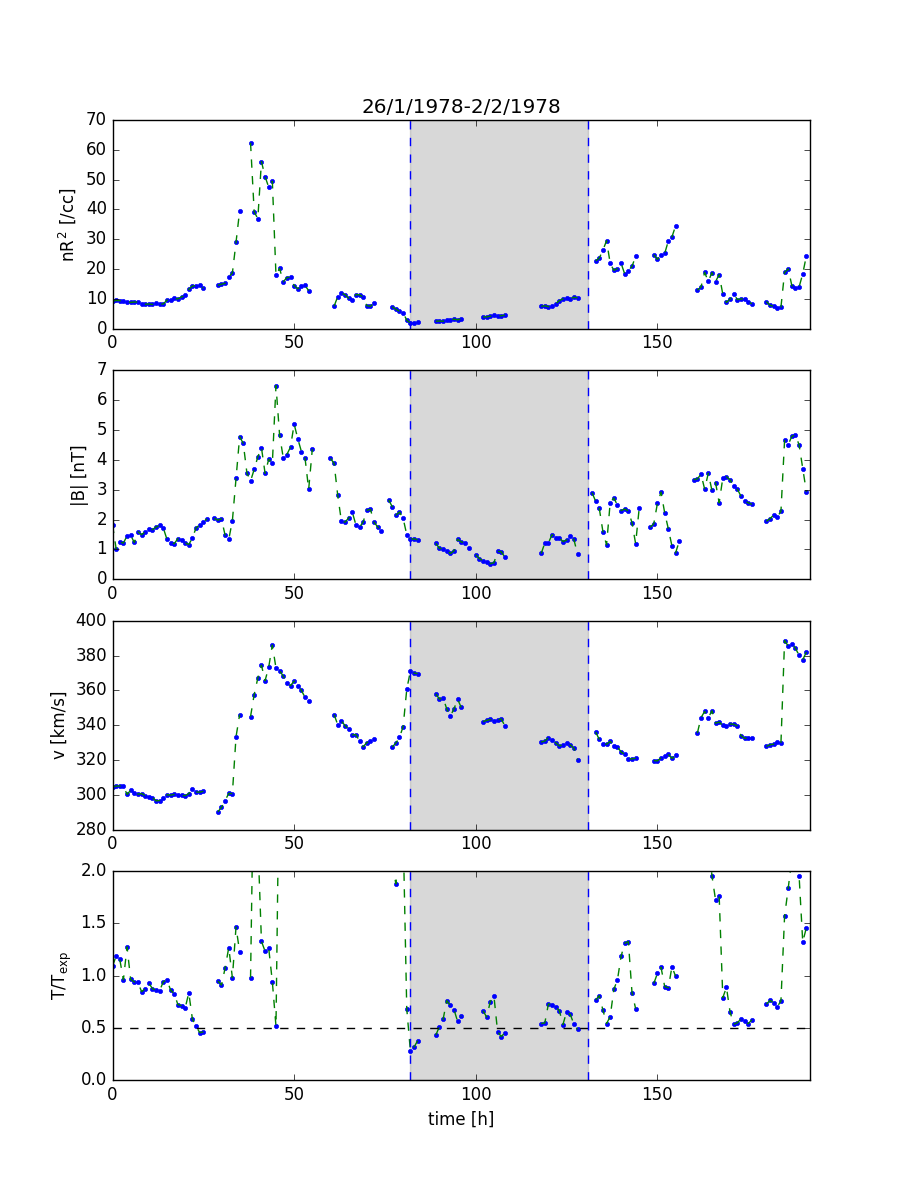}
  \includegraphics[trim=1cm 1.5cm 1.5cm 2cm, clip=true,width=0.495\linewidth]{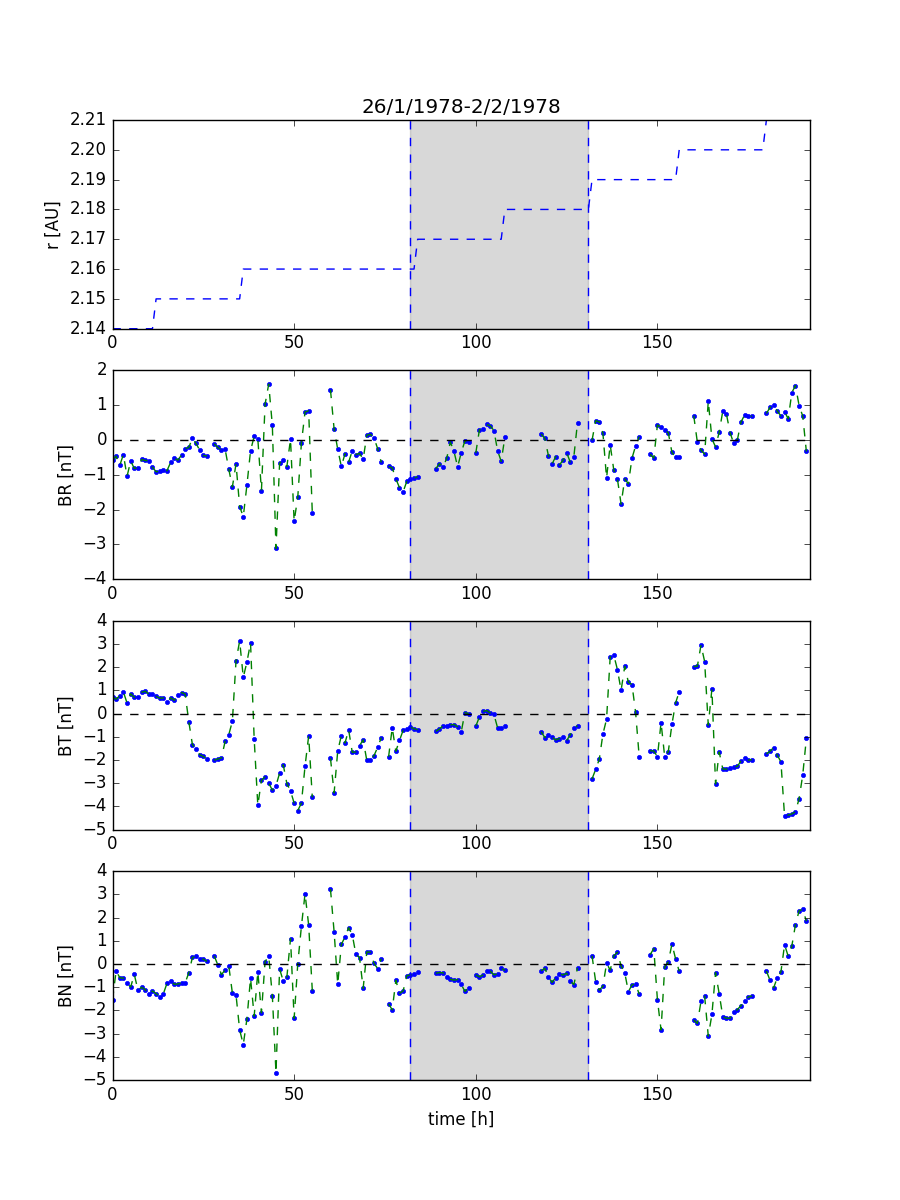} 
\caption{Voyager 1 (upper row) and 2 (lower row) time series for Event C.}\label{fig:eventC}
\end{figure*}

Event C was measured by VOY1 at 19:00 UT on 30/01/1978 at a radial distance of 2.21~AU, and by VOY2 at 10~a.m. at a distance of 2.16~AU. 
\Needspace{100pt}
\begin{wraptable}{r}{0.52\linewidth}
\begin{center}
\begin{footnotesize}
\vspace{-20pt}
\begin{tabular}{c|c|c}
 & VOY1 & VOY2 \\ \cline{1-3}
r [AU] & 2.21 & 2.16 \\
$\mathrm{N_{avg} [cm^{-3}]}$ & 0.88 & 1.15 \\
$\mathrm{B_{avg} [nT]}$ & 0.72 & 1.00 \\
$\mathrm{V_{avg} [km/s]}$ & 319.5 & 342.7  \\
$\mathrm{V_{beg} [km/s]}$ & 325.6 & 371.2 \\
width [AU]& 0.22 & 0.40 \\
\end{tabular}
\caption{Table of properties for event C}\label{tab:eventC}
\vspace{-20pt}
\end{footnotesize}
\end{center}
\end{wraptable}
There is a slight disparity for the properties listed in Table \ref{tab:eventC}. Figure \ref{fig:eventC} shows that the $\mathrm{T/T_{exp}}$ time series for both spacecraft are quite dissimilar, especially the VOY2 time series fluctuates around the threshold value of 0.5 making it difficult to identify the ICME boundaries. This results in a much higher VOY2 ICME width than listed for VOY1, affecting the average ICME values severely. Moreover, when comparing the bulk velocity time series of both spacecraft, the shock-sheath-magnetic cloud structure is visible in both but the values for VOY2 are higher, resulting in for example a frontal velocity difference of over 45~km/s. Although there are some data gaps, it is unlikely that it is the cause of the differences, but rather that the spacecraft are hit by different parts of the ICME with different properties.

\subsubsection*{\textbf{EVENT D}}

\begin{figure*}[!htb]
\centering
  \centering
  \includegraphics[trim=1cm 1.5cm 1.5cm 2cm, clip=true,width=0.495\linewidth]{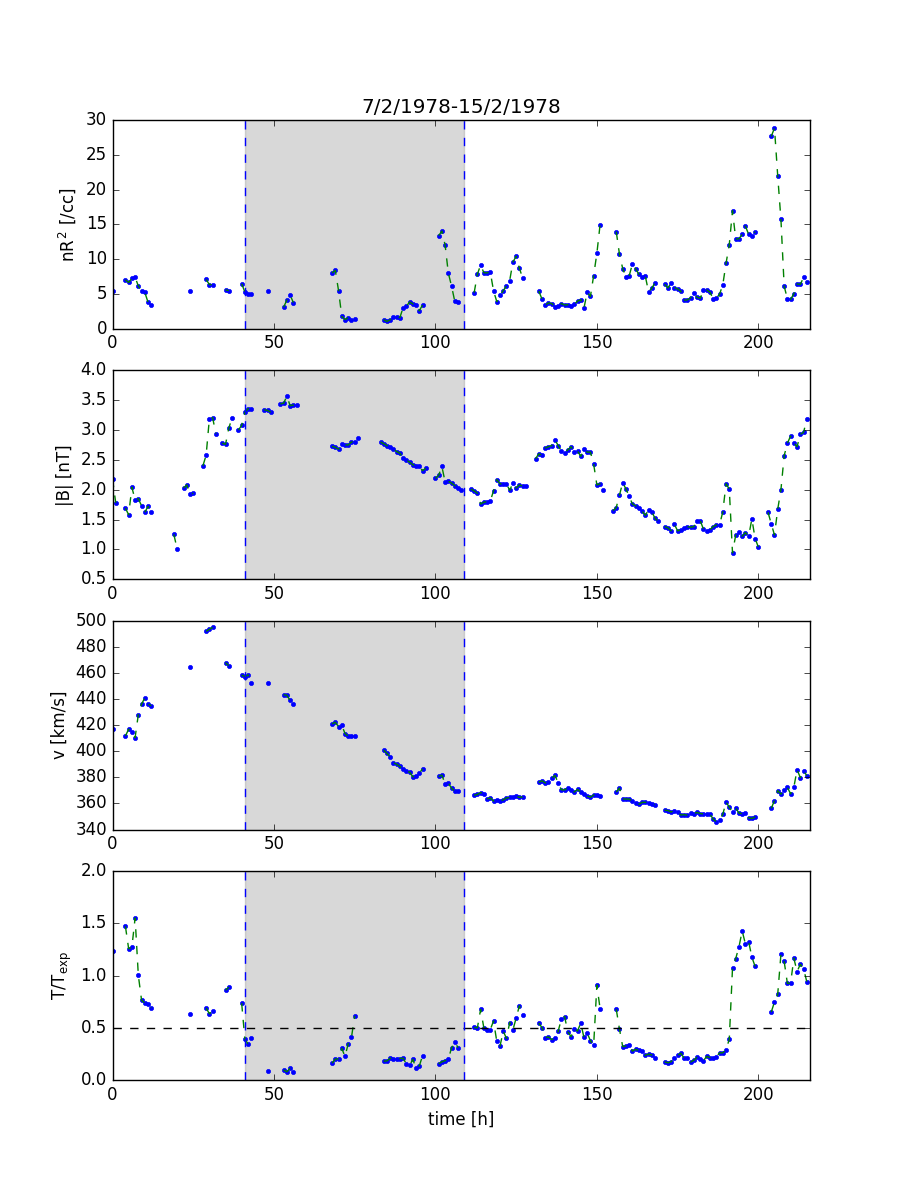} 
  \includegraphics[trim=1cm 1.5cm 1.5cm 2cm, clip=true,width=0.495\linewidth]{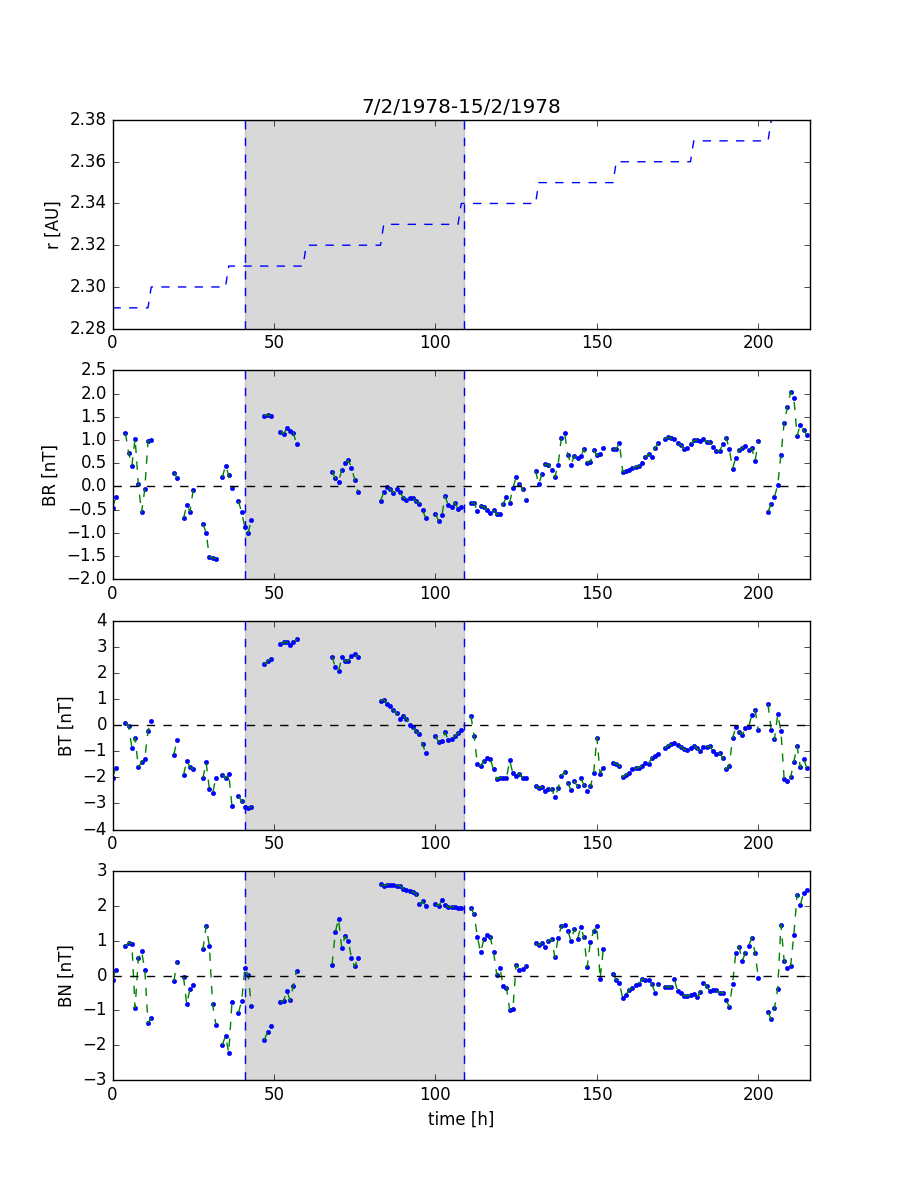}
  \includegraphics[trim=1cm 1.5cm 1.5cm 2cm, clip=true,width=0.495\linewidth]{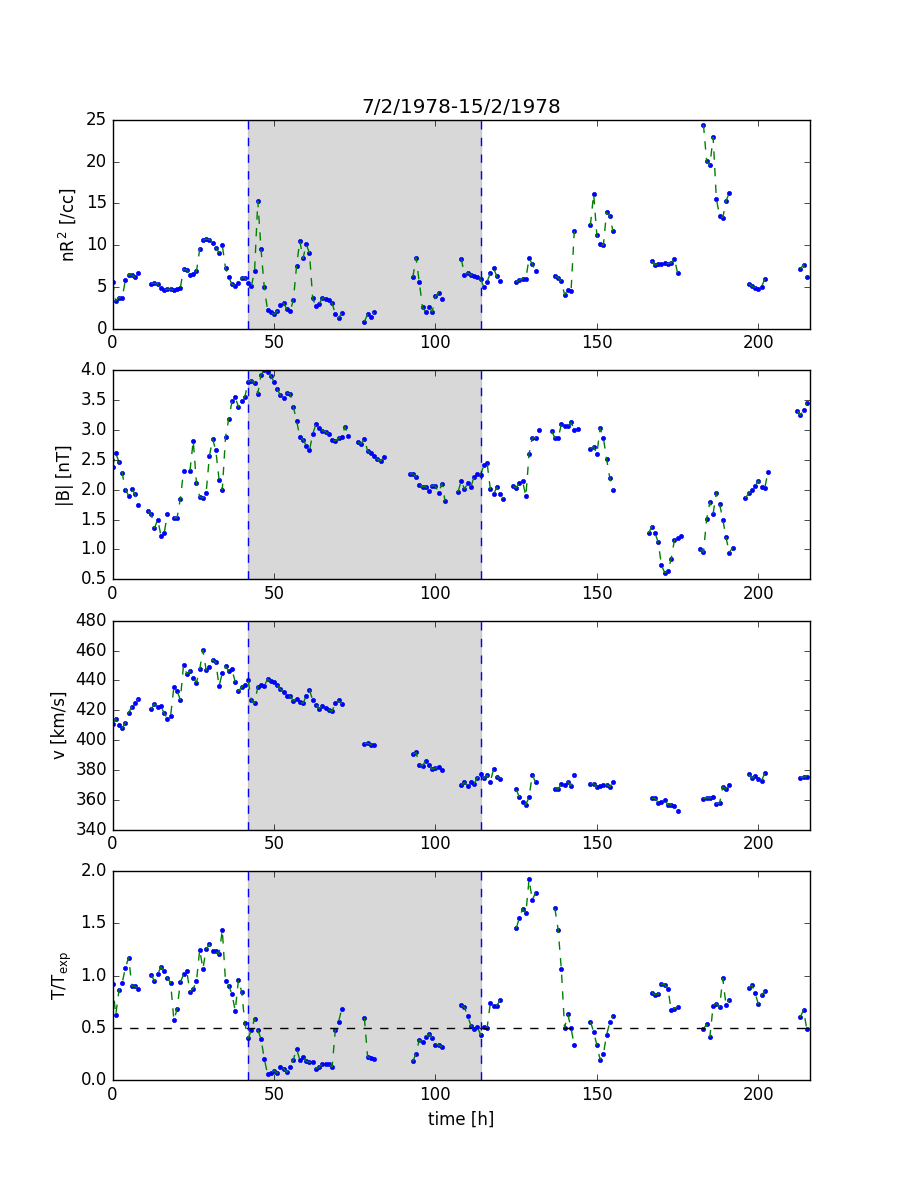}
  \includegraphics[trim=1cm 1.5cm 1.5cm 2cm, clip=true,width=0.495\linewidth]{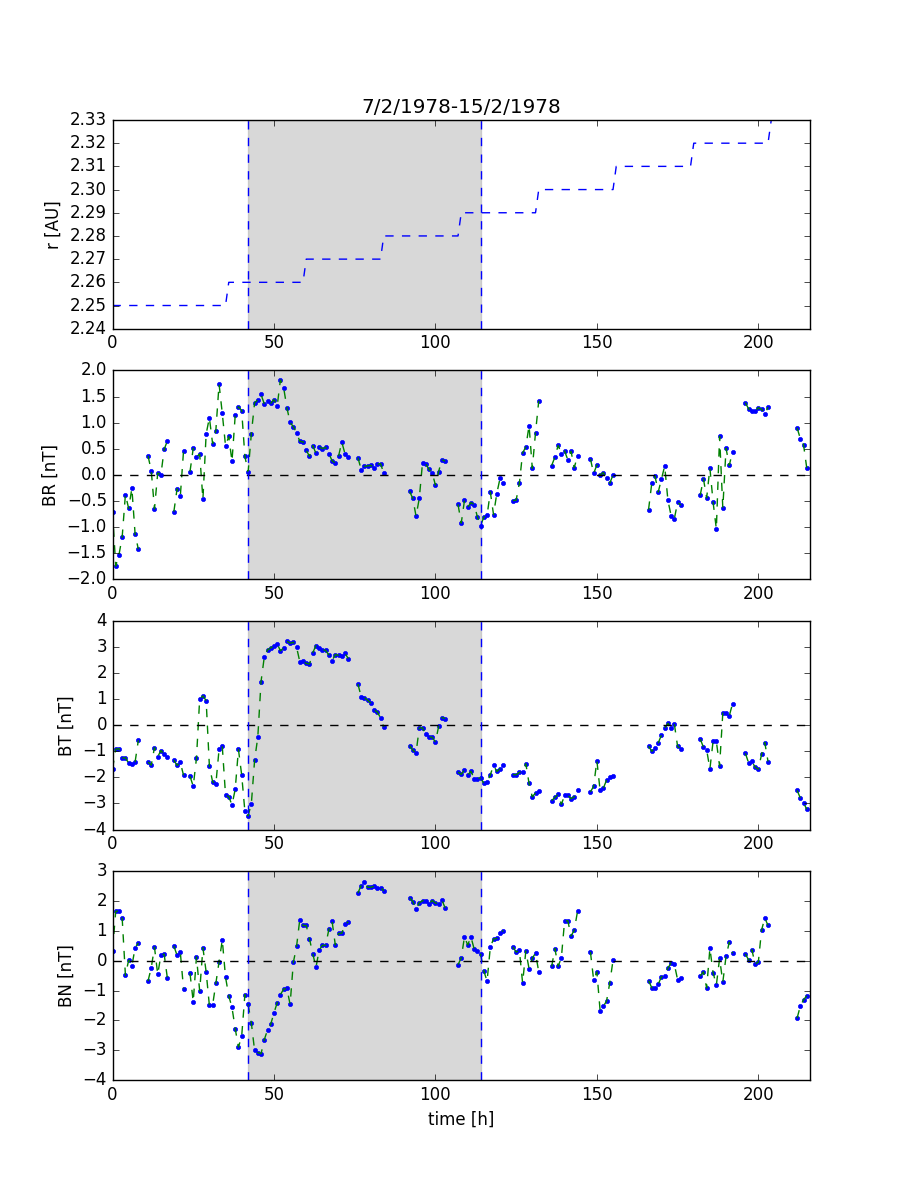} 
\caption{Voyager 1 (upper row) and 2 (lower row) time series for Event D.}\label{fig:eventD}
\end{figure*}

VOY1 and VOY2 were hit by event D on 02/09/1978 at 5~p.m. and at 6~p.m., respectively at a radial distance of 2.31 and 2.26~AU, respectively. The later time of detection by VOY2 is surprising considering its lower distance from the Sun.
\Needspace{100pt}
\begin{wraptable}{r}{0.52\linewidth}
\begin{center}
\begin{footnotesize}
\vspace{-20pt}
\begin{tabular}{c|c|c}
 & VOY1 & VOY 2 \\ \cline{1-3}
r [AU] & 2.31 & 2.26 \\
$\mathrm{N_{avg} [cm^{-3}]}$ & 0.85 & 0.84 \\
$\mathrm{B_{avg} [nT]}$ & 2.79 & 2.70 \\
$\mathrm{V_{avg} [km/s]}$ & 412.6 & 405.7  \\
$\mathrm{V_{beg} [km/s]}$ & 457.3 & 440.2 \\
width [AU]& 0.67 & 0.70 \\
\end{tabular}
\caption{Table of properties for event D}\label{tab:eventD}
\vspace{-20pt}
\end{footnotesize}
\end{center}
\end{wraptable}
This is possibly due to a faulty assignment of the ICME front, due to the ambiguity of the $\mathrm{T/T_{exp}}$ time series at the front in Figure~\ref{fig:eventD}. The VOY1 bulk velocity time series shows higher velocity magnitudes in the sheath and front part of the ICME compared to VOY2, also seen as a higher ICME front velocity in Table~\ref{tab:eventD}. Despite these discrepancies, the magnetic field components show similar smooth rotation for both spacecraft and similar average magnetic field strength and average number density.

\subsubsection*{\textbf{EVENT E}}
 
\begin{figure*}[!htb]
\centering
  \centering
  \includegraphics[trim=1cm 1.5cm 1.5cm 2cm, clip=true,width=0.495\linewidth]{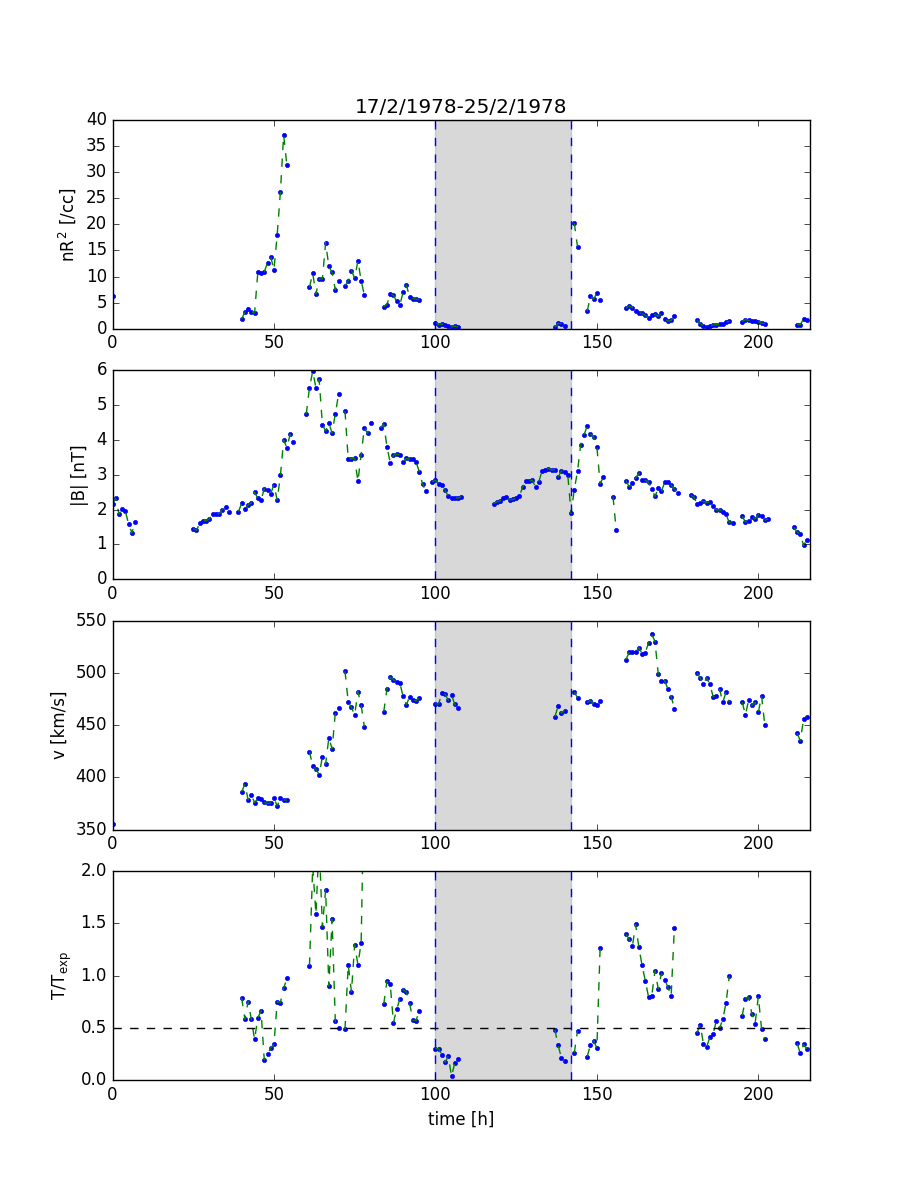} 
  \includegraphics[trim=1cm 1.5cm 1.5cm 2cm, clip=true,width=0.495\linewidth]{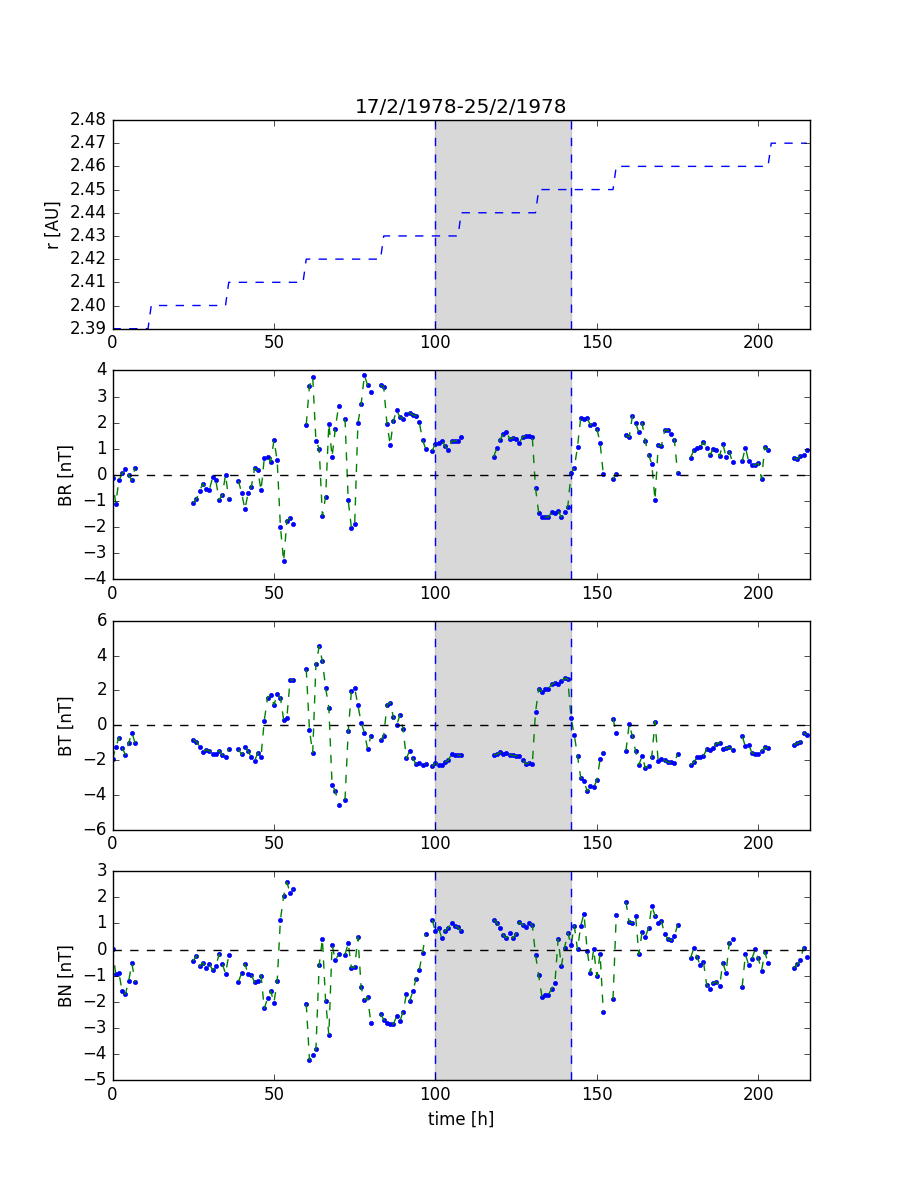}
  \includegraphics[trim=1cm 1.5cm 1.5cm 2cm, clip=true,width=0.495\linewidth]{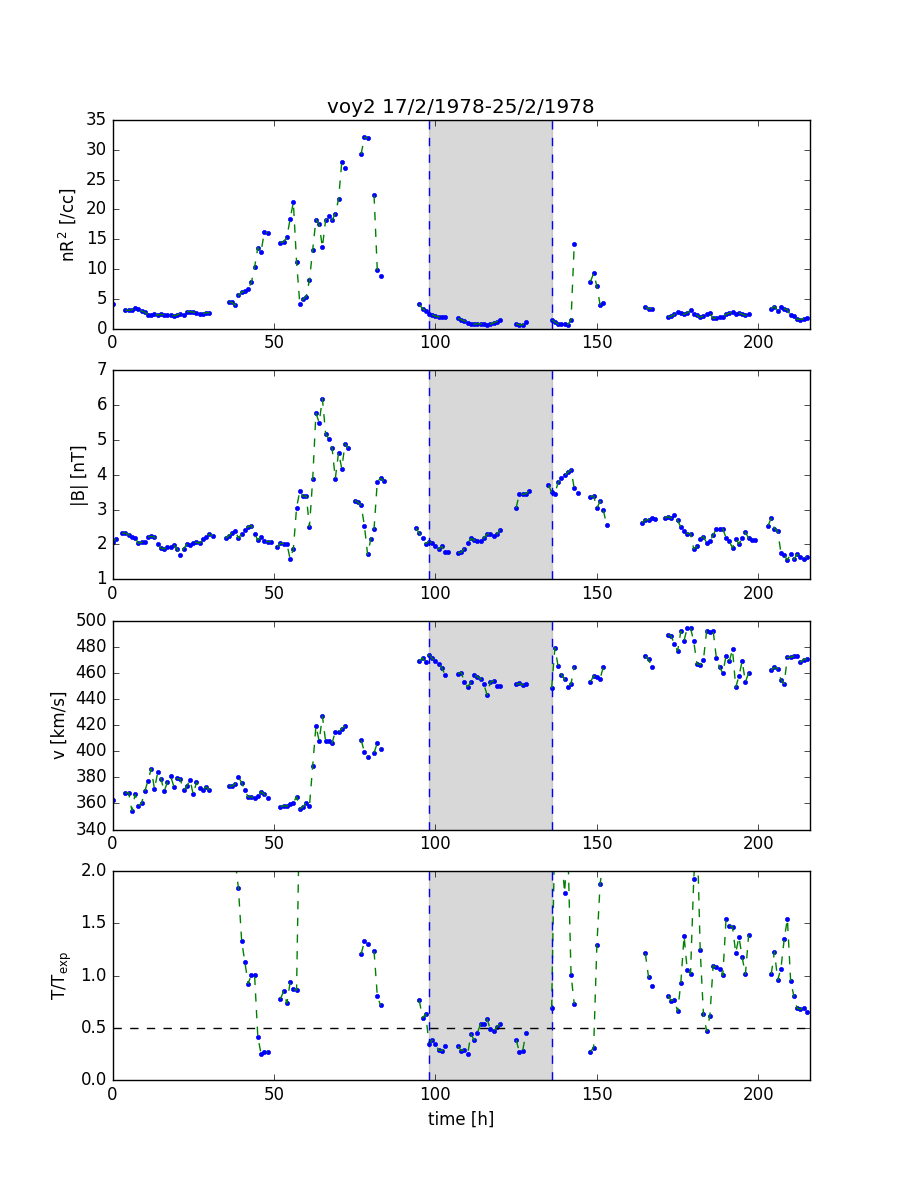}
  \includegraphics[trim=1cm 1.5cm 1.5cm 2cm, clip=true,width=0.495\linewidth]{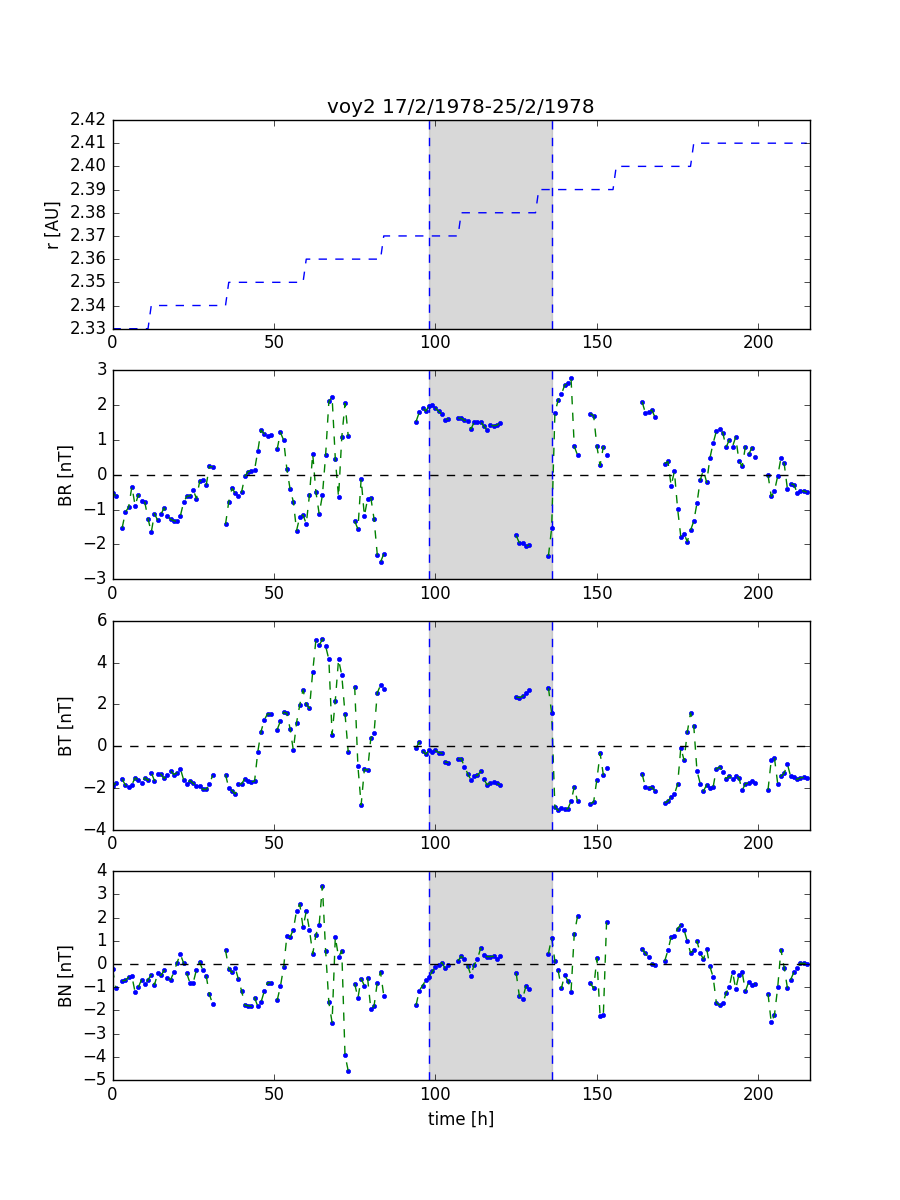} 
\caption{Voyager 1 (upper row) and 2 (lower row) time series for Event E.}\label{fig:eventE}
\end{figure*}

Event E was detected by VOY1 on 22/02/1978 at 04:00 UT at a radial distance of 2.43~AU and by VOY2 at 01:00 UT on the same day at 2.37~AU. A summary of its properties can be seen in Table \ref{tab:eventE}.
\Needspace{100pt}
\begin{wraptable}{r}{0.52\linewidth}
\begin{center}
\begin{footnotesize}
\vspace{-20pt}
\begin{tabular}{c|c|c}
 & VOY1 & VOY 2 \\ \cline{1-3}
r [AU] & 2.43 & 2.37 \\
$\mathrm{N_{avg} [cm^{-3}]}$ & 0.10 & 0.23 \\
$\mathrm{B_{avg} [nT]}$ & 2.51 & 2.51 \\
$\mathrm{V_{avg} [km/s]}$ & 464.6 & 455.0  \\
$\mathrm{V_{beg} [km/s]}$ & 469.9 & 448.9 \\
width [AU]& 0.47 & 0.42 \\
\end{tabular}
\caption{Table of properties for event E}\label{tab:eventE}
\vspace{-20pt}
\end{footnotesize}
\end{center}
\end{wraptable}
Both voyagers are still relatively close together (\textasciitilde0.06~AU), which results in very similar measurements for this event. The average number density for VOY1 is lower than that for VOY2 though, and since the width and the bulk velocity time series (see Figure \ref{fig:eventE}) show no signs of severe expansion, it is more likely the result of the large data gap at the centre of the ICME in the VOY1 plasma time series.

\subsubsection*{\textbf{EVENT G}}

Event G impacted VOY1 at 3.38~AU on 07/06/1978 at 02:00 UT, and VOY2 at 3.21~AU on 06/06/1978 at 10:00 UT. The radial separation between the two spacecraft was \textasciitilde0.17~AU. A summary of its properties can be seen in Table \ref{tab:eventG}.

\Needspace{100pt}
\begin{wraptable}{r}{0.52\linewidth}
\begin{center}
\begin{footnotesize}
\vspace{-20pt}
\begin{tabular}{c|c|c}
 & VOY1 & VOY 2 \\ \cline{1-3}
r [AU] & 3.38 & 3.21 \\
$\mathrm{N_{avg} [cm^{-3}]}$ & 0.25 & 0.31 \\
$\mathrm{B_{avg} [nT]}$ & 0.61 & 0.74 \\
$\mathrm{V_{avg} [km/s]}$ & 405.6 & 402.5  \\
$\mathrm{V_{beg} [km/s]}$ & 430 & 424.4 \\
width [AU]& 0.63 & 0.62 \\
\end{tabular}
\caption{Table of properties for event G}\label{tab:eventG}
\vspace{-20pt}
\end{footnotesize}
\end{center}
\end{wraptable}

The magnetic field components show in Figure~\ref{fig:eventG} a significant decrease in variability but not a smooth rotation. Average magnetic field strength and average number density are low for both spacecraft. A monotonically decreasing velocity profile is present, with the average and leading edge velocities being similar. The back ICME boundary for VOY1 is determined by the magnetic field components since the $\mathrm{T/T_{exp}}$ ratio is going above the threshold at approximately the middle of the ICME.

\subsubsection*{\textbf{EVENT H}}

Event H was detected by VOY1 and VOY2 on 24/09/1978 at 22:00 UT and 23/09/1978 at 08:00 UT, respectively. The radial separation between the two measurements was \textasciitilde0.29~AU. A summary of its properties can be seen in Table \ref{tab:eventH}.

\Needspace{100pt}
\begin{wraptable}{r}{0.52\linewidth}
\begin{center}
\begin{footnotesize}
\vspace{-20pt}
\begin{tabular}{c|c|c}
 & VOY1 & VOY 2 \\ \cline{1-3}
r [AU] & 4.22 & 3.93 \\
$\mathrm{N_{avg} [cm^{-3}]}$ & 0.05 & 0.05 \\
$\mathrm{B_{avg} [nT]}$ & 0.86 & 1.16 \\
$\mathrm{V_{avg} [km/s]}$ & 452.0 & 469.7  \\
$\mathrm{V_{beg} [km/s]}$ & 502.7 & 538.3 \\
width [AU]& 1.05 & 1.10 \\
\end{tabular}
\caption{Table of properties for event H}\label{tab:eventH}
\vspace{-20pt}
\end{footnotesize}
\end{center}
\end{wraptable}

Figure~\ref{fig:eventH} shows that the event H time series contain almost all considered ICME signatures. The number density is extremely low, the velocity profile is monotonically decreasing and the magnetic field has low variability and the components show a smoothly rotating field. At the leading edge, there is a large jump in plasma speed (\textasciitilde150~km/s), that decreases smoothly after. According to Equation~\ref{eq:v_avg}, the width decreases by 0.05~AU and the average B-field strength decreases with 0.3~nT, showing that the ICME might have eroded during its propagation from VOY2 to VOY1.

\subsubsection*{\textbf{EVENT I}}

On 17/11/1978 at 21:00 UT event I was detected by VOY1. VOY2 detected this event at 13:00 UT on 16/11/1978. At the time of detection, VOY1 and VOY2 were at 4.60 and 4.25~AU, respectively. A summary of its properties can be seen in Table \ref{tab:eventI}.
\Needspace{100pt}
\begin{wraptable}{r}{0.52\linewidth}
\begin{center}
\begin{footnotesize}
\vspace{-20pt}
\begin{tabular}{c|c|c}
 & VOY1 & VOY 2 \\ \cline{1-3}
r [AU] & 4.60 & 4.25 \\
$\mathrm{N_{avg} [cm^{-3}]}$ & 0.16 & 0.19 \\
$\mathrm{B_{avg} [nT]}$ & 0.53 & 0.41 \\
$\mathrm{V_{avg} [km/s]}$ & 395.7 & 396.6  \\
$\mathrm{V_{beg} [km/s]}$ & 415.3 & 422.3 \\
width [AU]& 1.10 & 0.93 \\
\end{tabular}
\caption{Table of properties for event I}\label{tab:eventI}
\vspace{-20pt}
\end{footnotesize}
\end{center}
\end{wraptable}
Both VOY1 and VOY2 show data gaps at the front boundary in the plasma time series, as shown in Figure~\ref{fig:eventI}. The number density is low for both spacecraft, the velocity decreases linearly and variability of the B-field components is small. The VOY1 density time series shows some fine structure of increased density, B-field, and velocity close to the rear end.

\subsubsection*{\textbf{EVENT J}}

VOY1 and VOY2 were hit by event J on 10/12/1978 at 04:00 UT and on 08/12/1978 at 22:00 UT, respectively. Their radial distance was 4.74~AU for VOY1 and 4.36~AU for VOY2. A summary of its properties can be seen in Table \ref{tab:eventJ}.
\Needspace{100pt}
\begin{wraptable}{r}{0.52\linewidth}
\begin{center}
\begin{footnotesize}
\vspace{-20pt}
\begin{tabular}{c|c|c}
 & VOY1 & VOY 2 \\ \cline{1-3}
r [AU] & 4.74 & 4.36 \\
$\mathrm{N_{avg} [cm^{-3}]}$ & 0.09 & 0.08 \\
$\mathrm{B_{avg} [nT]}$ & 0.21 & 0.66 \\
$\mathrm{V_{avg} [km/s]}$ & 479.9 & 482.8  \\
$\mathrm{V_{beg} [km/s]}$ & 661.4 & 677.2 \\
width [AU]& 3.51 & 3.46 \\
\end{tabular}
\caption{Table of properties for event J}\label{tab:eventJ}
\vspace{-20pt}
\end{footnotesize}
\end{center}
\end{wraptable}
Figure~\ref{fig:eventJ} shows that event J is another strong ICME with most ICME signatures present in the time series for both spacecraft. There is a long time period where $\mathrm{T/T_{exp}}$ falls below the threshold, though data gaps obscure both VOY1 ICME boundaries. The plasma speed at the front of the ICME just before the boundary is very high for an ICME at this radial distance (\textasciitilde700~km/s), after which there is a long monotonic decrease to \textasciitilde350~km/s. Both beginning and average velocity do not alter significantly between both spacecraft. The variability is strongly decreased, though a rotating B-field is difficult to identify. The average number density stays approximately the same while the average magnetic field strength decreases by over two thirds. This decrease in B-field strength does not seem to be the result of expansion (the width increases by 0.05~AU), but perhaps the ICME was subjected to strong magnetic erosion.

\subsubsection*{\textbf{EVENT K}}

Event K was detected by both voyagers at distance of 4.91~AU on 04/01/1979 at 23:00 UT for VOY1 and at 4.36~AU on 3/01/1979 at 21:00 UT for VOY2. A summary of its properties can be seen in Table \ref{tab:eventK}.
\Needspace{100pt}
\begin{wraptable}{r}{0.52\linewidth}
\begin{center}
\begin{footnotesize}
\vspace{-20pt}
\begin{tabular}{c|c|c}
 & VOY1 & VOY 2 \\ \cline{1-3}
r [AU] & 4.91 & 4.36 \\
$\mathrm{N_{avg} [cm^{-3}]}$ & 0.05 & 0.05 \\
$\mathrm{B_{avg} [nT]}$ & 0.29 & 0.43 \\
$\mathrm{V_{avg} [km/s]}$ & 493.3 &  488.9 \\
$\mathrm{V_{beg} [km/s]}$ & 612.3 & 630.3 \\
width [AU]& 2.46 & 2.47 \\
\end{tabular}
\caption{Table of properties for event K}\label{tab:eventK}
\vspace{-20pt}
\end{footnotesize}
\end{center}
\end{wraptable}
Following event J, event K is another strong ICME with a very high front speed, which decreases with approximately 18.1~km/s from VOY2 to VOY1. This means the front of the ICME slows down, likely due to a large velocity difference between the ICME and the solar wind which increases the drag. The boundaries of the event are well defined for both spacecraft due to a clear shift in time series behaviour for all ICME parameters at the same time. The ICME is not expanding, shown by the constant width and average number density in Table~\ref{tab:eventK}, thus the decrease in magnetic field strength implies magnetic field erosion.

\subsubsection*{\textbf{EVENT L}}

On 18/02/1979 at 20:00 UT VOY1 detected event L at a radial distance of 5.19~AU. The same event was captured by VOY2 at 12:00 UT on 17/02/1979 at 4.71~AU. A summary of its properties can be seen in Table \ref{tab:eventL}.
\Needspace{100pt}
\begin{wraptable}{r}{0.52\linewidth}
\begin{center}
\begin{footnotesize}
\vspace{-20pt}
\begin{tabular}{c|c|c}
 & VOY1 & VOY 2 \\ \cline{1-3}
r [AU] & 5.19 & 4.71 \\
$\mathrm{N_{avg} [cm^{-3}]}$ & 0.15 & 0.21 \\
$\mathrm{B_{avg} [nT]}$ & 0.61 & 0.73 \\
$\mathrm{V_{avg} [km/s]}$ & 426.6 &  425 \\
$\mathrm{V_{beg} [km/s]}$ & 470.3 & 463.4 \\
width [AU]& 1.56 & 1.49 \\
\end{tabular}
\caption{Table of properties for event L}\label{tab:eventL}
\vspace{-20pt}
\end{footnotesize}
\end{center}
\end{wraptable}
In contrast to most events in the list, event~L does not show a low number density compared to the solar wind in the time series shown in Figure~\ref{fig:eventL}. The ICME does have high density in the sheath and immediately following the ejecta. A somewhat rotating magnetic field is present, though not smooth, with the magnetic field having a low variability. The velocity time series shows a monotonic decreasing profile. The average ICME speed is approximately the same, but the higher duration of the VOY1 detection, which was 6~h longer than the VOY2 time series, results in a width increase of 0.07~AU. This expansion causes the average number density and the average magnetic field strength decreasing by 29\% and 16\%, respectively.

\subsubsection*{\textbf{EVENT M}}

Event M was captured by VOY1 at 21:00 UT on 17/04/1979 at a radial distance of 5.43~AU, and by VOY2 at 23:00 UT on 15/04/1979 at a distance of 4.96~AU. A summary of its properties can be seen in Table \ref{tab:eventM}.
\Needspace{100pt}
\begin{wraptable}{r}{0.52\linewidth}
\begin{center}
\begin{footnotesize}
\vspace{-20pt}
\begin{tabular}{c|c|c}
 & VOY1 & VOY 2 \\ \cline{1-3}
r [AU] & 5.43 & 4.96 \\
$\mathrm{N_{avg} [cm^{-3}]}$ & 0.15 & 0.18 \\
$\mathrm{B_{avg} [nT]}$ & 0.80 & 0.69 \\
$\mathrm{V_{avg} [km/s]}$ & 403.2 &  403.9 \\
$\mathrm{V_{beg} [km/s]}$ & 425.1 & 424.0 \\
width [AU]& 0.94 & 0.95 \\
\end{tabular}
\caption{Table of properties for event M}\label{tab:eventM}
\vspace{-20pt}
\end{footnotesize}
\end{center}
\end{wraptable}
The legitimacy of event M being an ICME is debatable. The $\mathrm{T/T_{exp}}$ ratio time series goes below the threshold for both spacecraft for an extended period of time and the variability of the magnetic field is low. However, the magnetic field is not smoothly rotating. The velocity time series is decreasing but it is not monotonic between the ICME boundaries. Both the average and front speed remain approximately constant, and due to the duration of the event in the VOY2 time series being only 1~h longer, the ICME width is also very similar. The average magnetic field increases while the average magnetic decreases, but both only slightly and considering the low confidence in this event no reasonable conclusions can be drawn.

\subsubsection*{\textbf{EVENT N AND EVENT O}}

Event N was detected by VOY1 on 14/05/1979 at 02:00 UT at a radial distance of 5.53~AU, and by VOY2 on 12/05/1979 at 07:00 UT at 5.07~AU. Event N is quickly followed by event O. VOY1 and VOY2 detected this event at 26/05/1979 at 01:00 UT and on 22/05/1979 at 20:00 UT, respectively, when the spacecraft were at and 5.58~AU and 5.12, respectively. A summary of these events properties can be seen in Table \ref{tab:eventNO}.

  \begin{table}[htb!]
    \begin{minipage}{.22\textwidth}
      \centering
      \begin{footnotesize}
\begin{tabular}{c|c|c}
 & VOY1 & VOY 2 \\ \cline{1-3}
r [AU] & 5.53 & 5.07 \\
$\mathrm{N_{avg} [cm^{-3}]}$ & 0.05 & 0.07 \\
$\mathrm{B_{avg} [nT]}$ & 0.41 & 0.31 \\
$\mathrm{V_{avg} [km/s]}$ & 404.9 &  397.0 \\
$\mathrm{V_{beg} [km/s]}$ & 444.5 & 450.3 \\
width [AU] & 1.33 & 1.32 \\
\end{tabular}
\end{footnotesize}
    \end{minipage}
    \begin{minipage}{.22\textwidth}
      \centering
      \begin{footnotesize}
\begin{tabular}{c|c|c}
 & VOY1 & VOY 2 \\ \cline{1-3}
r [AU] & 5.58 & 5.12 \\
$\mathrm{N_{avg} [cm^{-3}]}$ & 0.07 & 0.05 \\
$\mathrm{B_{avg} [nT]}$ & 0.32 & 0.21 \\
$\mathrm{V_{avg} [km/s]}$ & 452.6 &  473.5 \\
$\mathrm{V_{beg} [km/s]}$ & 479.6 & 518.1 \\
width [AU]& 0.91 & 0.99 \\
\end{tabular}
\end{footnotesize}
    \end{minipage}\caption{Table of properties for event N (left) and event O (right)}\label{tab:eventNO}
  \end{table}

Event N and event O are of particular interest. For event N, the number density and magnetic field variability is very low, and the plasma speed is monotonically decreasing. The width and average number density stay approximately constant while the average magnetic field strength increases. Event O also shows a low number density, low B-field variability and a linearly decreasing velocity profile. The front speed is much higher than that of event N, implying that event O presses against event N. This results in a region of very high density and B-field strength between the two events, due to the compression of plasma. The front and average speed of the event decrease by 38.5~km/s and 20.9~km/s, respectively. The width decreases, though this might be the result of the front boundary for the VOY1 detection being chosen too late. Although the $\mathrm{T/T_{exp}}$ ratio is below the threshold for a few hours before the shaded region in Figure~\ref{fig:eventNO}, but if the boundary would be chosen there the measured front velocity would drop significantly. Event~O may have collided with event~N at some point during their propagation. This could be due to a higher initial ICME speed of event~O, or event~N sweeping up the solar wind in front of event~O, resulting in a higher decelerating drag force for event~N than for event~O. The average number density for both events is similar, though the rear of event~O shows a high density region. When both events are measured by VOY1 and VOY2, event O is slowing down as a result of the collision, while the velocity of event N stays constant. The time difference between the detection of the two ICMEs for VOY1 and VOY2 is 11.96~days and 10.54~days, respectively, which further suggests that the high pressure in between the two ICMEs pushes event~O away from event~N.

\subsubsection*{\textbf{EVENT P}}

Event P impacted VOY1 at 5.68~AU on 16/06/1979 at 02:00 UT, and VOY2 at 5.21~AU on 15/06/1979 at 12:00 UT. A summary of its properties can be seen in Table \ref{tab:eventP}.

\Needspace{100pt}
\begin{wraptable}{r}{0.52\linewidth}
\begin{center}
\begin{footnotesize}
\vspace{-20pt}
\begin{tabular}{c|c|c}
 & VOY1 & VOY 2 \\ \cline{1-3}
r [AU] & 5.68 & 5.21 \\
$\mathrm{N_{avg} [cm^{-3}]}$ & 0.06 & 0.09 \\
$\mathrm{B_{avg} [nT]}$ & 0.28 & 0.49 \\
$\mathrm{V_{avg} [km/s]}$ & 451.9 &  439.5 \\
$\mathrm{V_{beg} [km/s]}$ & 502.8 & 486.9 \\
width [AU]& 2.02 & 1.89 \\
\end{tabular}
\caption{Table of properties for event P}\label{tab:eventP}
\vspace{-20pt}
\end{footnotesize}
\end{center}
\end{wraptable}

As most events in the list, this events shows a low number density, low B-variability and (largely) monotonic decreasing velocity profile. The $\mathrm{T/T_{exp}}$ ratio time series goes above the 0.5 threshold value for the VOY1 measurements inside the shaded region (see Figure~\ref{fig:eventP}), since the latter boundary was chosen according to the other parameters. Still, the data gap at the rear of this event make the location of this boundary ambiguous. The ICME's increasing width, decreasing average number density, and decreasing average B-field magnitude imply that the ICME is still expanding significantly.

\subsubsection*{\textbf{EVENT Q}}

Event Q was detected by both voyagers at distance of 4.91~AU on 04/01/1979 at 23:00 UT for VOY1 and at 4.36~AU on 3/01/1979 at 21:00 UT for VOY2. A summary of its properties can be seen in Table \ref{tab:eventQ}.
\Needspace{100pt}
\begin{wraptable}{r}{0.52\linewidth}
\begin{center}
\begin{footnotesize}
\vspace{-20pt}
\begin{tabular}{c|c|c}
 & VOY1 & VOY 2 \\ \cline{1-3}
r [AU] & 7.15 & 6.2 \\
$\mathrm{N_{avg} [cm^{-3}]}$ & 0.03 & 0.06 \\
$\mathrm{B_{avg} [nT]}$ & 0.35 & 0.46 \\
$\mathrm{V_{avg} [km/s]}$ & 419.8 &  414.9 \\
$\mathrm{V_{beg} [km/s]}$ & 503.4 & 482 \\
width [AU]& 3.86 & 3.36 \\
\end{tabular}
\caption{Table of properties for event Q}\label{tab:eventQ}
\vspace{-20pt}
\end{footnotesize}
\end{center}
\end{wraptable}
There is a large uncertainty on the location of the ICME boundaries for the VOY2 time series. Especially at the front boundary the $\mathrm{T/T_{exp}}$ ratio fluctuates close to the threshold, as shown in Figure~\ref{fig:eventQ}. Other ICME signatures such as the plasma velocity and the magnetic field components were also used to estimate the boundary, though the boundary was not chosen earlier because there is a strong increase in $\mathrm{T/T_{exp}}$ there. However, this might explain the apparent increasing front velocity and width in Table~\ref{tab:eventQ}. The VOY1 boundaries are better defined. Both the average number density and the average magnetic field decrease significantly (50\% and 24\%, respectively), suggesting that though the width increase might be influenced by an erroneous choice of the ICME boundaries, the ICME is still expanding.

\subsubsection*{\textbf{EVENT R}}

Event R was detected by VOY1 and VOY2 on 13/05/1980 at 16:00 UT and on 07/05/1980 at 20:00 UT, respectively. Their radial distance at their time of measurement was 7.94~AU for VOY1 and 6.70~AU for VOY2. A summary of its properties can be seen in Table \ref{tab:eventR}.
\Needspace{100pt}
\begin{wraptable}{r}{0.52\linewidth}
\begin{center}
\begin{footnotesize}
\vspace{-20pt}
\begin{tabular}{c|c|c}
 & VOY1 & VOY 2 \\ \cline{1-3}
r [AU] & 7.94 & 6.70 \\
$\mathrm{N_{avg} [cm^{-3}]}$ & 0.05 & 0.06 \\
$\mathrm{B_{avg} [nT]}$ & 0.20 & 0.29 \\
$\mathrm{V_{avg} [km/s]}$ & 367.9 &  363.5 \\
$\mathrm{V_{beg} [km/s]}$ & 377.3 & 380.9\\
width [AU]& 1.63 & 1.61 \\
\end{tabular}
\caption{Table of properties for event R}\label{tab:eventR}
\vspace{-20pt}
\end{footnotesize}
\end{center}
\end{wraptable}
This event is characterised by a long period of low $\mathrm{T/T_{exp}}$, but with a small peak showing in the middle of the ICME boundaries for both spacecraft, seen in the bottom left panels of both rows of Figure~\ref{fig:eventR}. Since this bump shows up for both spacecraft, it is unlikely that this is a fine structure inside of the ejection. This might imply that this event is the result of two ICMEs merging at some point before they reached Voyager 2. On the other hand, Figure~\ref{fig:voy_comp} shows that the event does not have a larger width compared to other events at a similar distance. In fact, it is relatively small. Thus the nature of this bump is still undetermined, it could be a simple discontinuity. The event itself stays relatively constant over a distance of \textasciitilde1.24~AU as it propagates from VOY1 to VOY2.

\subsubsection*{\textbf{EVENT S AND EVENT T}}

Event S and event T were detected by VOY1 on 05/06/1980 at 19:00 UT and on 29/06/1980 at 01:00 UT, respectively. VOY2 was hit by these events on 01/06/1980 at 19:00 UT and on 23/06/1980 at 17:00 UT, respectively. A summary of its properties can be seen in Table \ref{tab:eventST}.

  \begin{table}[htb!]
    \begin{minipage}{.22\textwidth}
      \centering
      \begin{footnotesize}
\begin{tabular}{c|c|c}
 & VOY1 & VOY 2 \\ \cline{1-3}
r [AU] & 8.15 & 6.84 \\
$\mathrm{N_{avg} [cm^{-3}]}$ & 0.04 & 0.07 \\
$\mathrm{B_{avg} [nT]}$ & 0.16 & 0.33 \\
$\mathrm{V_{avg} [km/s]}$ & 385.9 &  382.9 \\
$\mathrm{V_{beg} [km/s]}$ & 431.4 & 440.2 \\
width [AU] & 3.04 & 3.13 \\
\end{tabular}
\end{footnotesize}
    \end{minipage}
    \begin{minipage}{.22\textwidth}
      \centering
      \begin{footnotesize}
\begin{tabular}{c|c|c}
 & VOY1 & VOY 2 \\ \cline{1-3}
r [AU] & 8.32 & 6.96 \\
$\mathrm{N_{avg} [cm^{-3}]}$ & 0.03 & 0.06 \\
$\mathrm{B_{avg} [nT]}$ & 0.31 & 0.45 \\
$\mathrm{V_{avg} [km/s]}$ & 467.2 & 479.8 \\
$\mathrm{V_{beg} [km/s]}$ & 545.7 & 571.4 \\
width [AU]& 3.34 & 3.14 \\
\end{tabular}
\end{footnotesize}
    \end{minipage}\caption{Table of properties for event S (left) and event T (right)}\label{tab:eventST}
  \end{table}

As with event N and O, events S and T are two events following in quick succession. The front speed of event T is much larger compared to that of event S, meaning that it caught up and is now pressing against event S. This interaction results in event T decelerating more than event S, and the region in between the two events consisting of strongly compressed plasma. Event S stays relatively constant in radial width while event T increases. This event could very well be related to a CIR passing by VOY1 and VOY2.

\subsubsection*{\textbf{EVENT U}}

Event U was detected by VOY1 on 28/08/1980 at 08:00 UT and by VOY2 on 20/08/1980 at 22:00 UT. The difference in radial distance between the two detections is \textasciitilde1.55~AU. A summary of its properties can be seen in Table \ref{tab:eventU}.
\Needspace{100pt}
\begin{wraptable}{r}{0.52\linewidth}
\begin{center}
\begin{footnotesize}
\vspace{-20pt}
\begin{tabular}{c|c|c}
 & VOY1 & VOY 2 \\ \cline{1-3}
r [AU] & 8.84 & 7.29 \\
$\mathrm{N_{avg} [cm^{-3}]}$ & 0.04 & 0.06 \\
$\mathrm{B_{avg} [nT]}$ & 0.30 & 0.28 \\
$\mathrm{V_{avg} [km/s]}$ & 359.6 & 359.1 \\
$\mathrm{V_{beg} [km/s]}$ & 372.6 & 382.5 \\
width [AU]& 1.71 & 2.14 \\
\end{tabular}
\caption{Table of properties for event U}\label{tab:eventU}
\vspace{-20pt}
\end{footnotesize}
\end{center}
\end{wraptable}
This event is actually debatable because its $\mathrm{T/T_{exp}}$ time series does not drop below the threshold long enough. However, the other ICME signatures present, such as a monotonic velocity decrease, low density, rotating magnetic components and low B-field variability resulted in the event being considered as an ICME. The velocity of the event remains constant while the average B-field strength increases and the average density decreases. The differences in the time series for both spacecraft implies that the inconsistencies might be due to each spacecraft passing through a different part of the ICME.

\section{Events detected by Voyager 1 and Voyager 2}\label{appendix_time_series}

This appendix section contains the time series of all events listed in Table~\ref{table:CMEs}. CME detection was done primarily using the low $T/T_{exp}$ condition, which was also used to identify the boundaries. If the event showed other CME characteristics such as a smoothly rotating magnetic field and/or a monotonic decreasing velocity profile, they were also used in combination.

\begin{figure*}[!htb]
\centering
  \centering
  \includegraphics[trim=1cm 1.5cm 1.5cm 2cm, clip=true,width=0.495\linewidth]{voyager_figures/events/voy1/all1_30-5-1978_9-6-1978.png} 
  \includegraphics[trim=1cm 1.5cm 1.5cm 2cm, clip=true,width=0.495\linewidth]{voyager_figures/events/voy1/all2_30-5-1978_9-6-1978.png}
  \includegraphics[trim=1cm 1.5cm 1.5cm 2cm, clip=true,width=0.495\linewidth]{voyager_figures/events/voy2/all1_30-5-1978_9-6-1978.png}
  \includegraphics[trim=1cm 1.5cm 1.5cm 2cm, clip=true,width=0.495\linewidth]{voyager_figures/events/voy2/all2_30-5-1978_9-6-1978.png} 
\caption{Voyager 1 (upper row) and 2 (lower row) time series for Event F.}\label{fig:eventF}
\end{figure*}

\begin{figure*}[!htb]
\centering
  \centering
  \includegraphics[trim=1cm 1.5cm 1.5cm 2cm, clip=true,width=0.495\linewidth]{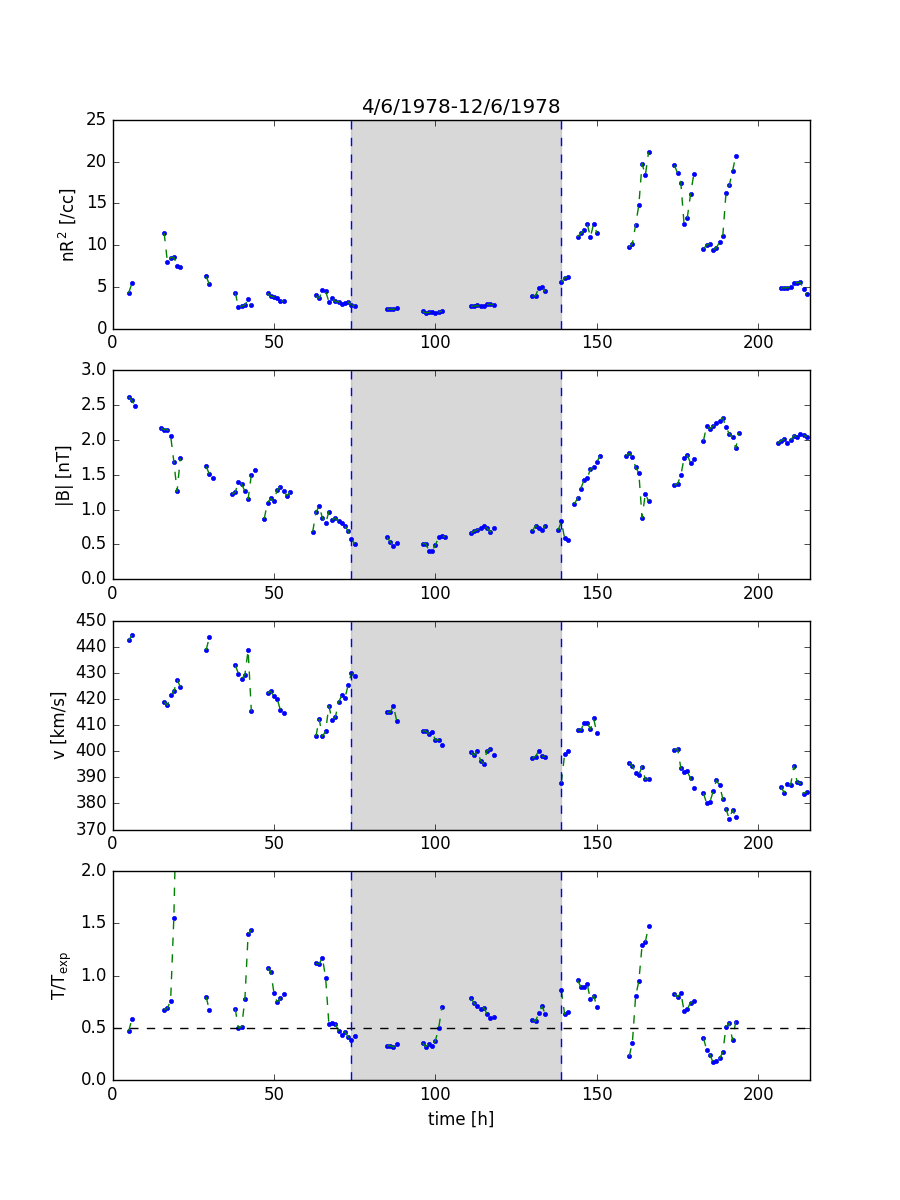} 
  \includegraphics[trim=1cm 1.5cm 1.5cm 2cm, clip=true,width=0.495\linewidth]{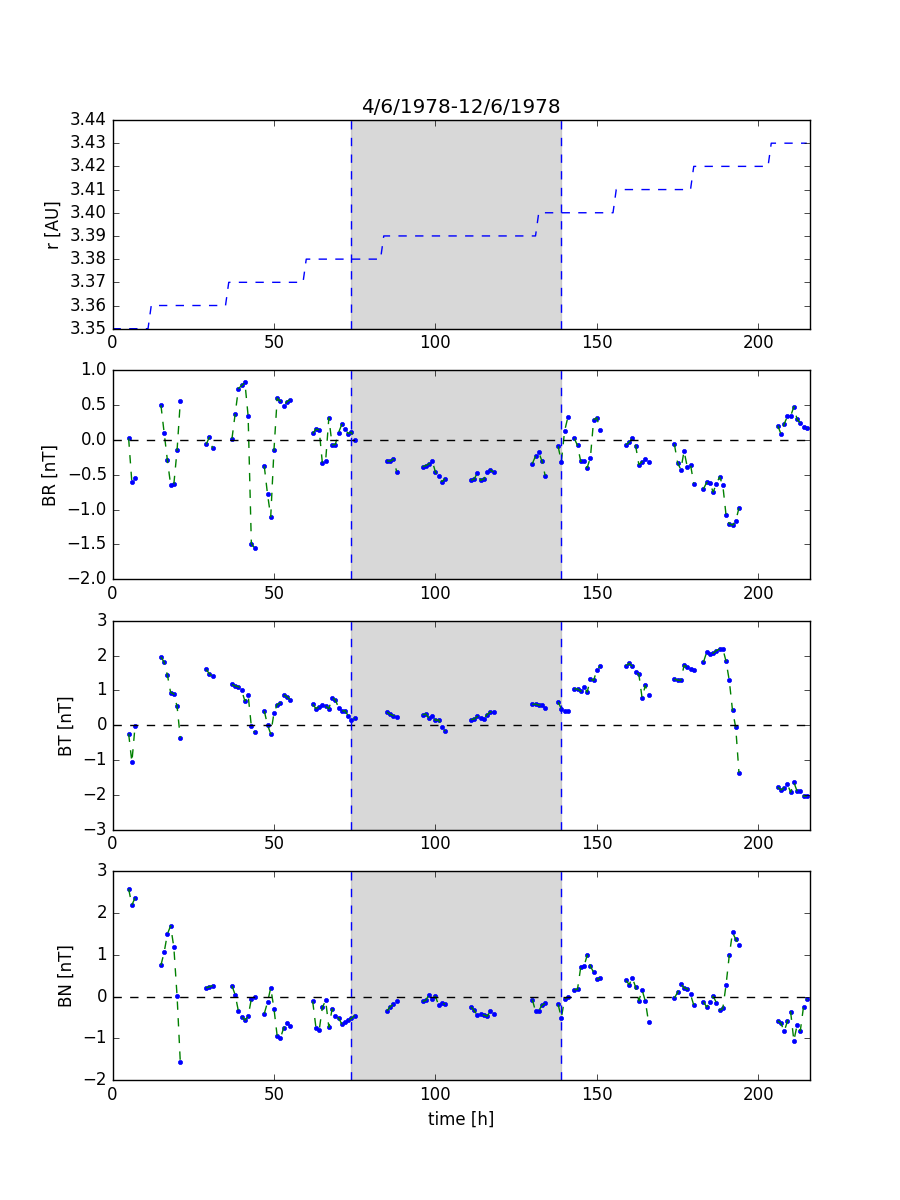}
  \includegraphics[trim=1cm 1.5cm 1.5cm 2cm, clip=true,width=0.495\linewidth]{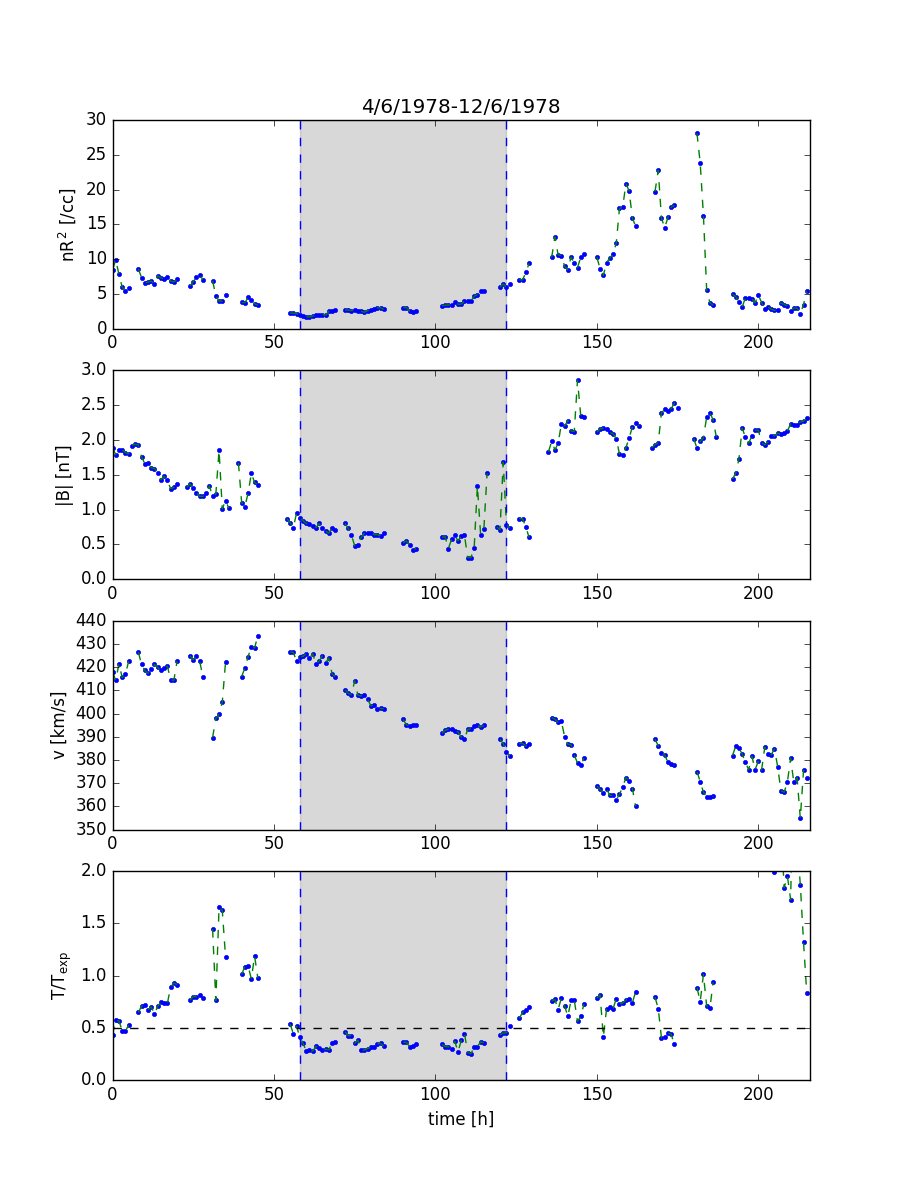}
  \includegraphics[trim=1cm 1.5cm 1.5cm 2cm, clip=true,width=0.495\linewidth]{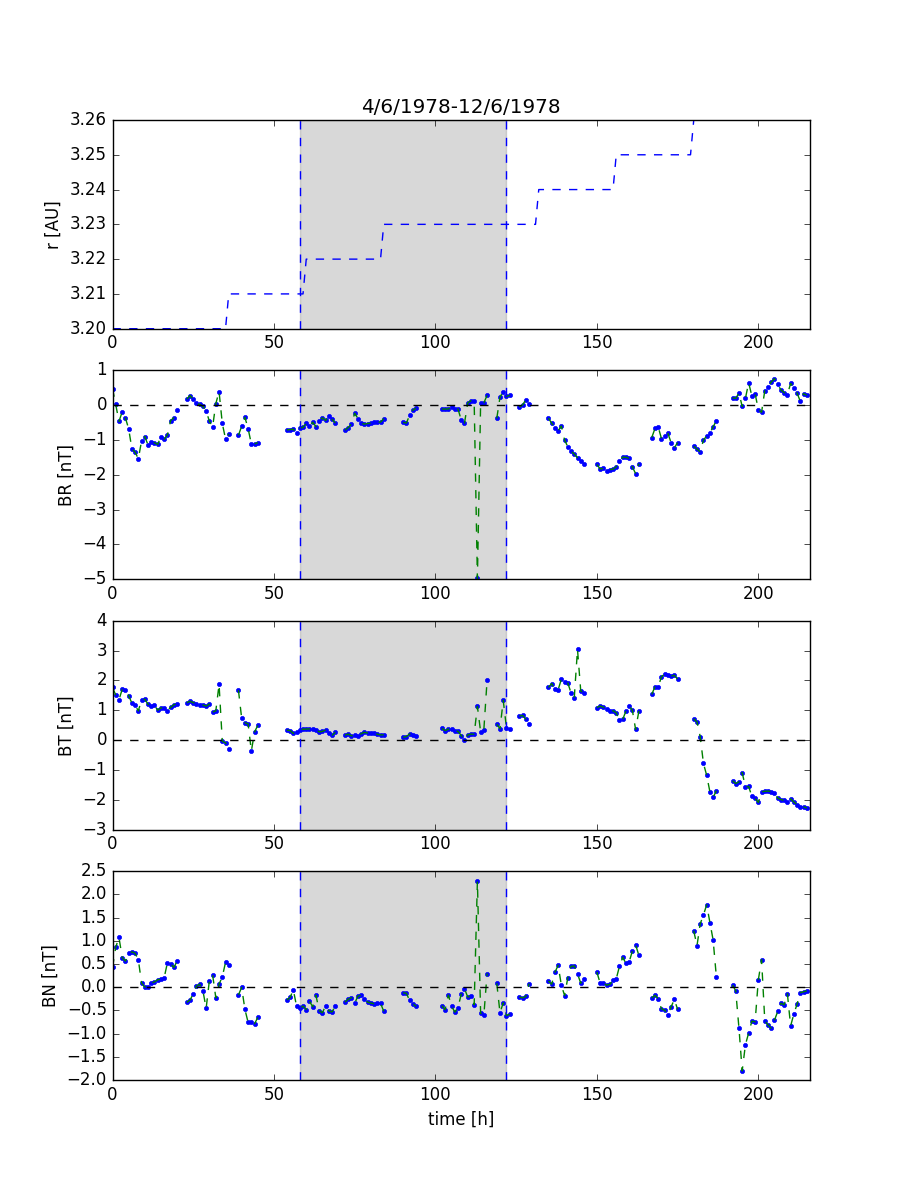} 
\caption{Voyager 1 (upper row) and 2 (lower row) time series for Event G.}\label{fig:eventG}
\end{figure*}

\begin{figure*}[!htb]
\centering
  \centering
  \includegraphics[trim=1cm 1.5cm 1.5cm 2cm, clip=true,width=0.495\linewidth]{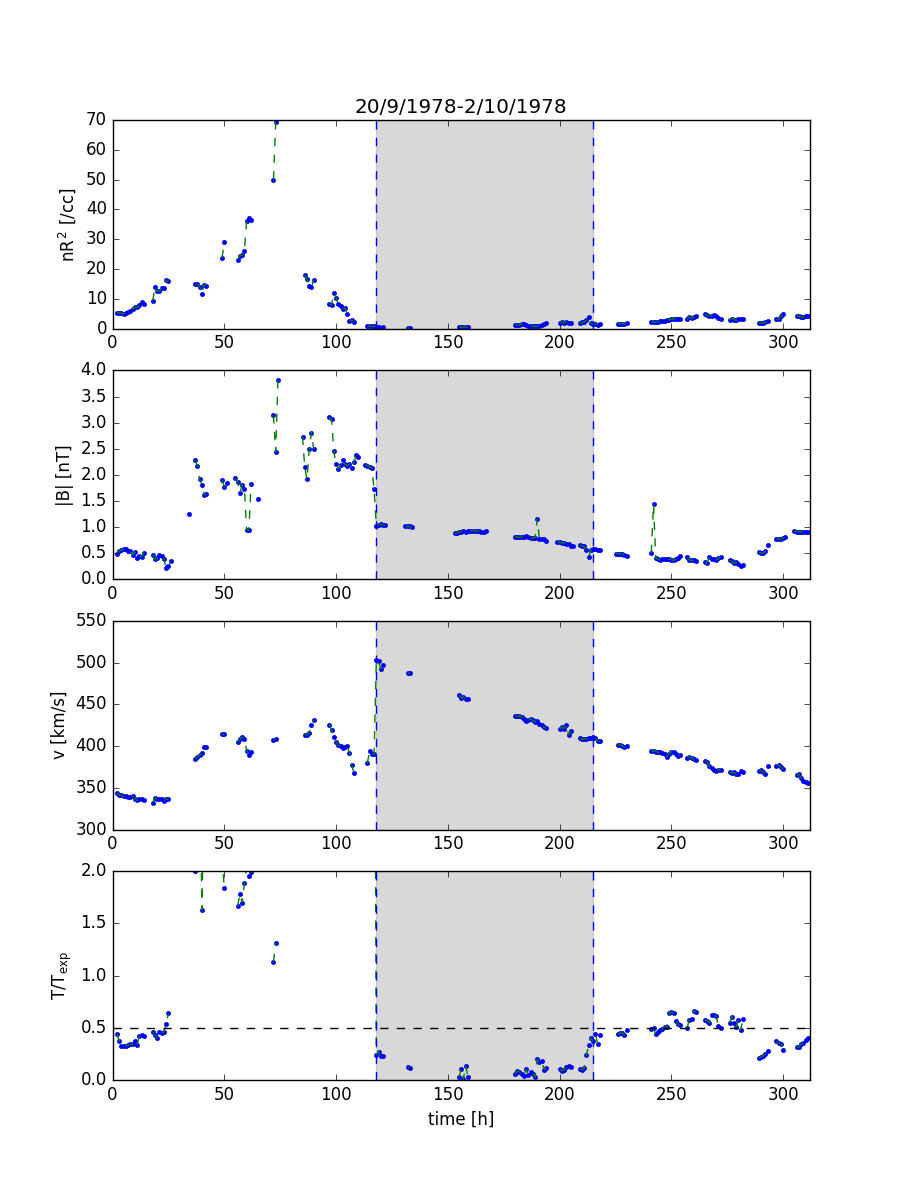} 
  \includegraphics[trim=1cm 1.5cm 1.5cm 2cm, clip=true,width=0.495\linewidth]{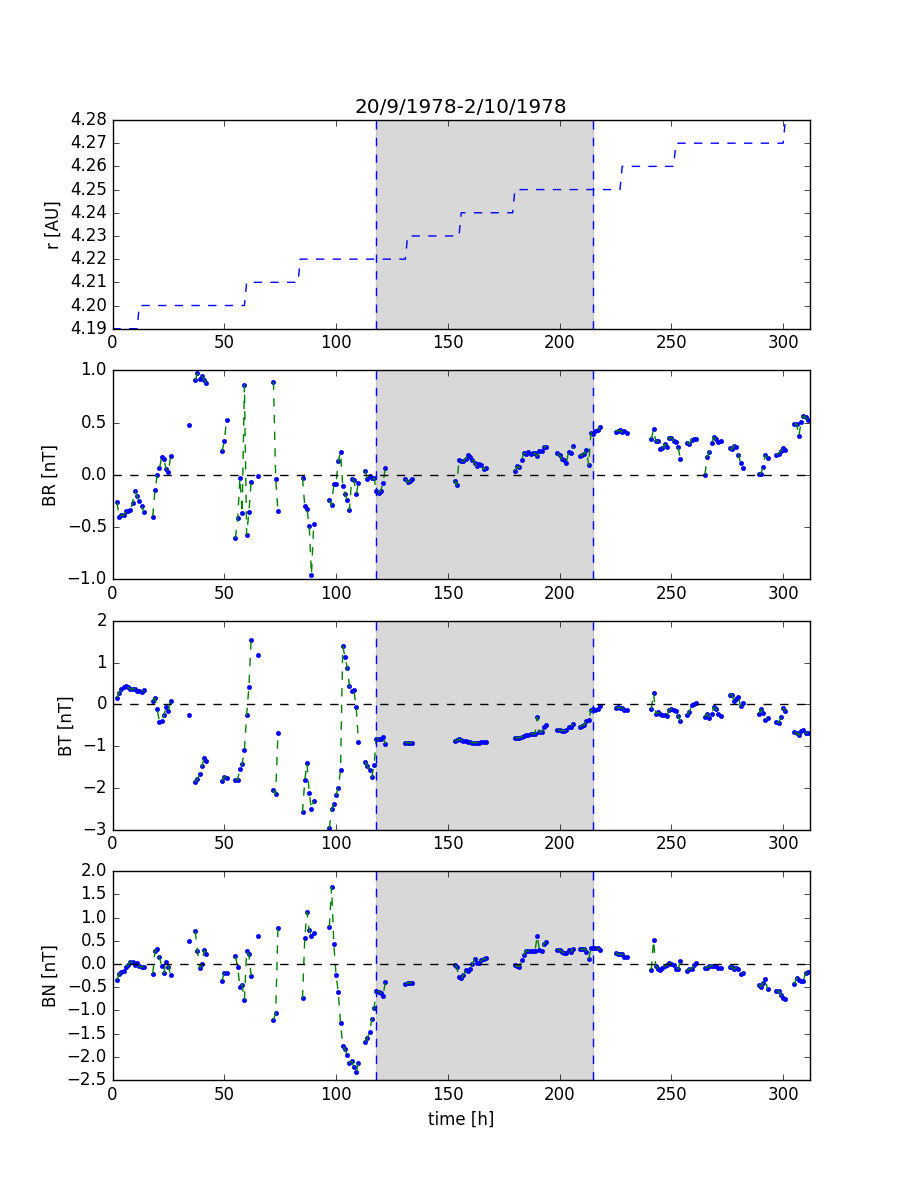}
  \includegraphics[trim=1cm 1.5cm 1.5cm 2cm, clip=true,width=0.495\linewidth]{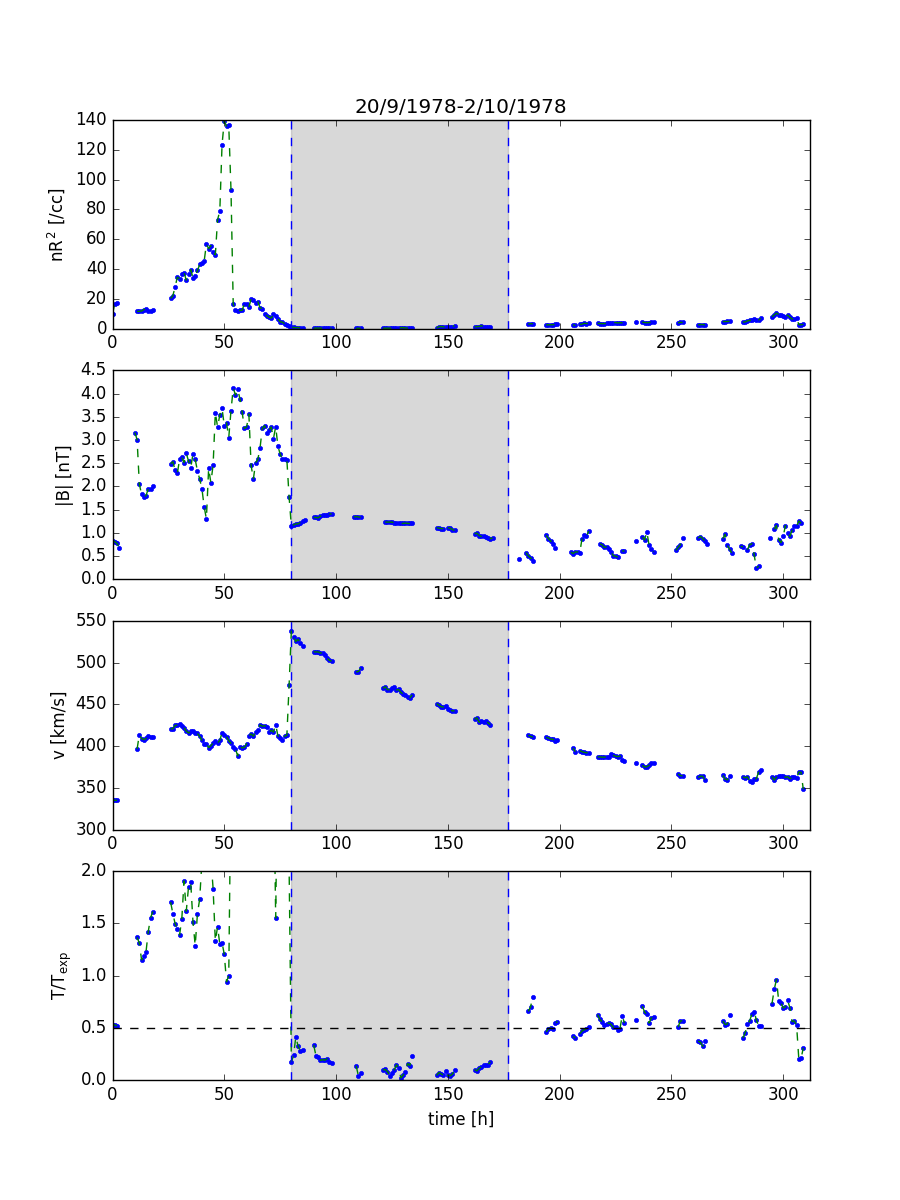}
  \includegraphics[trim=1cm 1.5cm 1.5cm 2cm, clip=true,width=0.495\linewidth]{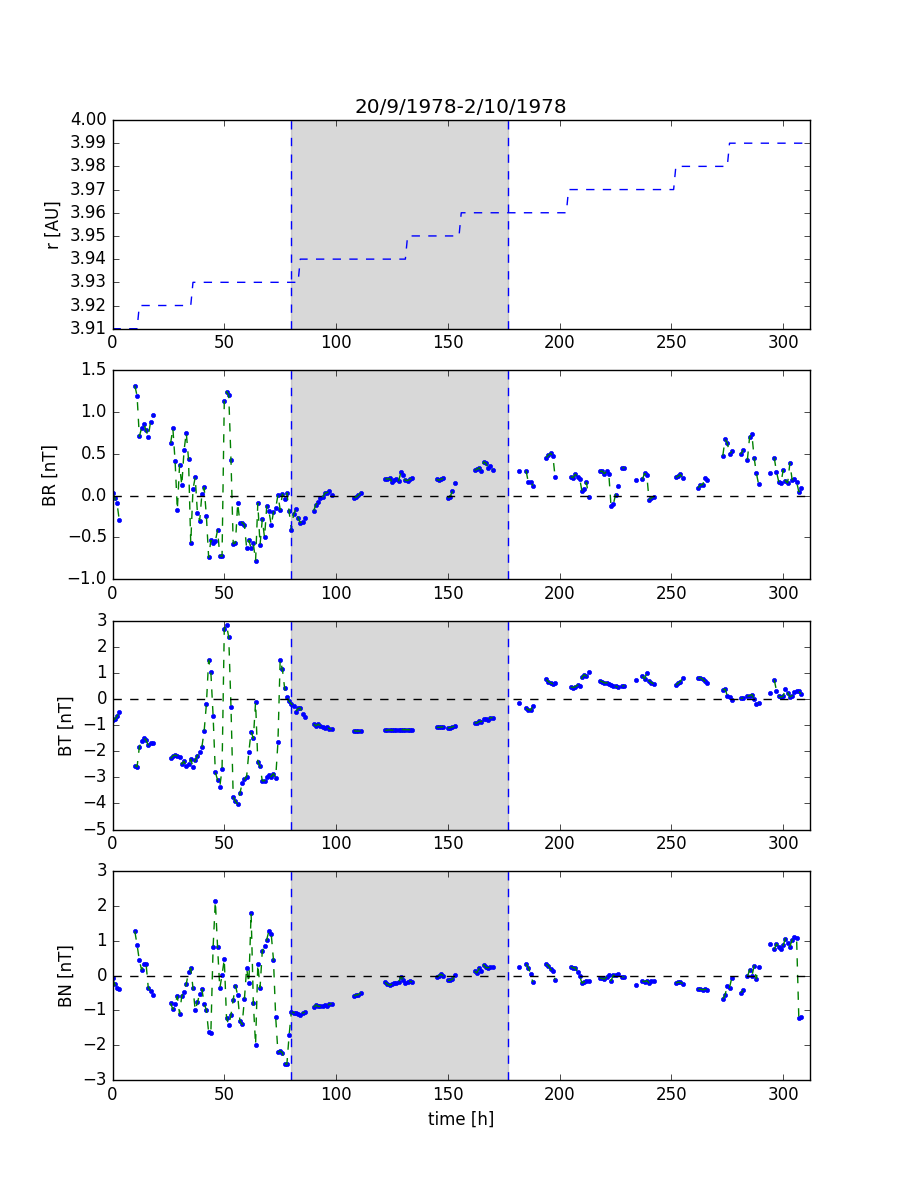} 
\caption{Voyager 1 (upper row) and 2 (lower row) time series for Event H.}\label{fig:eventH}
\end{figure*}

\begin{figure*}[!htb]
\centering
  \centering
  \includegraphics[trim=1cm 1.5cm 1.5cm 2cm, clip=true,width=0.495\linewidth]{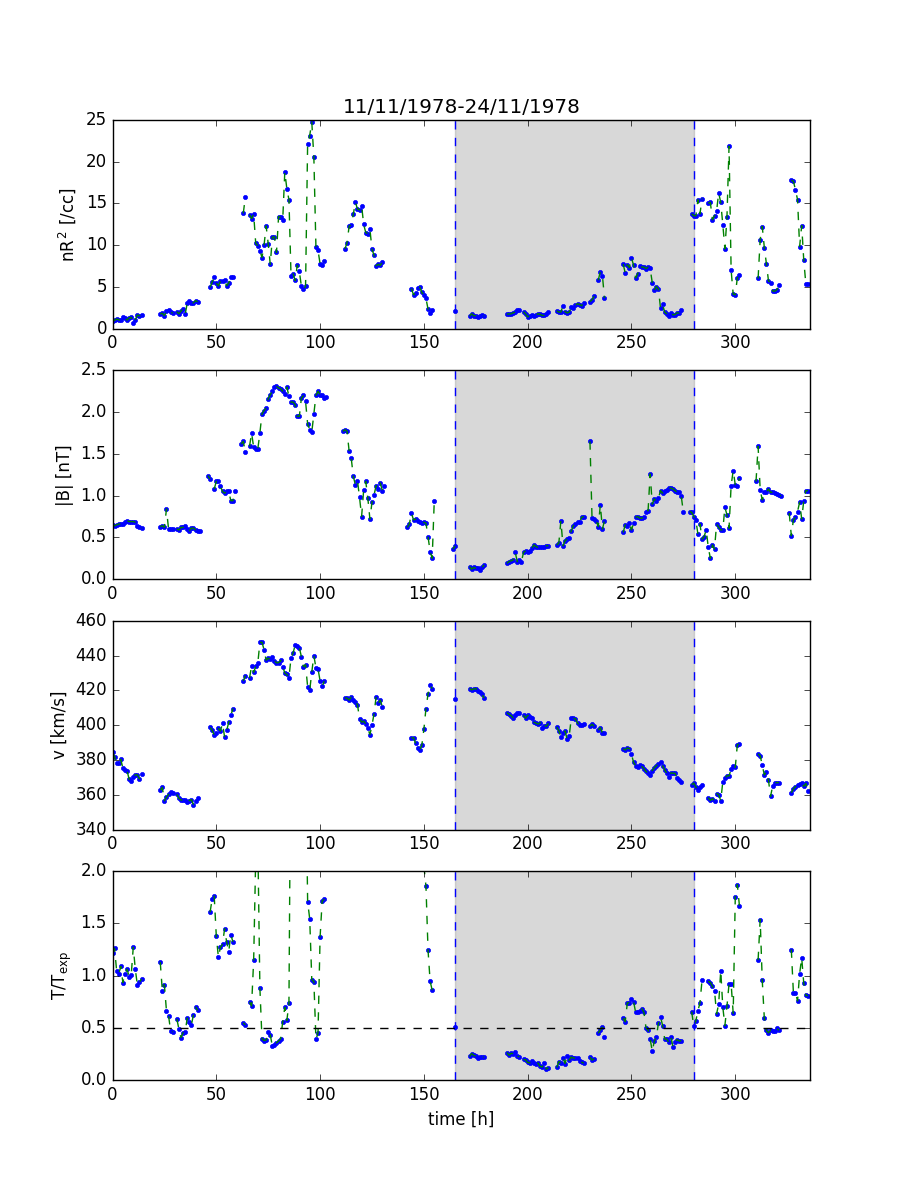} 
  \includegraphics[trim=1cm 1.5cm 1.5cm 2cm, clip=true,width=0.495\linewidth]{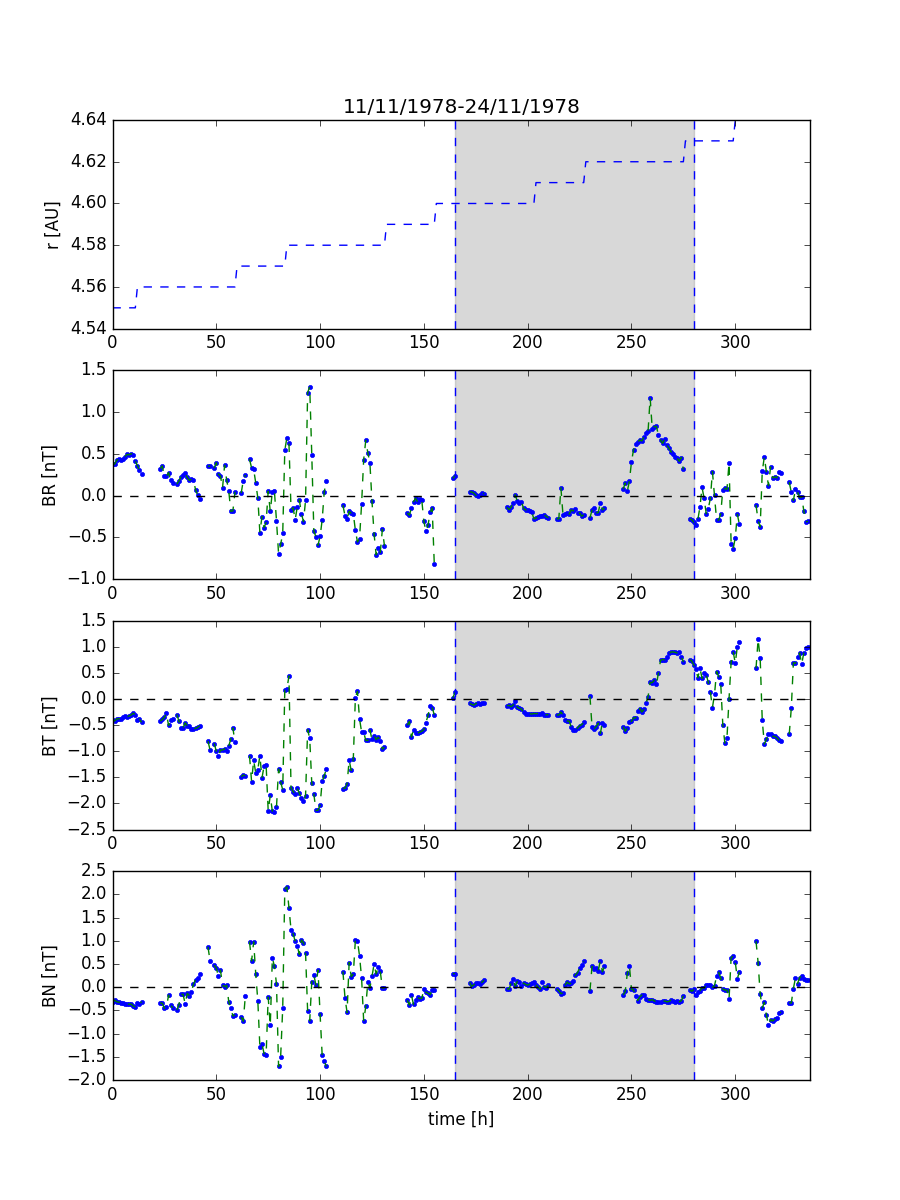}
  \includegraphics[trim=1cm 1.5cm 1.5cm 2cm, clip=true,width=0.495\linewidth]{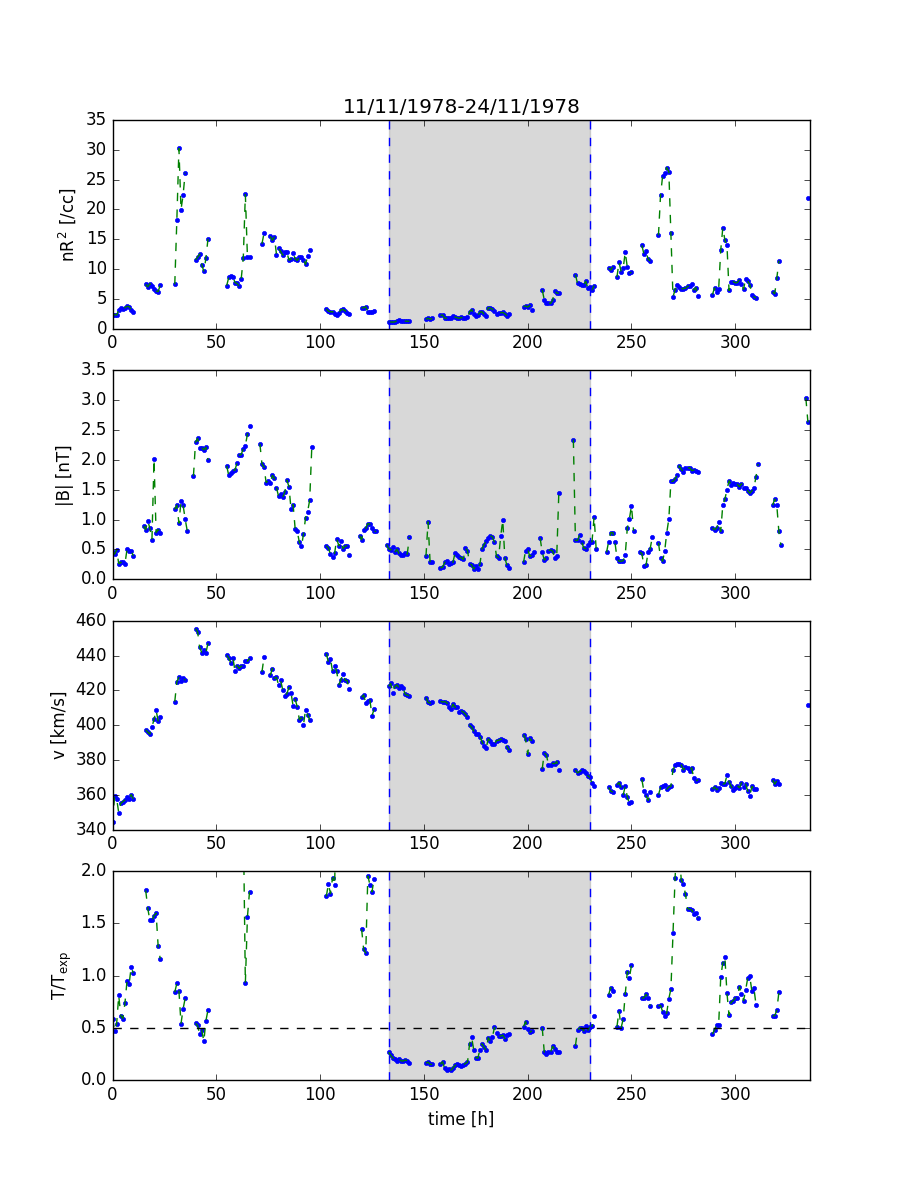}
  \includegraphics[trim=1cm 1.5cm 1.5cm 2cm, clip=true,width=0.495\linewidth]{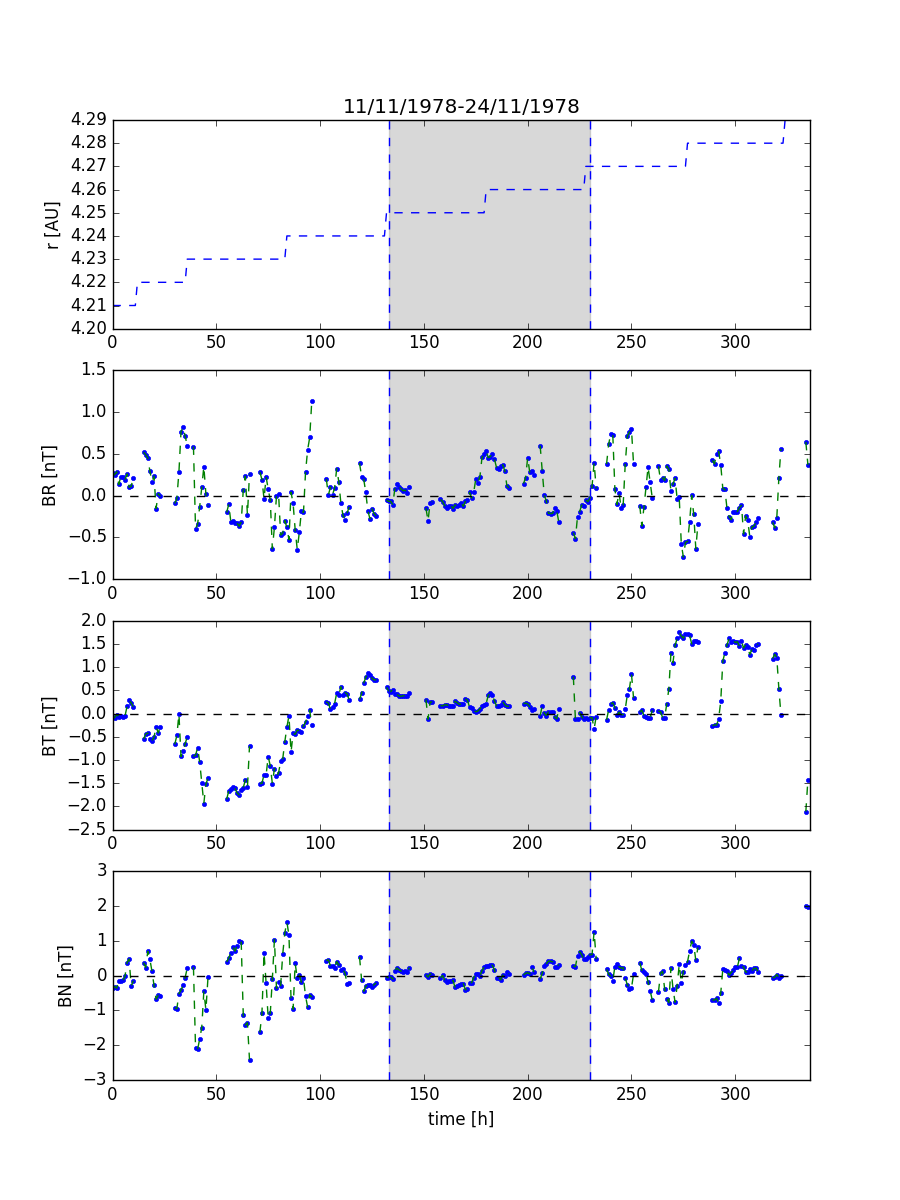} 
\caption{Voyager 1 (upper row) and 2 (lower row) time series for Event I.}\label{fig:eventI}
\end{figure*}

\begin{figure*}[!htb]
\centering
  \centering
  \includegraphics[trim=1cm 1.5cm 1.5cm 2cm, clip=true,width=0.495\linewidth]{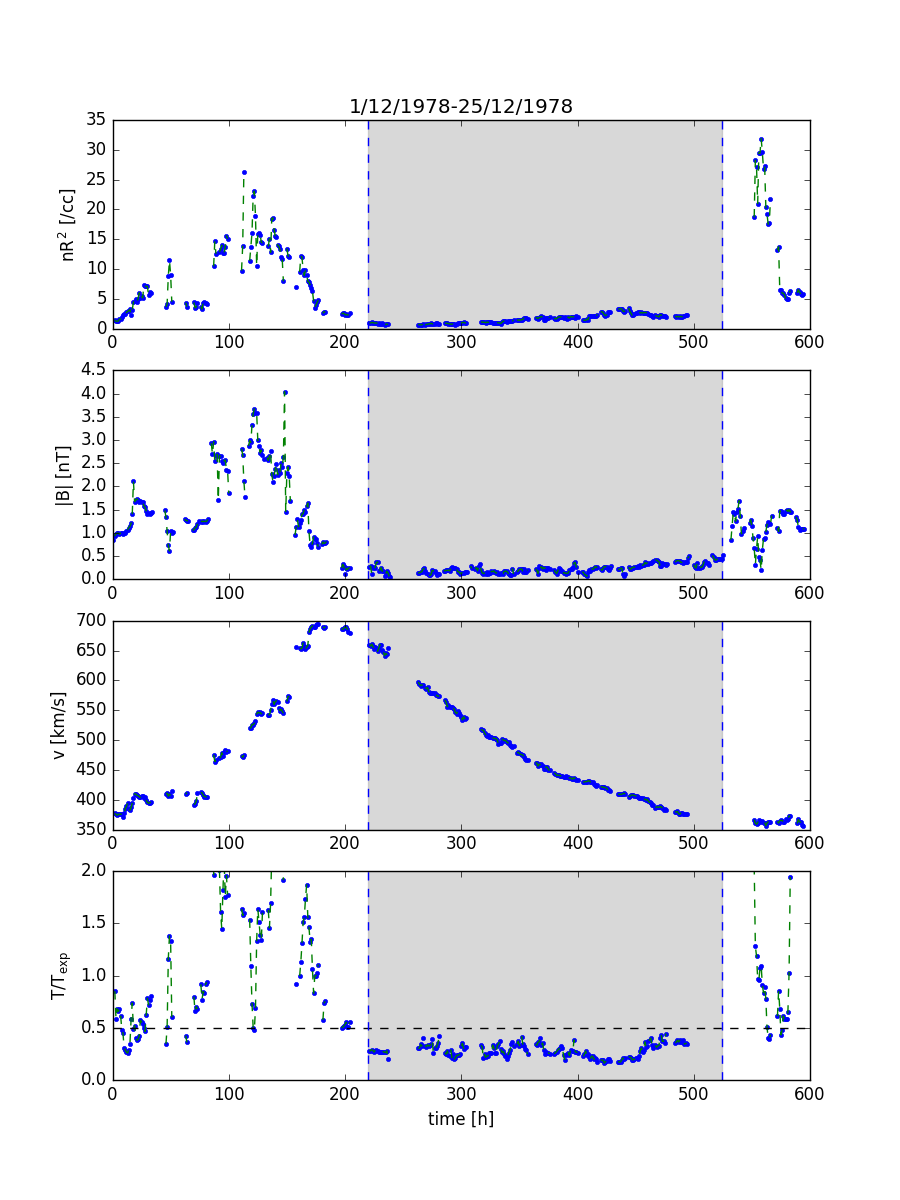} 
  \includegraphics[trim=1cm 1.5cm 1.5cm 2cm, clip=true,width=0.495\linewidth]{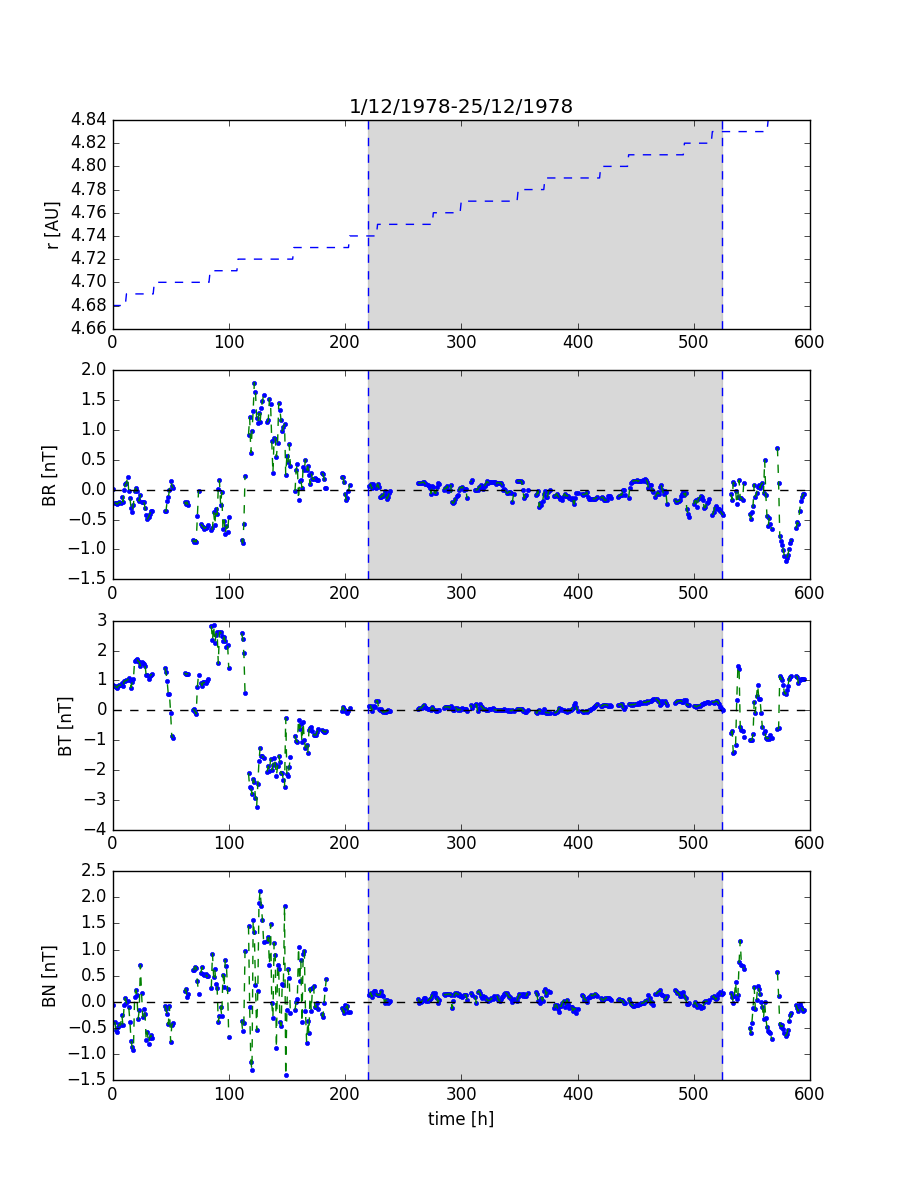}
  \includegraphics[trim=1cm 1.5cm 1.5cm 2cm, clip=true,width=0.495\linewidth]{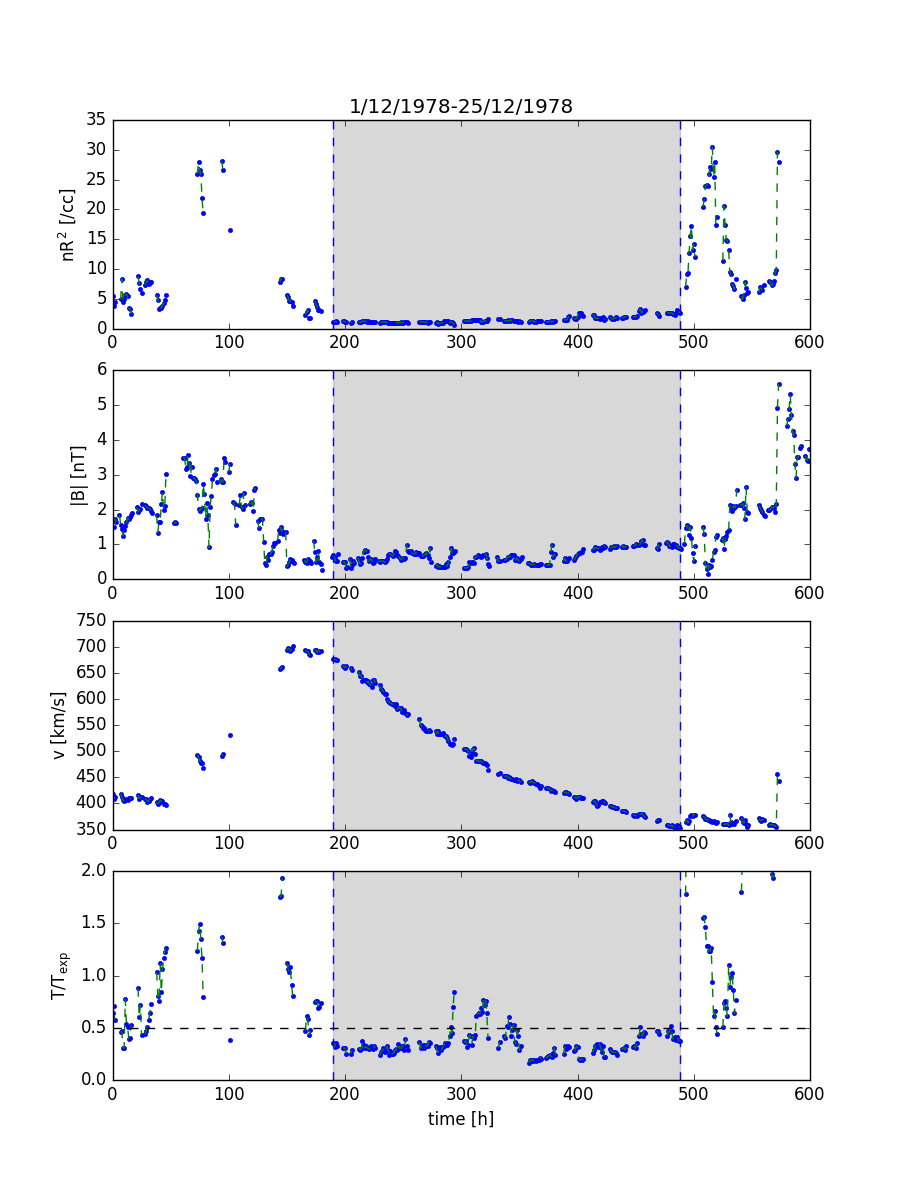}
  \includegraphics[trim=1cm 1.5cm 1.5cm 2cm, clip=true,width=0.495\linewidth]{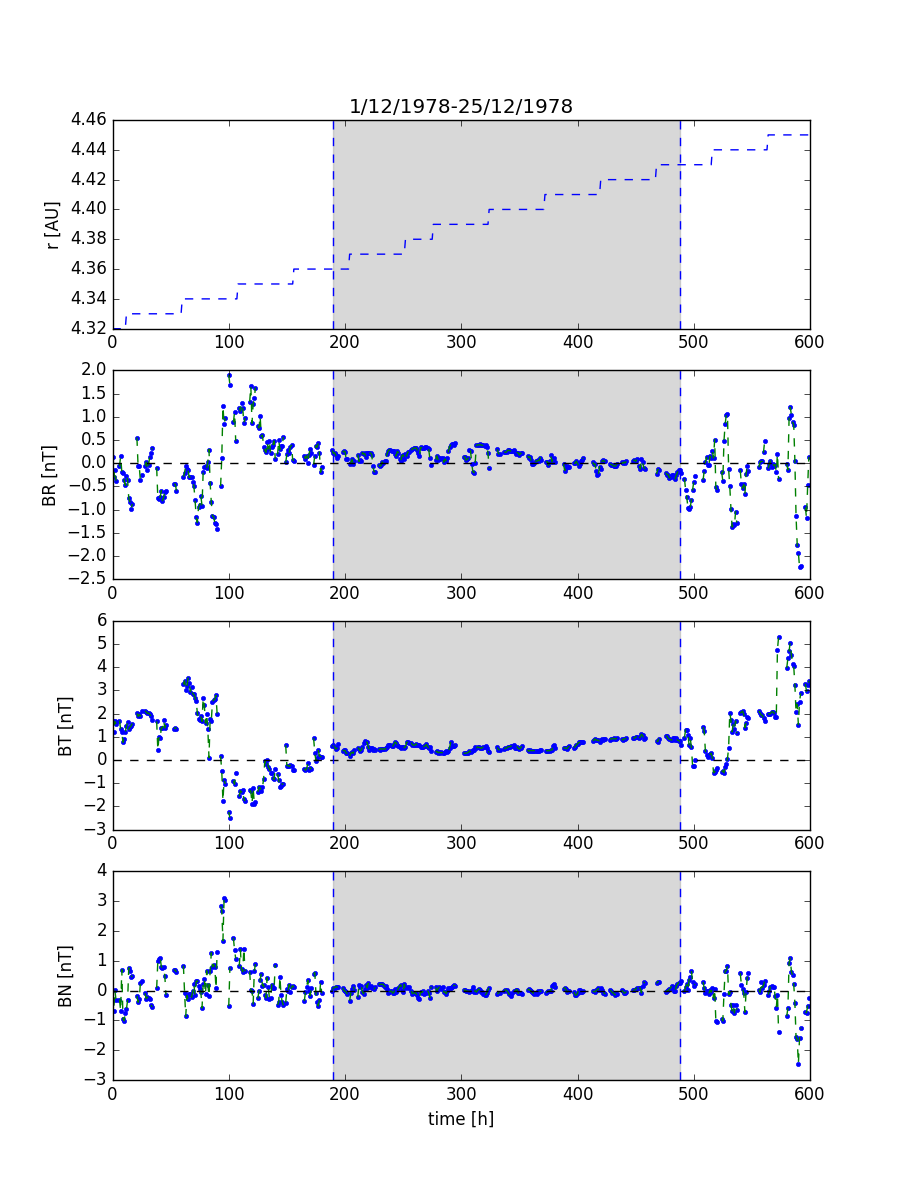} 
\caption{Voyager 1 (upper row) and 2 (lower row) time series for Event J.}\label{fig:eventJ}
\end{figure*}

\begin{figure*}[!htb]
\centering
  \centering
  \includegraphics[trim=1cm 1.5cm 1.5cm 2cm, clip=true,width=0.495\linewidth]{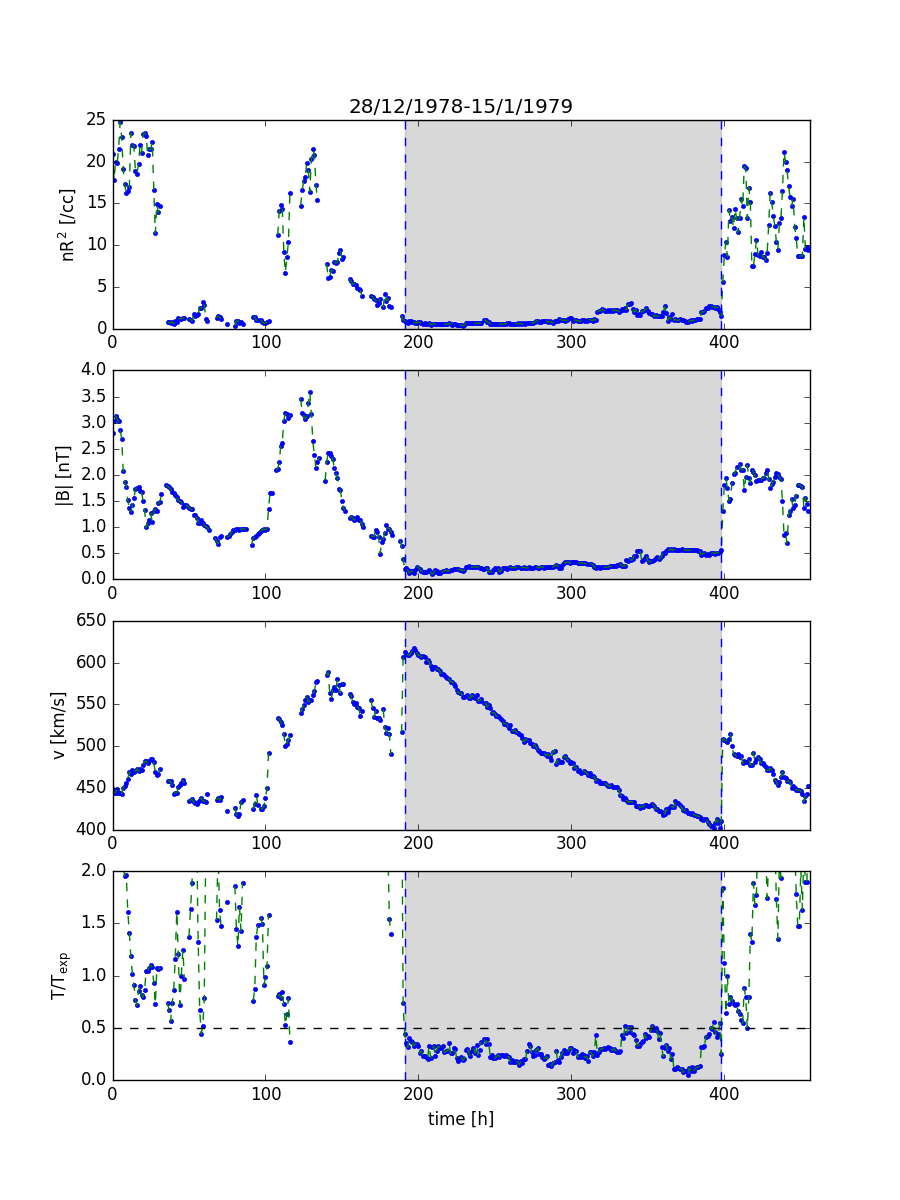} 
  \includegraphics[trim=1cm 1.5cm 1.5cm 2cm, clip=true,width=0.495\linewidth]{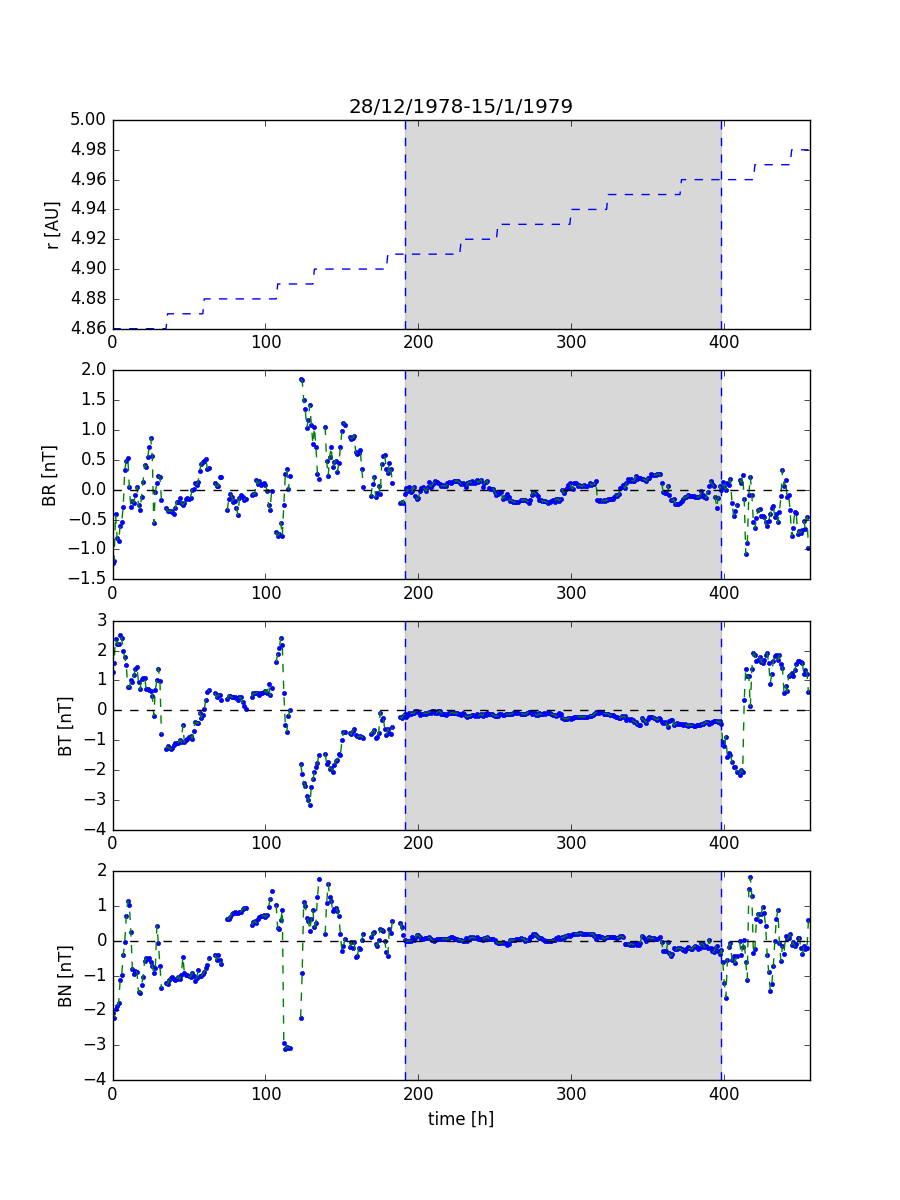}
  \includegraphics[trim=1cm 1.5cm 1.5cm 2cm, clip=true,width=0.495\linewidth]{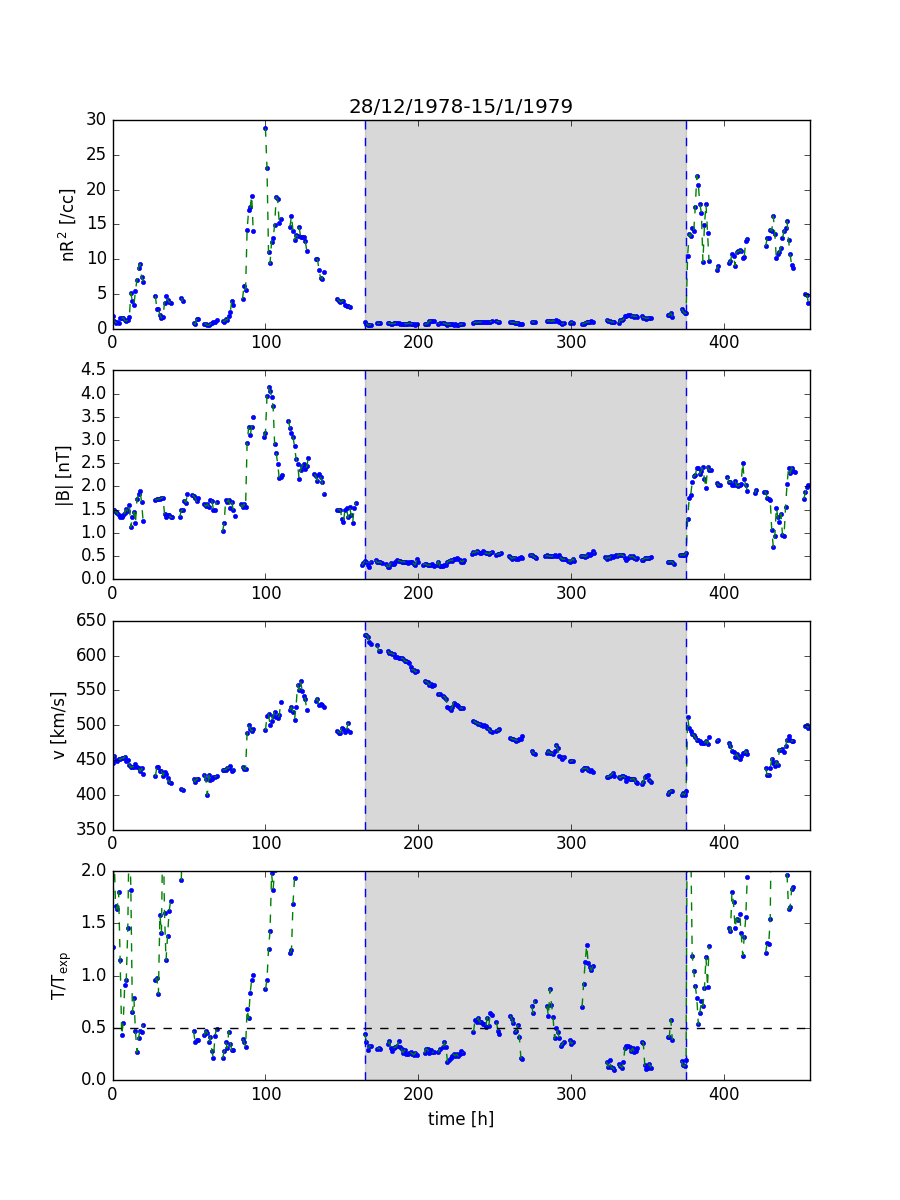}
  \includegraphics[trim=1cm 1.5cm 1.5cm 2cm, clip=true,width=0.495\linewidth]{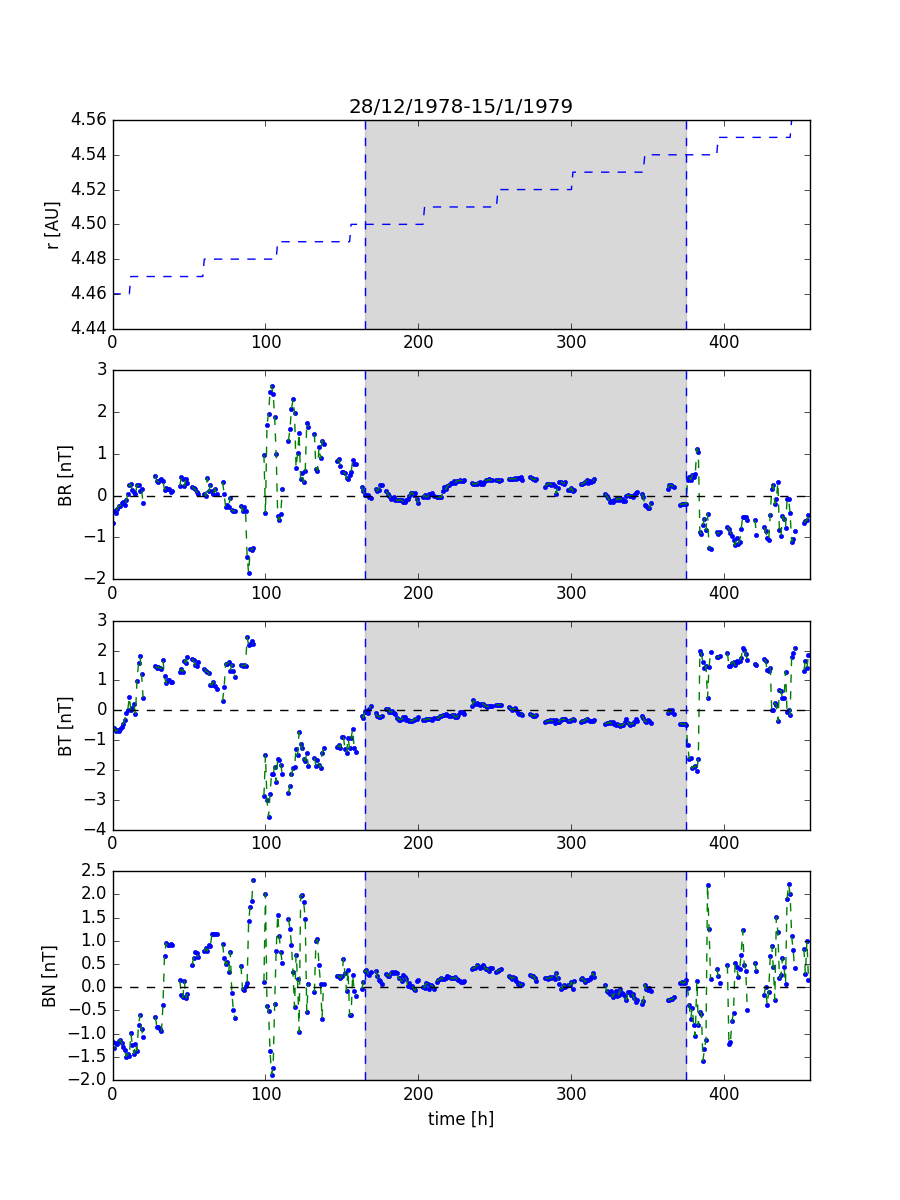} 
\caption{Voyager 1 (upper row) and 2 (lower row) time series for Event K.}\label{fig:eventK}
\end{figure*}

\begin{figure*}[!htb]
\centering
  \centering
  \includegraphics[trim=1cm 1.5cm 1.5cm 2cm, clip=true,width=0.495\linewidth]{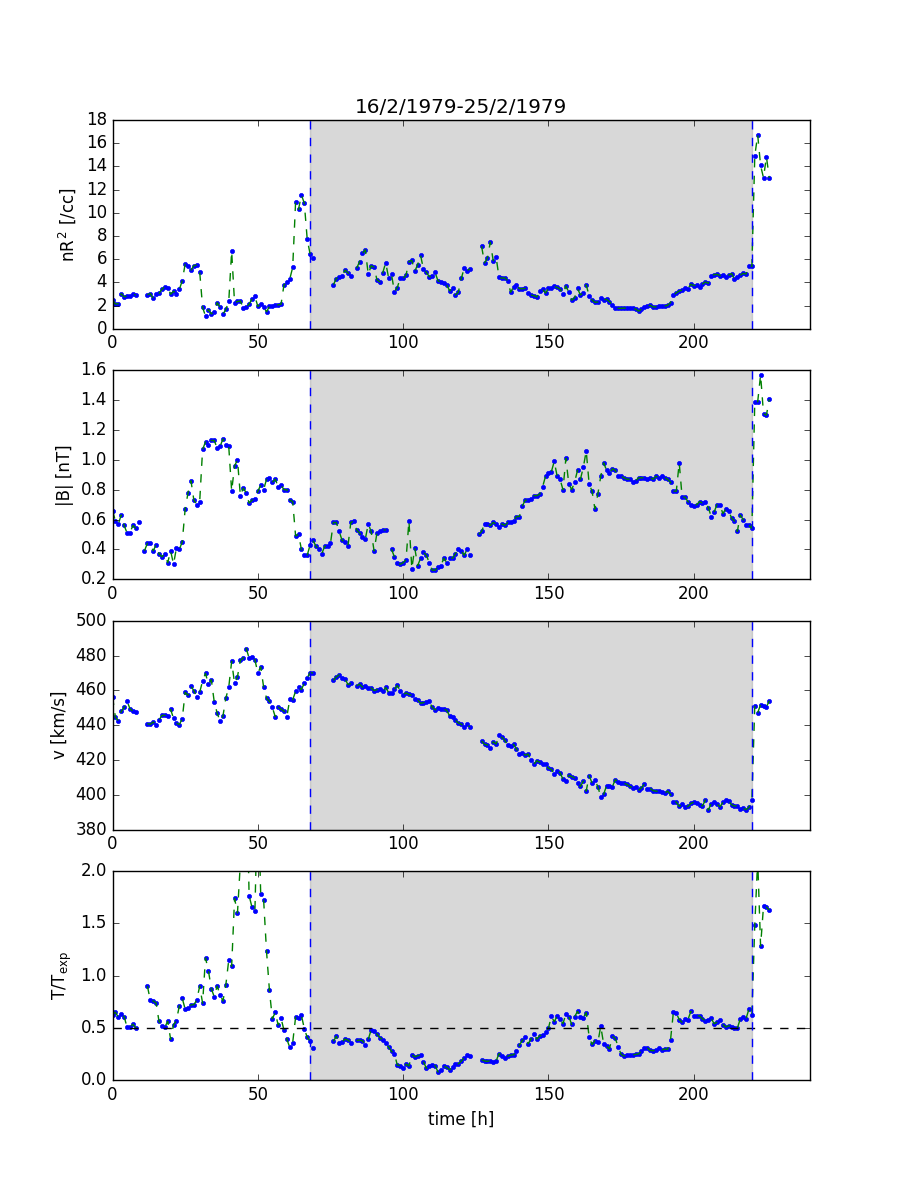} 
  \includegraphics[trim=1cm 1.5cm 1.5cm 2cm, clip=true,width=0.495\linewidth]{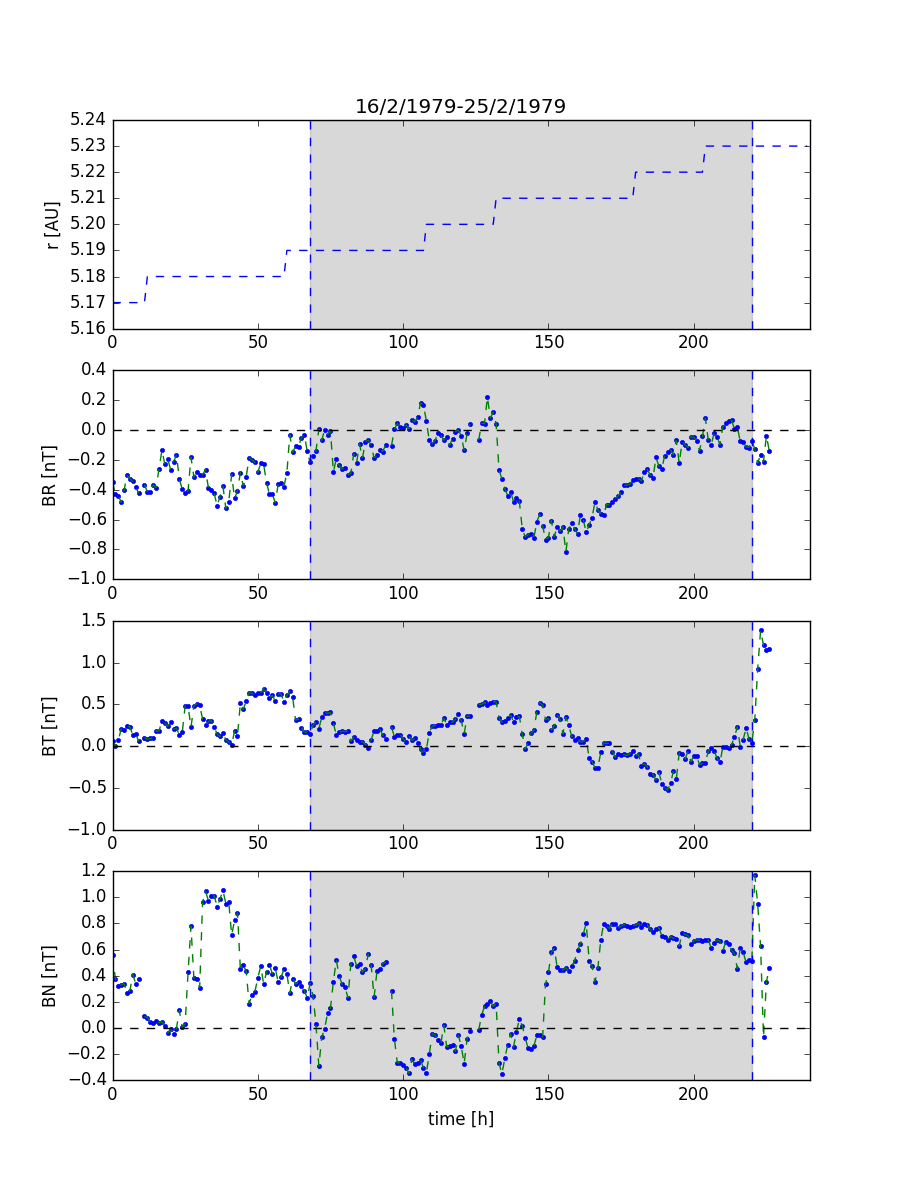}
  \includegraphics[trim=1cm 1.5cm 1.5cm 2cm, clip=true,width=0.495\linewidth]{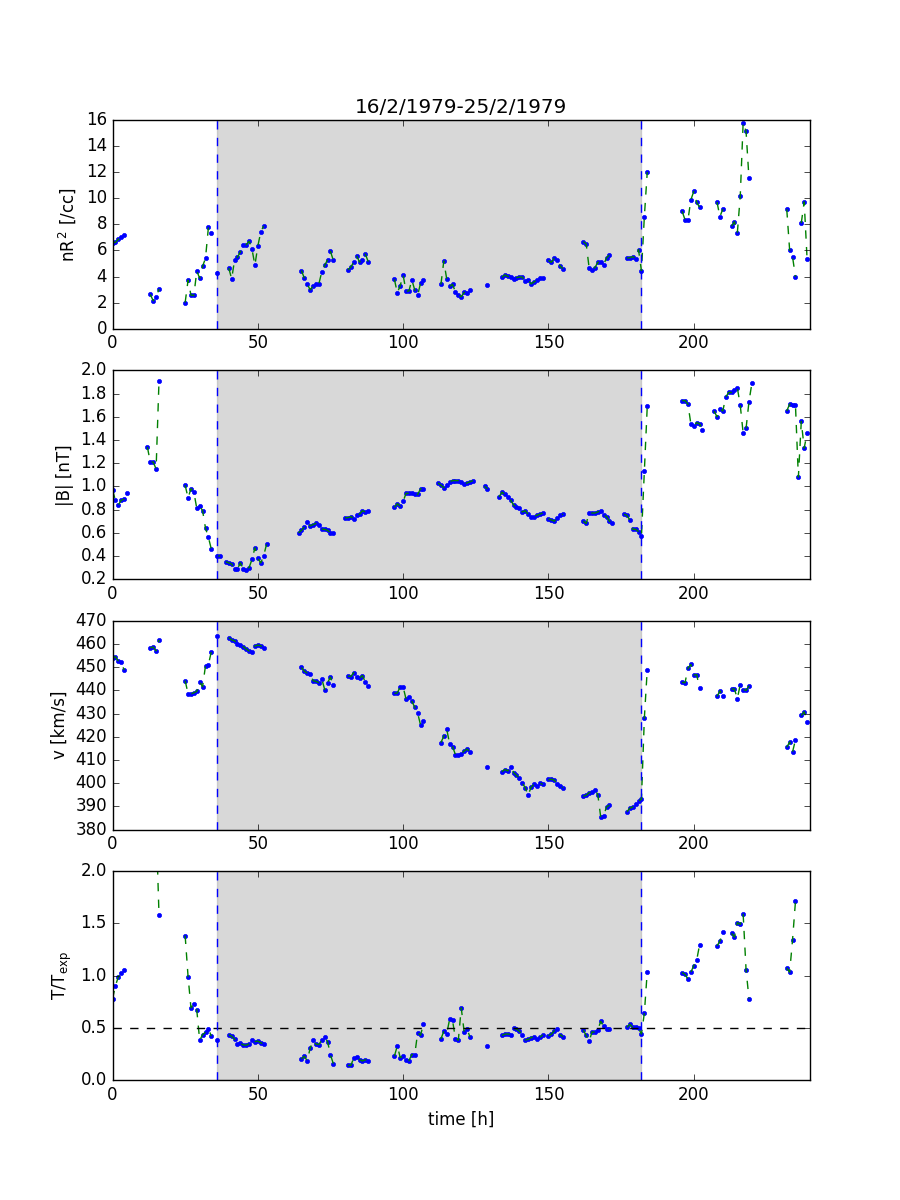}
  \includegraphics[trim=1cm 1.5cm 1.5cm 2cm, clip=true,width=0.495\linewidth]{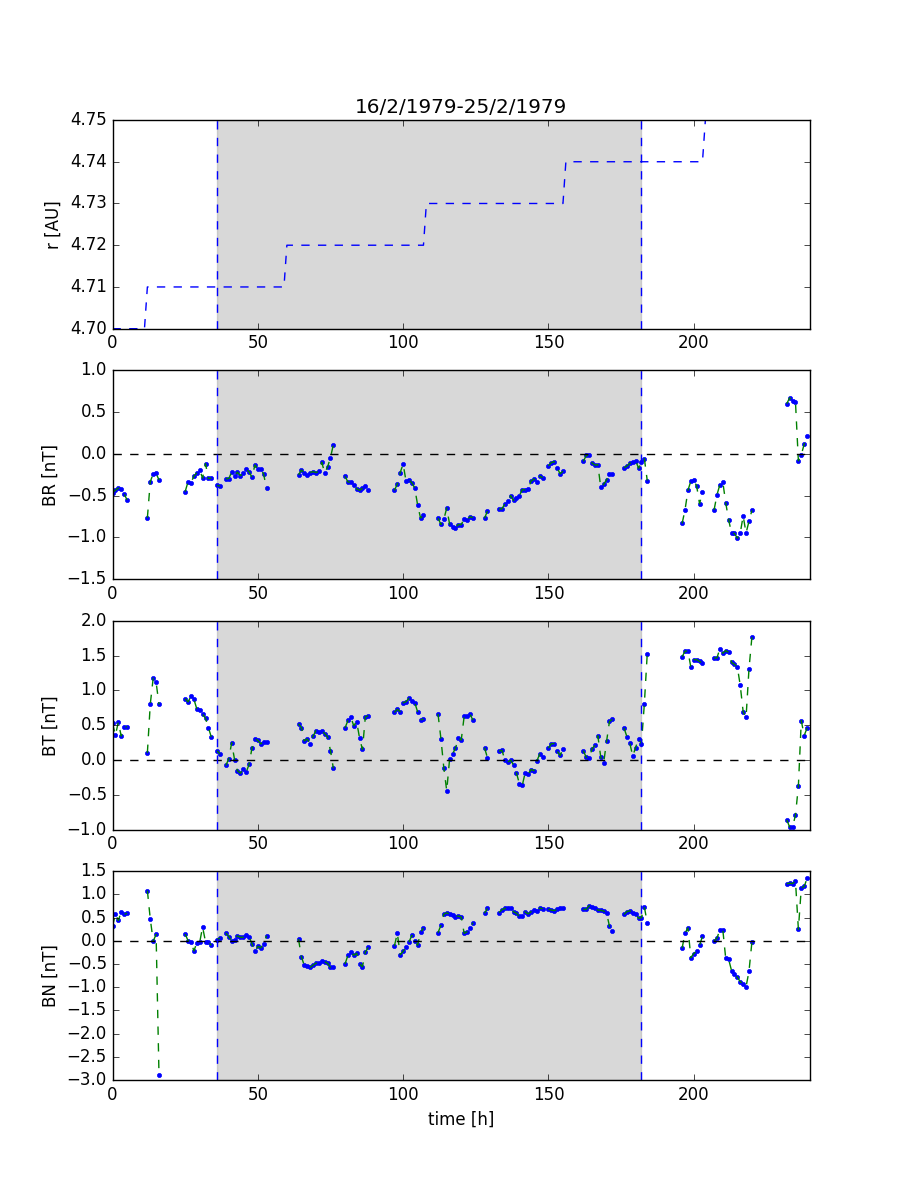} 
\caption{Voyager 1 (upper row) and 2 (lower row) time series for Event L.}\label{fig:eventL}
\end{figure*}

\begin{figure*}[!htb]
\centering
  \centering
  \includegraphics[trim=1cm 1.5cm 1.5cm 2cm, clip=true,width=0.495\linewidth]{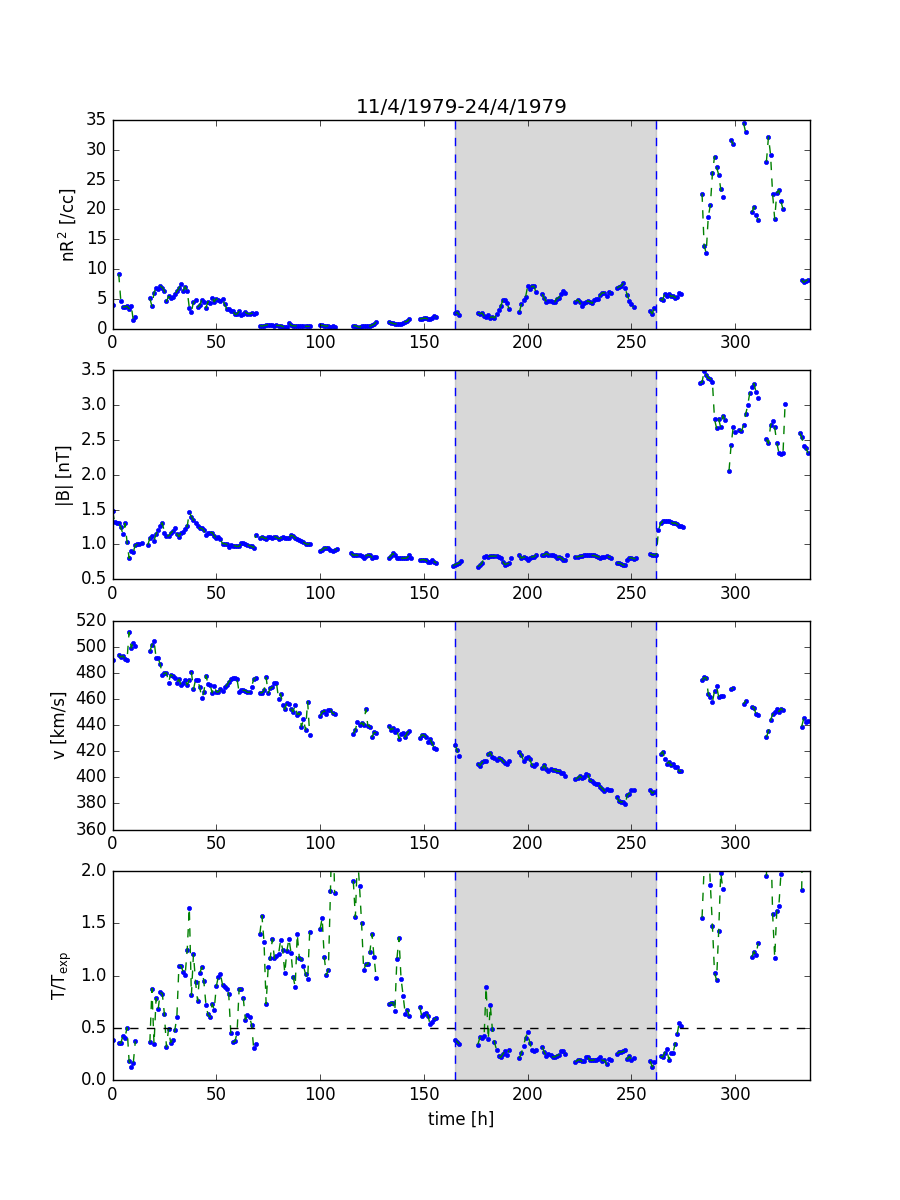} 
  \includegraphics[trim=1cm 1.5cm 1.5cm 2cm, clip=true,width=0.495\linewidth]{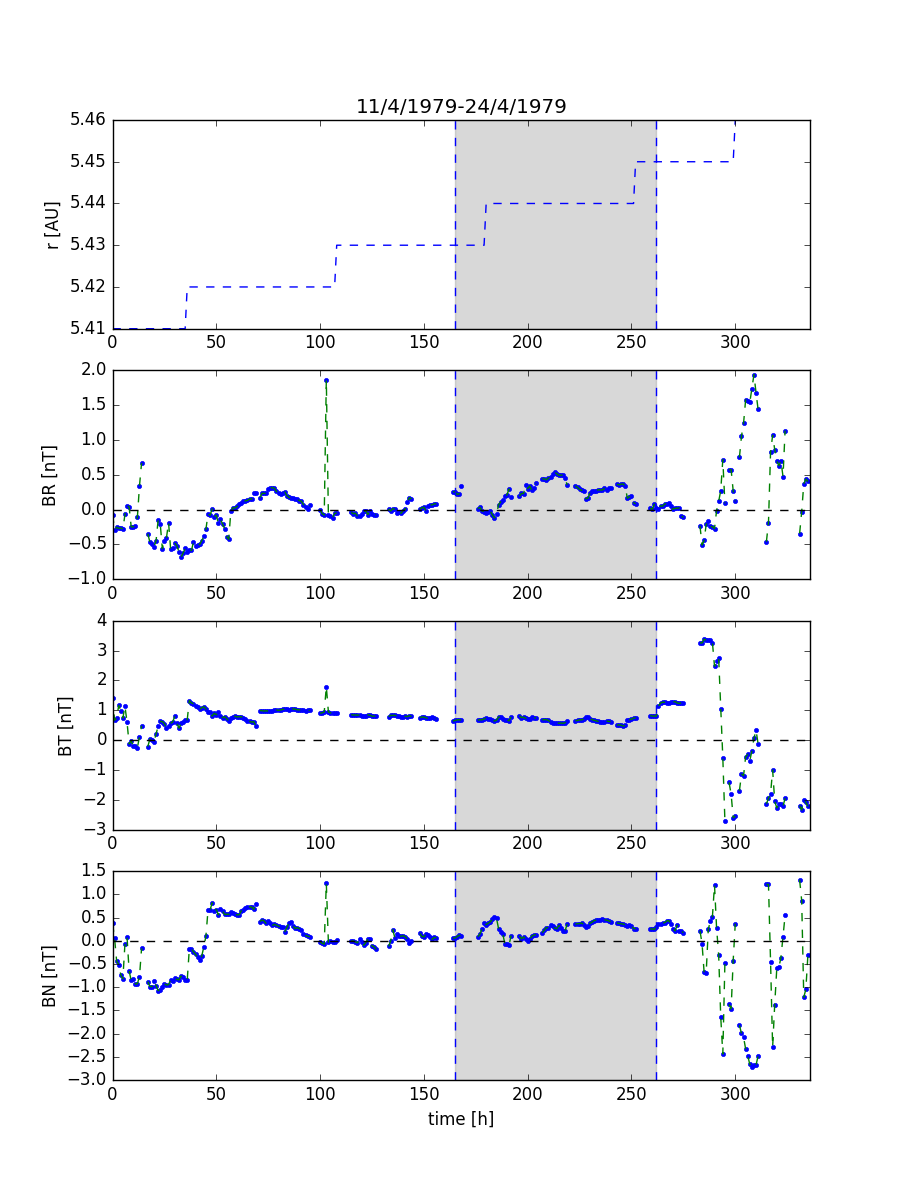}
  \includegraphics[trim=1cm 1.5cm 1.5cm 2cm, clip=true,width=0.495\linewidth]{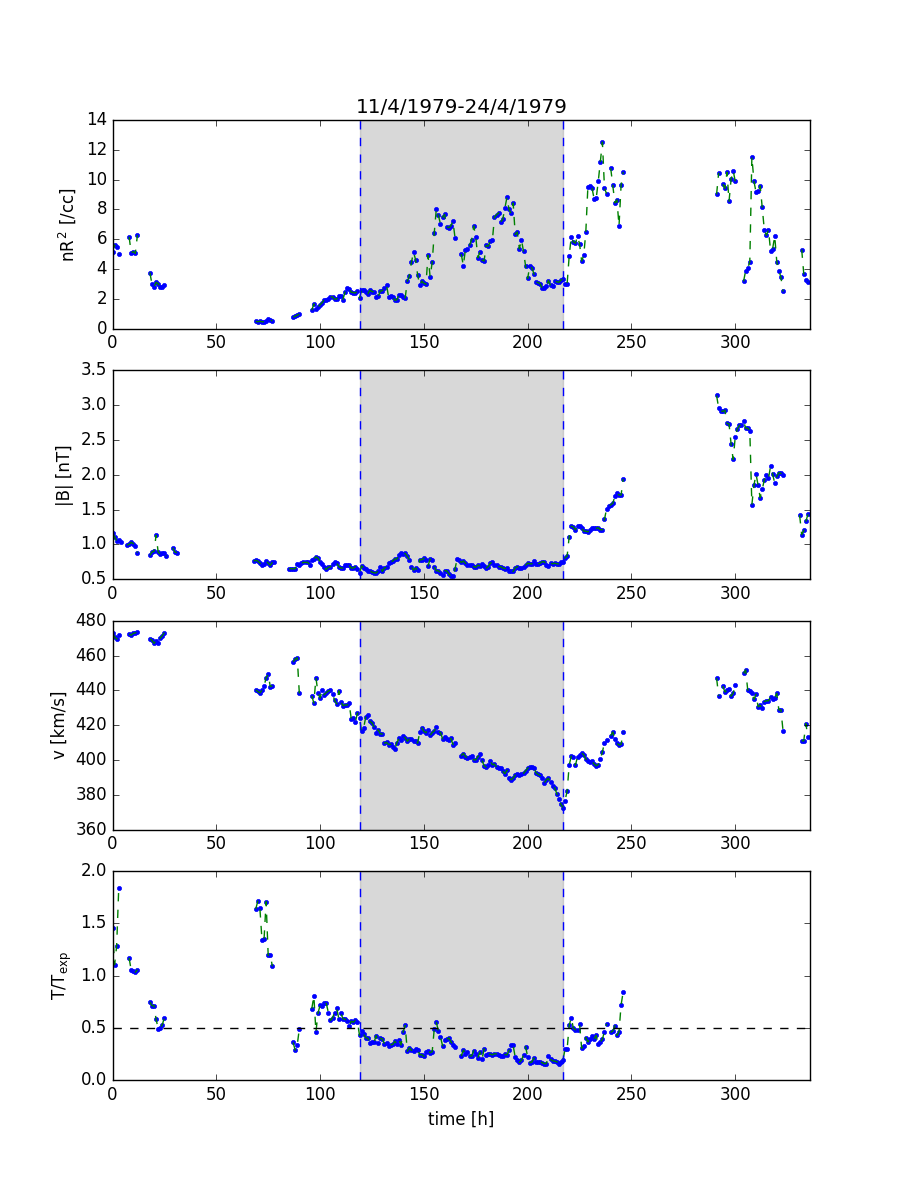}
  \includegraphics[trim=1cm 1.5cm 1.5cm 2cm, clip=true,width=0.495\linewidth]{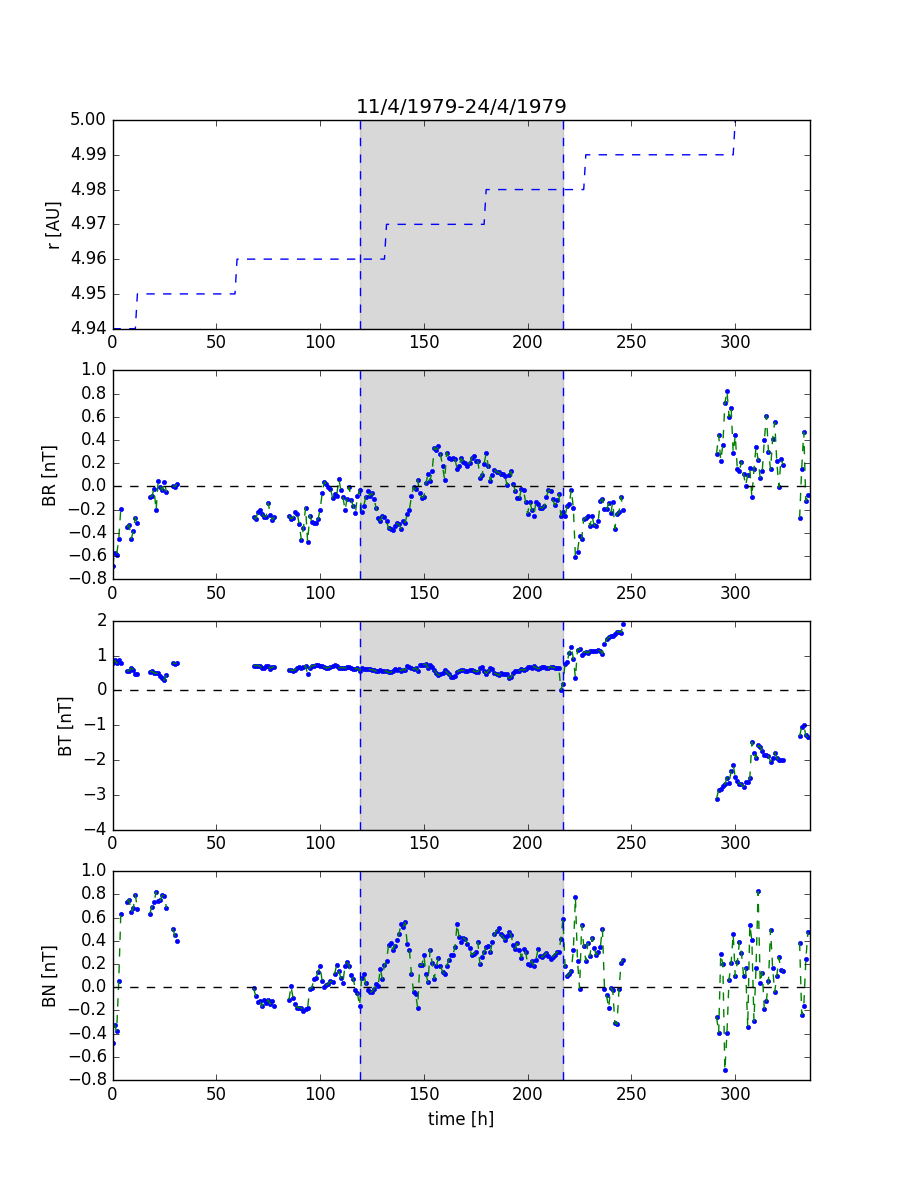} 
\caption{Voyager 1 (upper row) and 2 (lower row) time series for Event M.}\label{fig:eventM}
\end{figure*}

\begin{figure*}[!htb]
\centering
  \centering
  \includegraphics[trim=1cm 1.5cm 1.5cm 2cm, clip=true,width=0.495\linewidth]{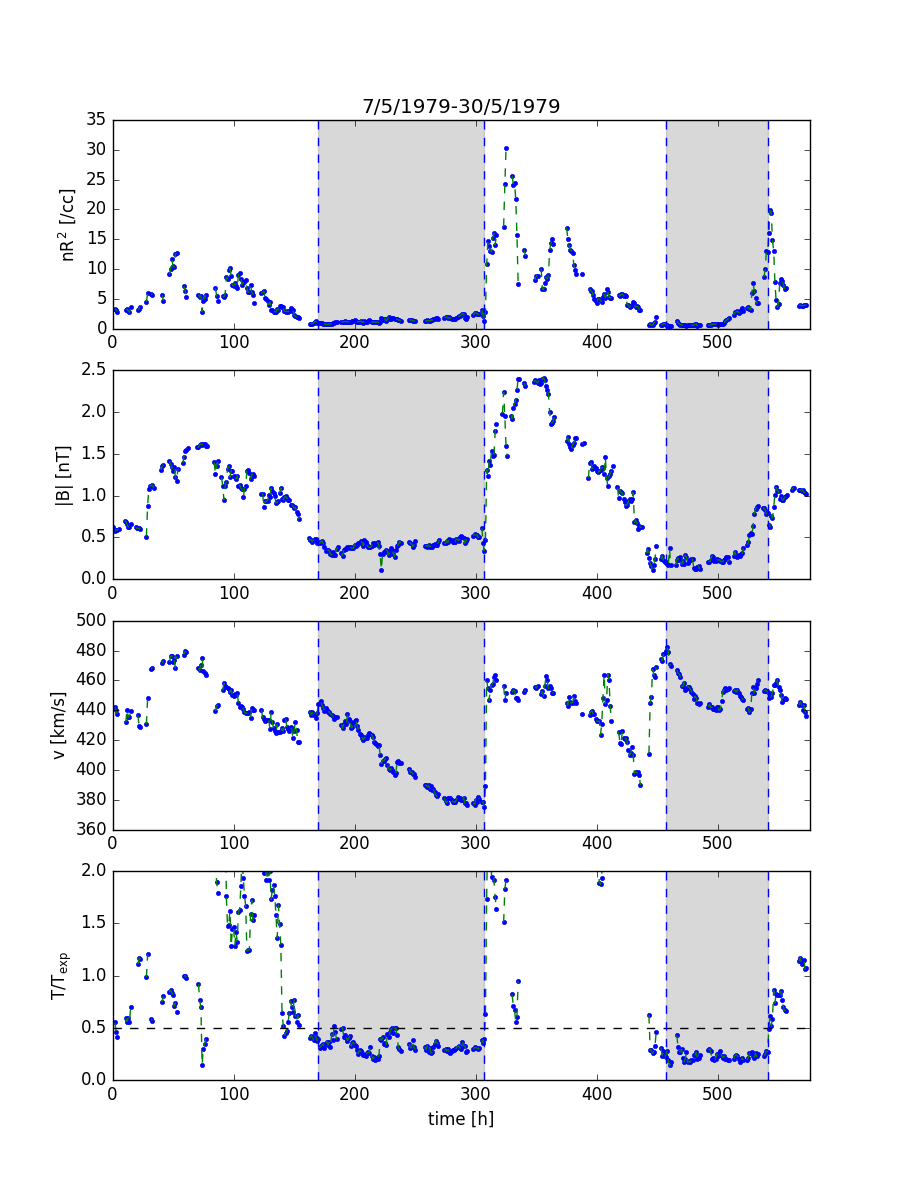} 
  \includegraphics[trim=1cm 1.5cm 1.5cm 2cm, clip=true,width=0.495\linewidth]{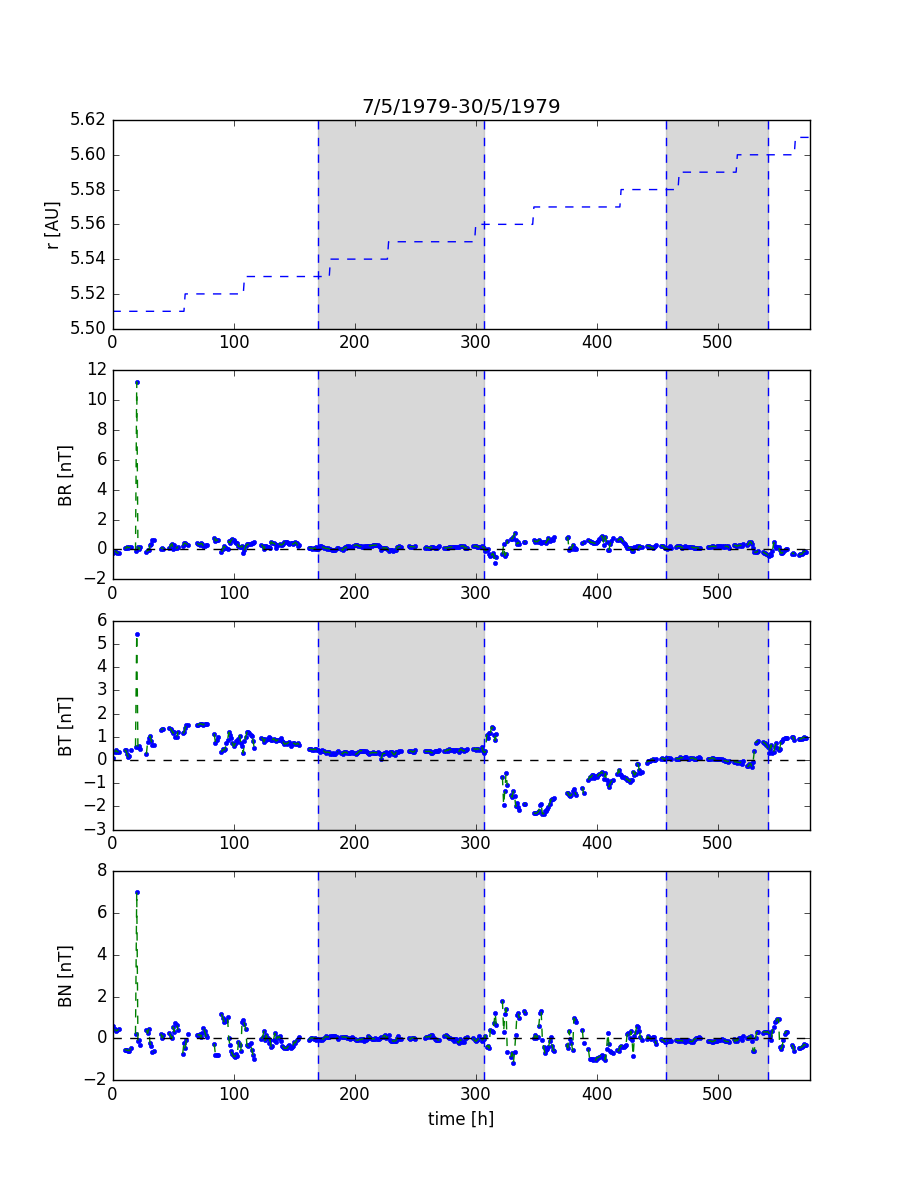}
  \includegraphics[trim=1cm 1.5cm 1.5cm 2cm, clip=true,width=0.495\linewidth]{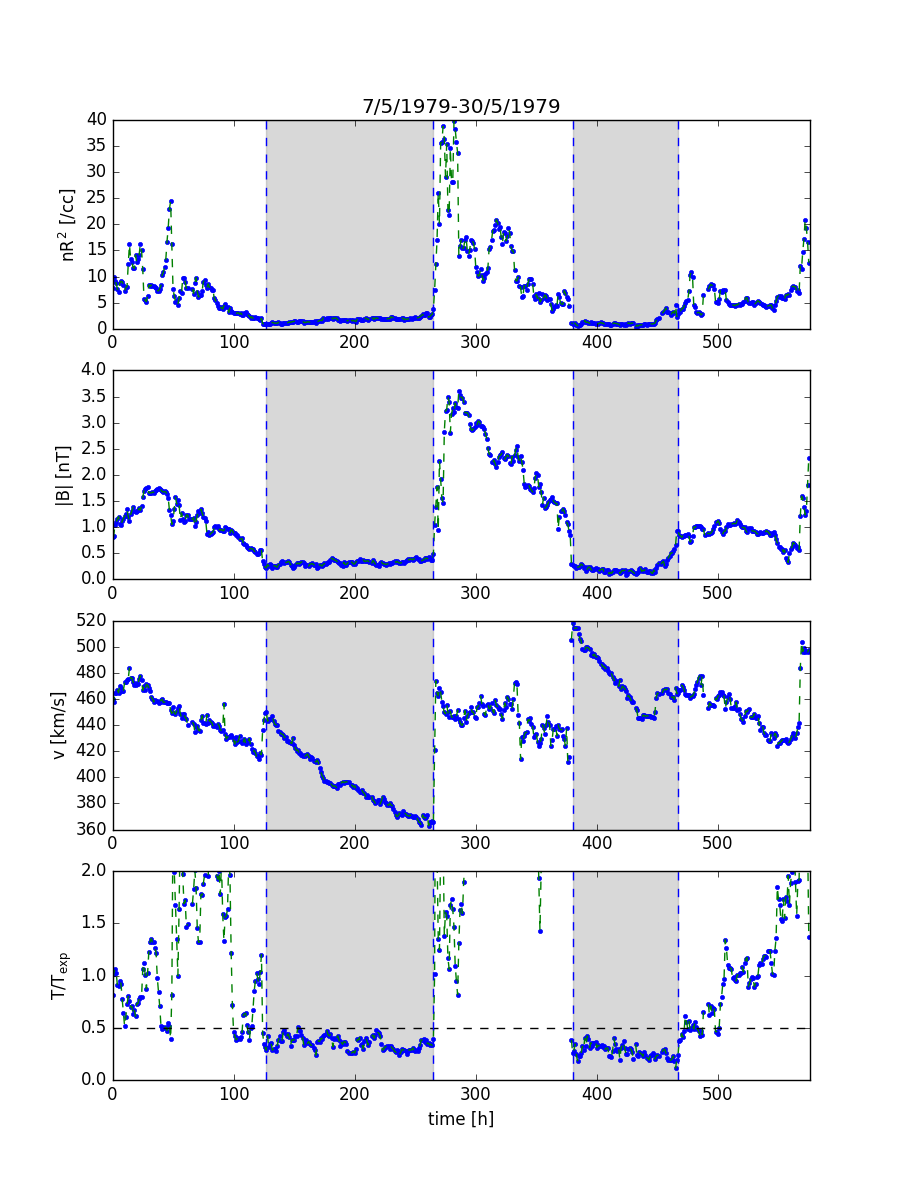}
  \includegraphics[trim=1cm 1.5cm 1.5cm 2cm, clip=true,width=0.495\linewidth]{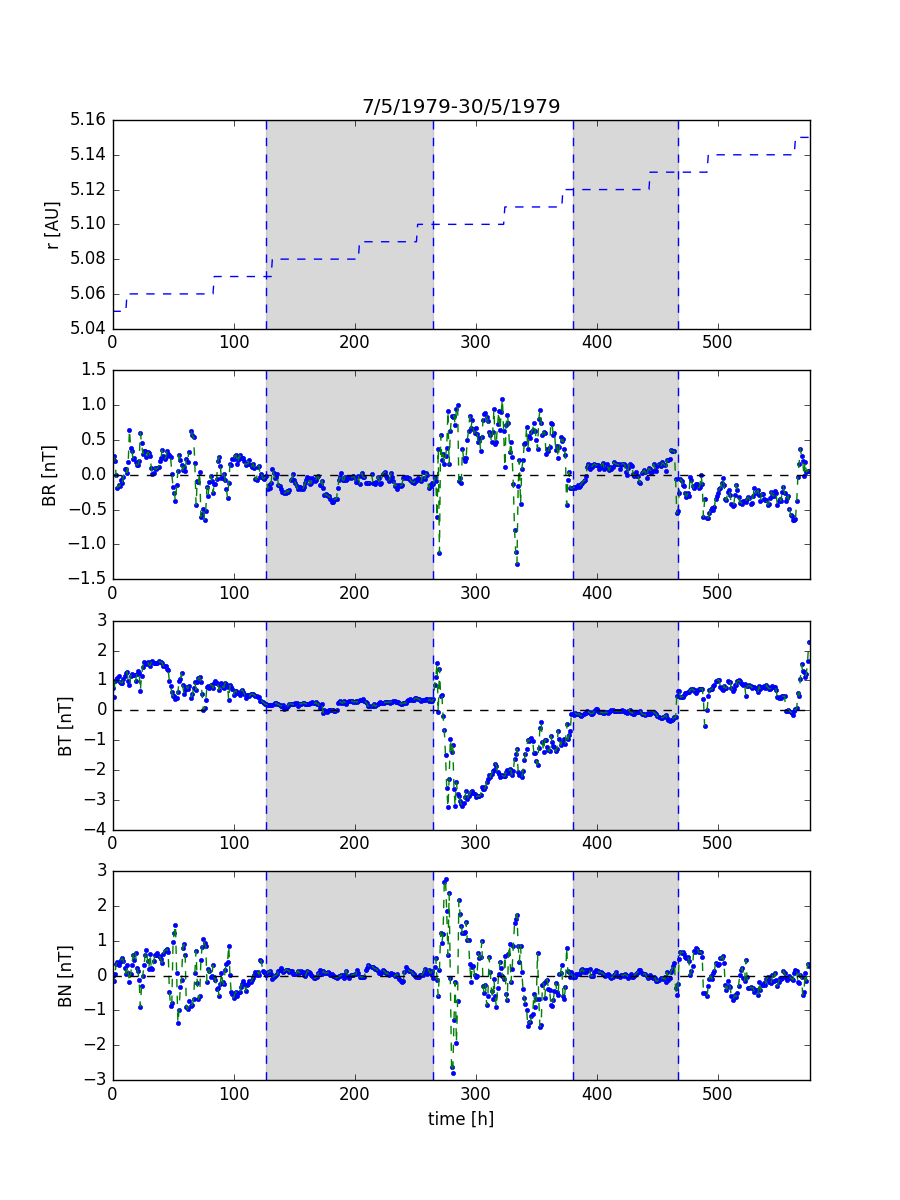} 
\caption{Voyager 1 (upper row) and 2 (lower row) time series for Events N (left shaded region) and O (right shaded region).}\label{fig:eventNO}
\end{figure*}


\begin{figure*}[!htb]
\centering
  \centering
  \includegraphics[trim=1cm 1.5cm 1.5cm 2cm, clip=true,width=0.495\linewidth]{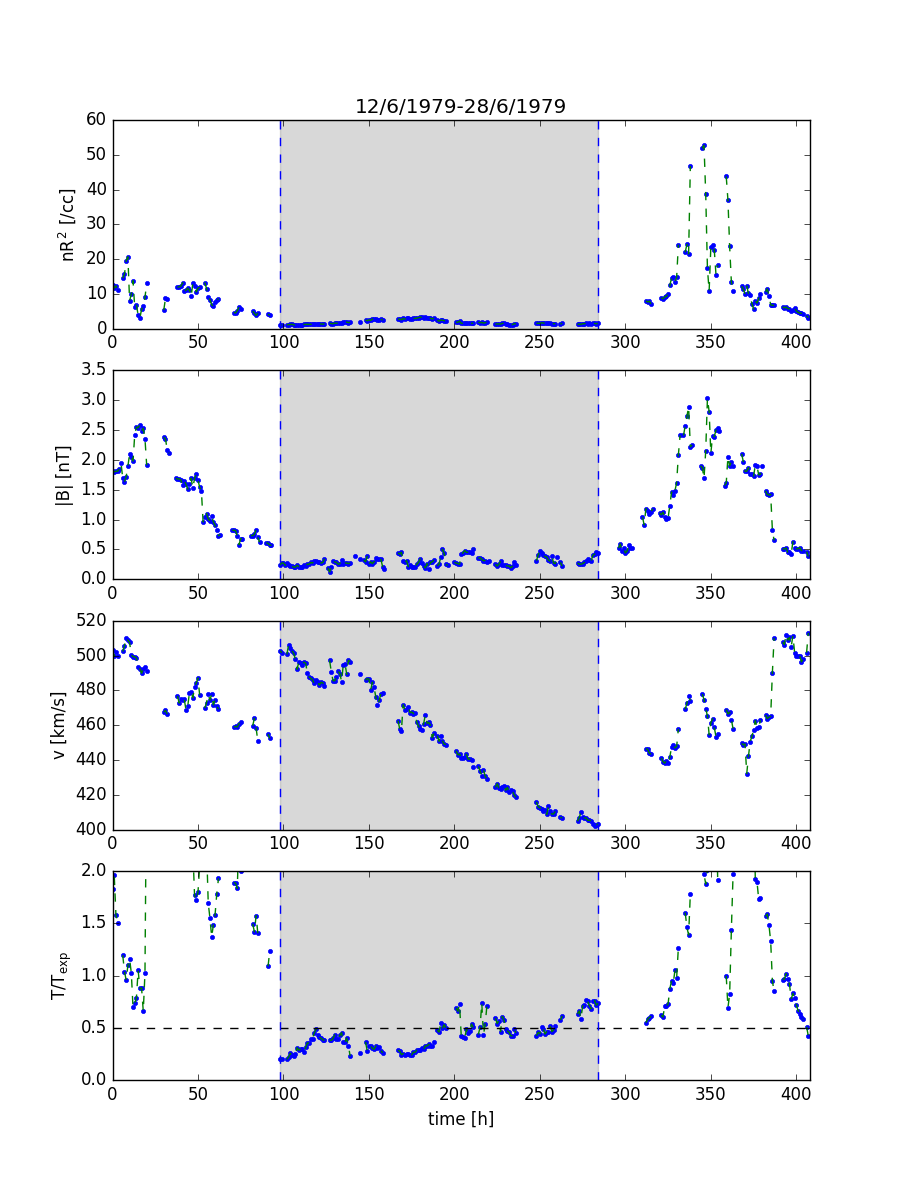} 
  \includegraphics[trim=1cm 1.5cm 1.5cm 2cm, clip=true,width=0.495\linewidth]{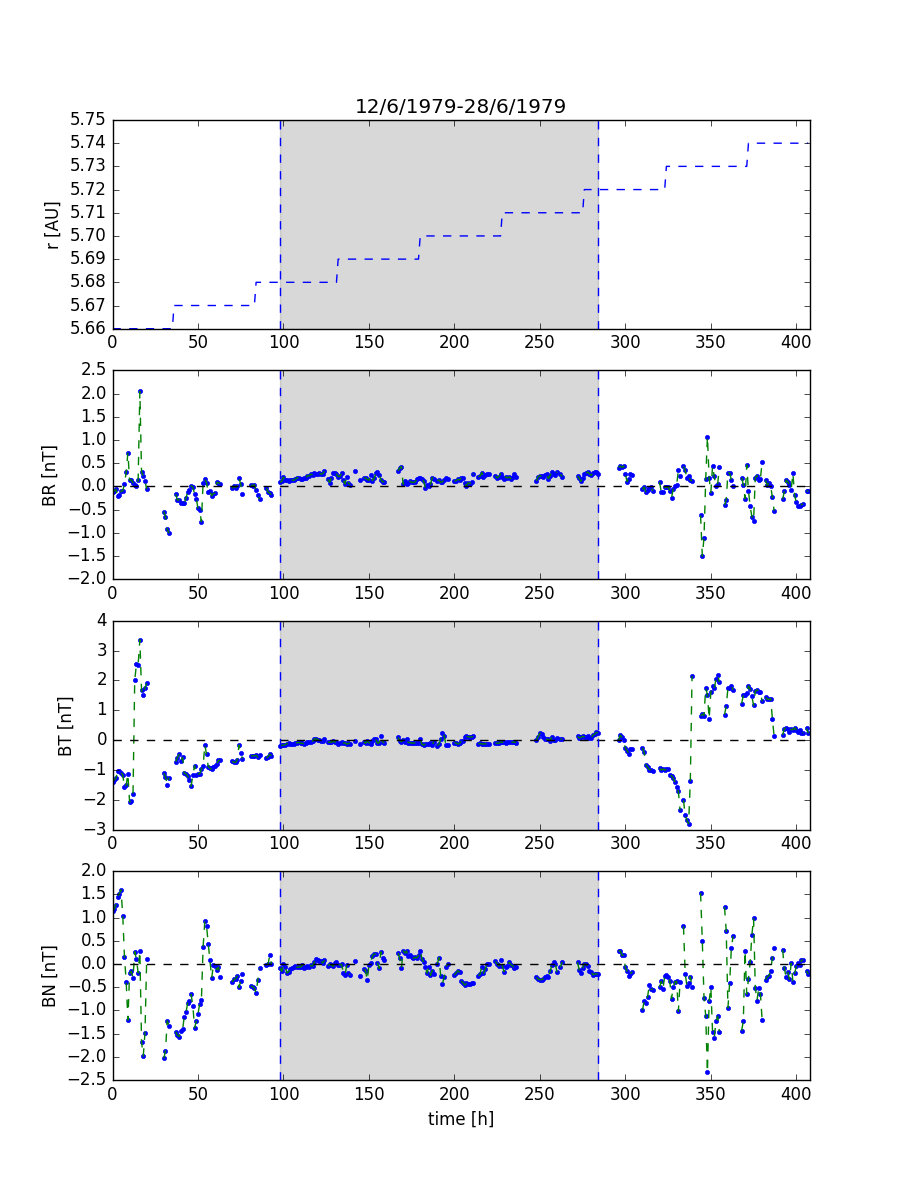}
  \includegraphics[trim=1cm 1.5cm 1.5cm 2cm, clip=true,width=0.495\linewidth]{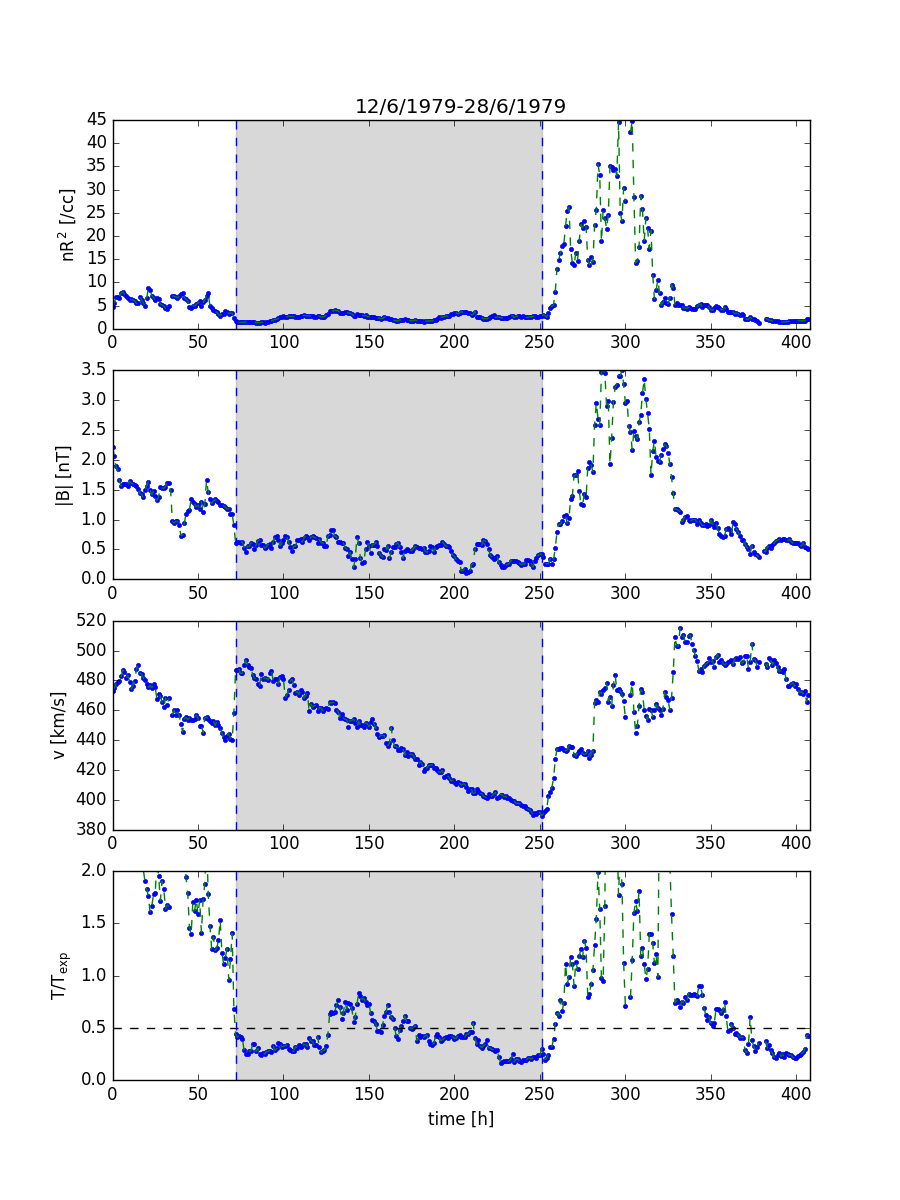}
  \includegraphics[trim=1cm 1.5cm 1.5cm 2cm, clip=true,width=0.495\linewidth]{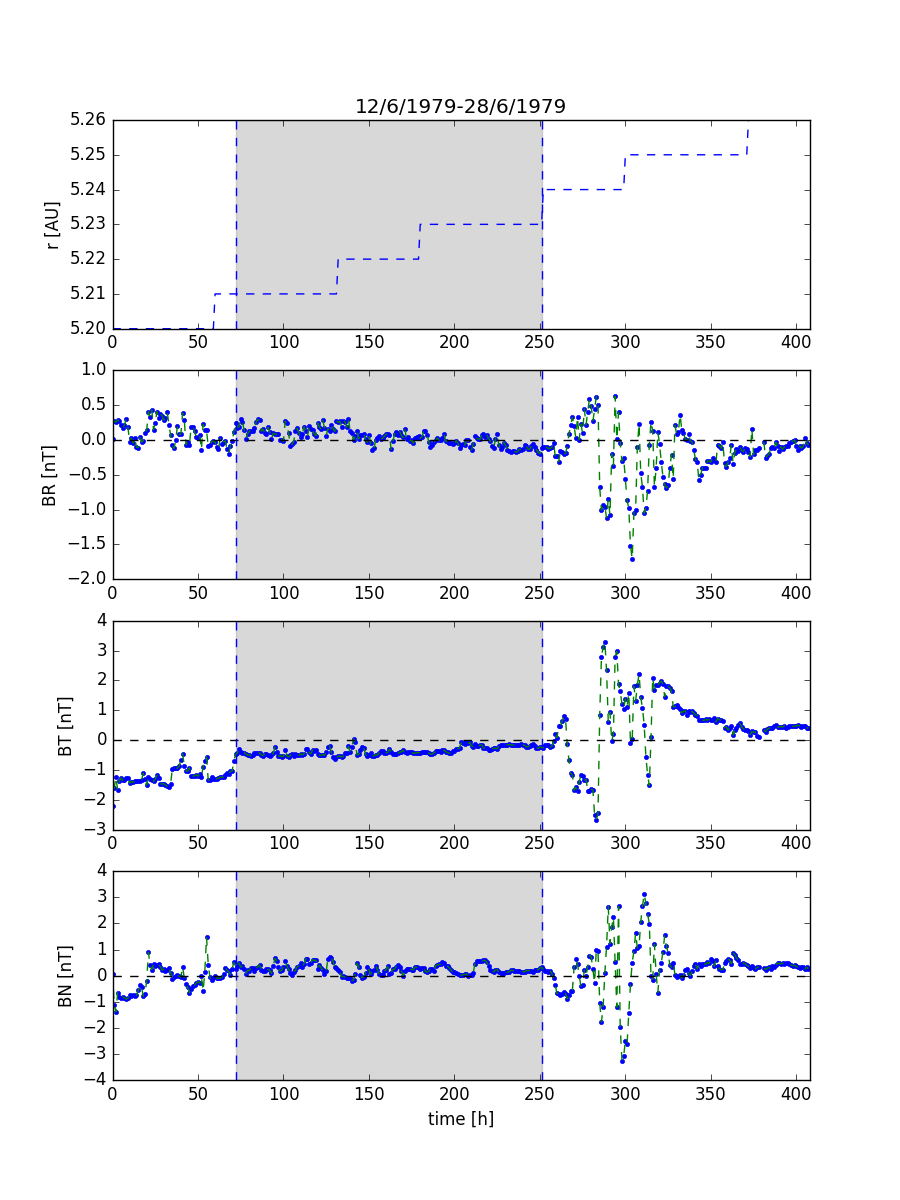} 
\caption{Voyager 1 (upper row) and 2 (lower row) time series for Event P.}\label{fig:eventP}
\end{figure*}

\begin{figure*}[!htb]
\centering
  \centering
  \includegraphics[trim=1cm 1.5cm 1.5cm 2cm, clip=true,width=0.495\linewidth]{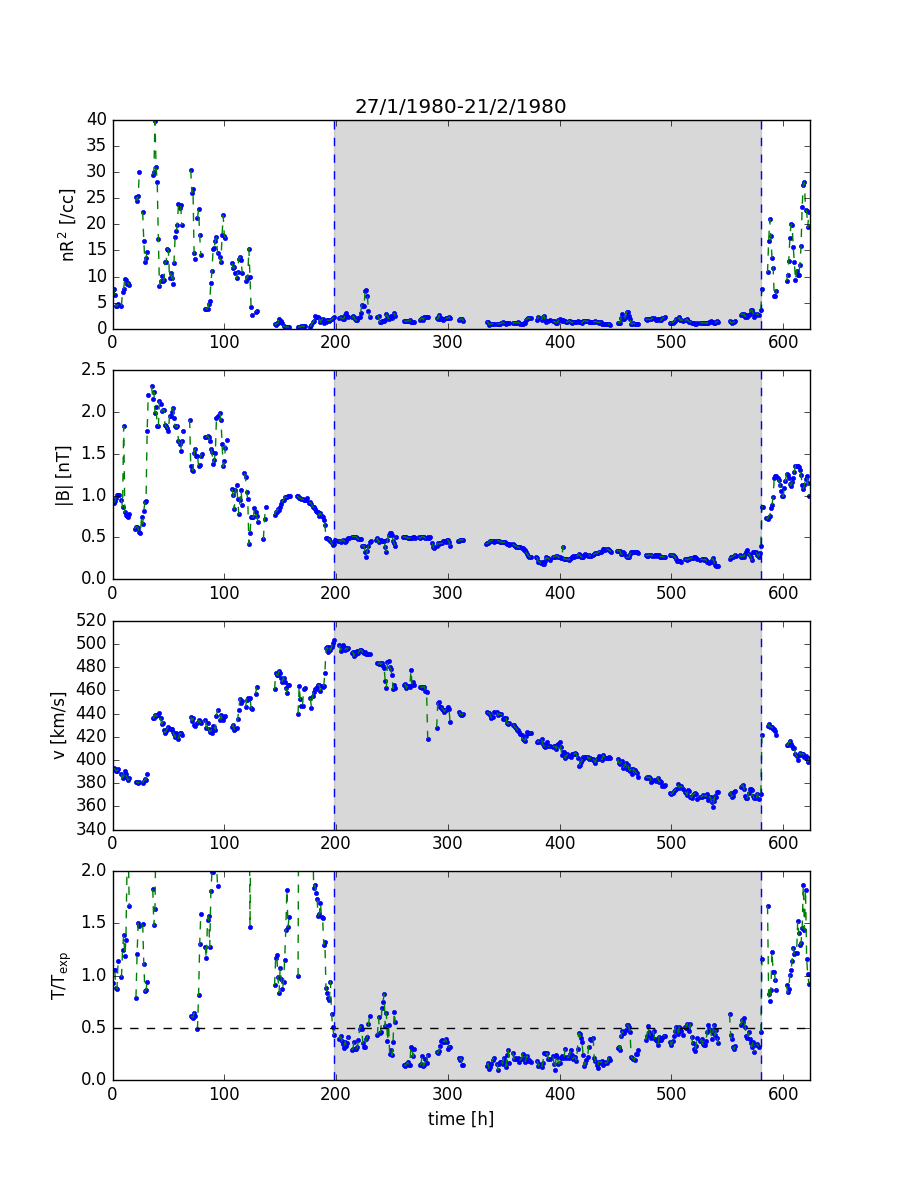} 
  \includegraphics[trim=1cm 1.5cm 1.5cm 2cm, clip=true,width=0.495\linewidth]{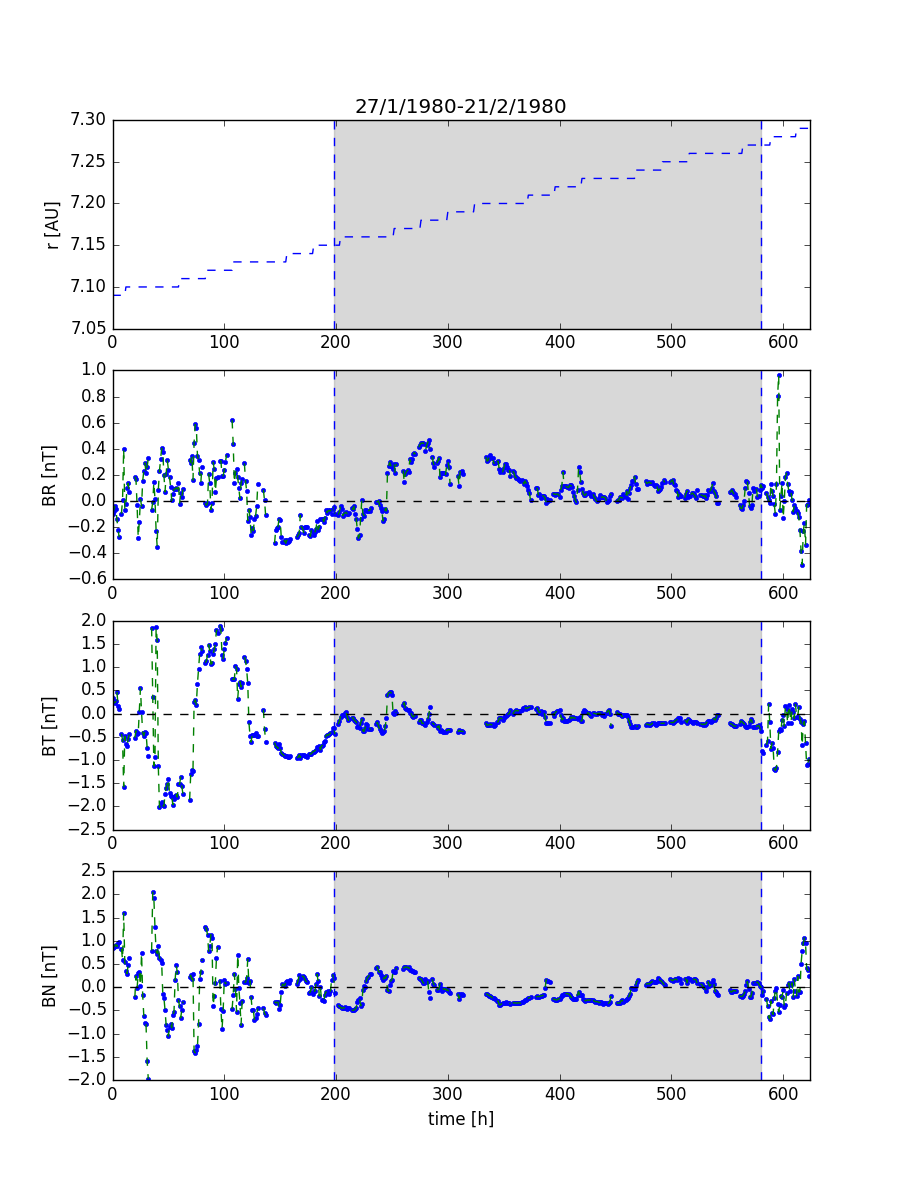}
  \includegraphics[trim=1cm 1.5cm 1.5cm 2cm, clip=true,width=0.495\linewidth]{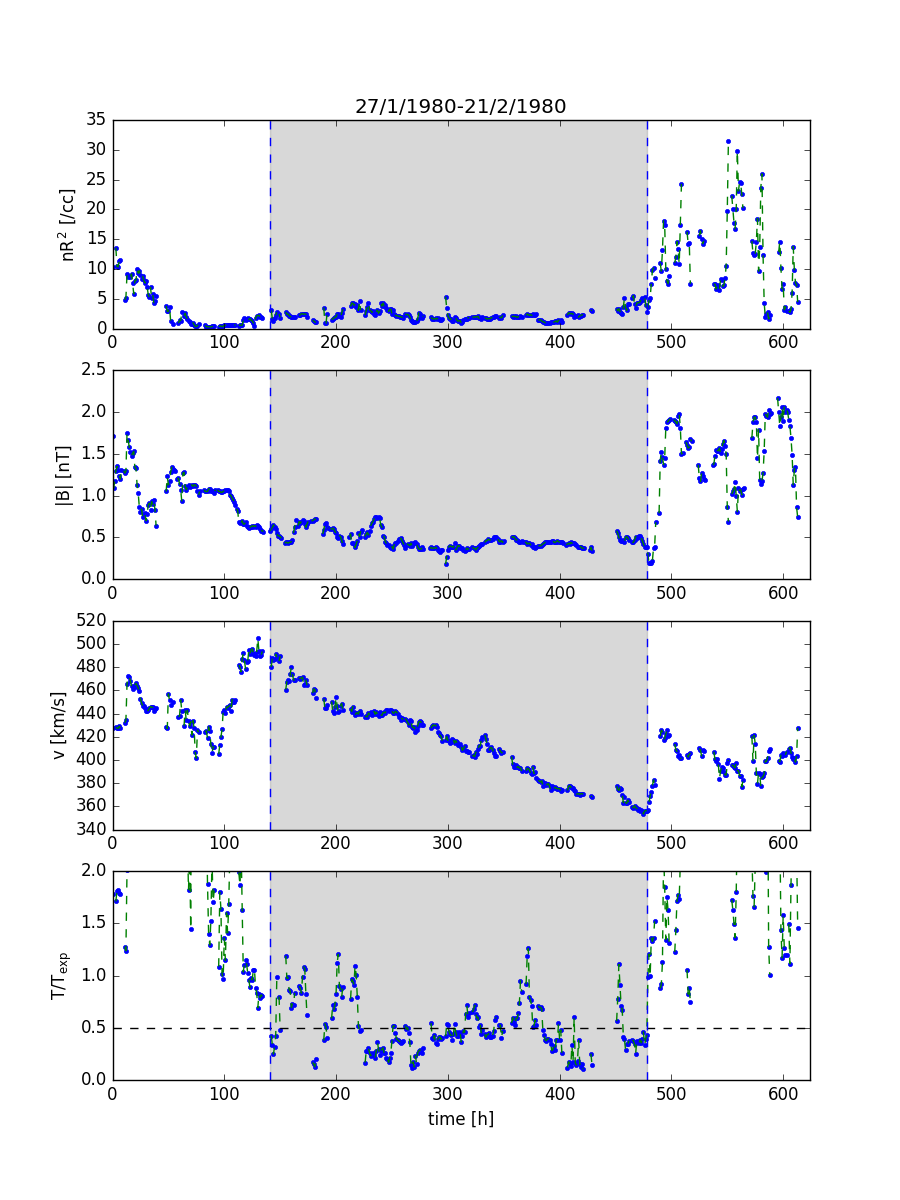}
  \includegraphics[trim=1cm 1.5cm 1.5cm 2cm, clip=true,width=0.495\linewidth]{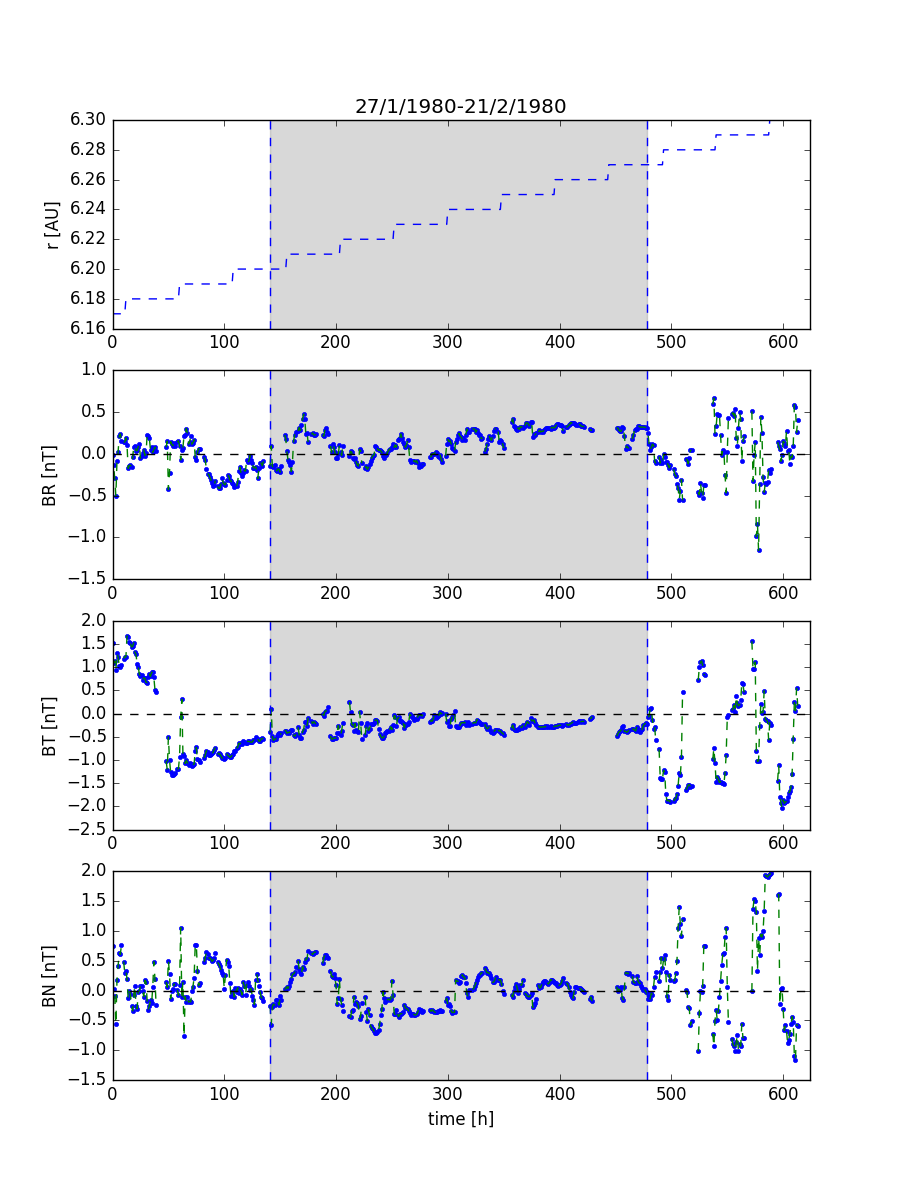} 
\caption{Voyager 1 (upper row) and 2 (lower row) time series for Event Q.}\label{fig:eventQ}
\end{figure*}

\begin{figure*}[!htb]
\centering
  \centering
  \includegraphics[trim=1cm 1.5cm 1.5cm 2cm, clip=true,width=0.495\linewidth]{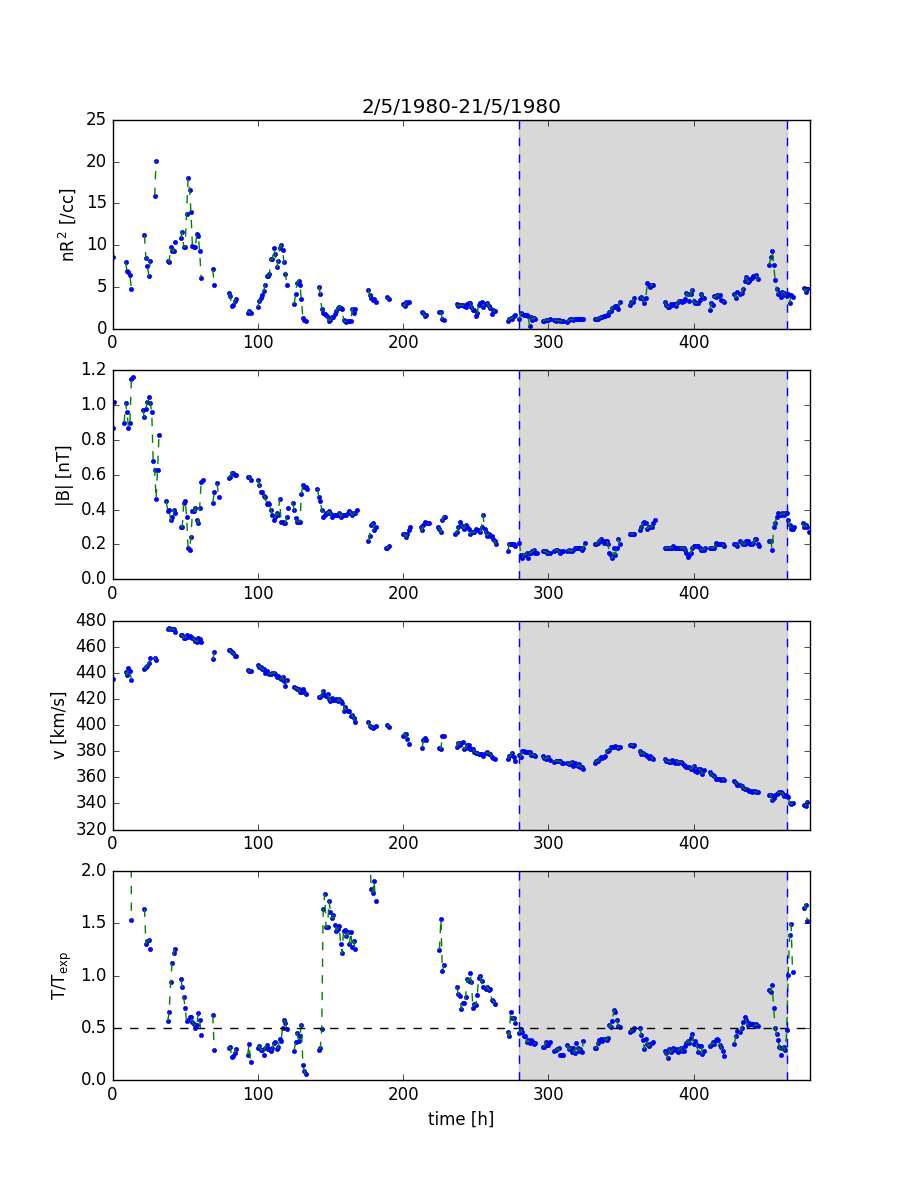} 
  \includegraphics[trim=1cm 1.5cm 1.5cm 2cm, clip=true,width=0.495\linewidth]{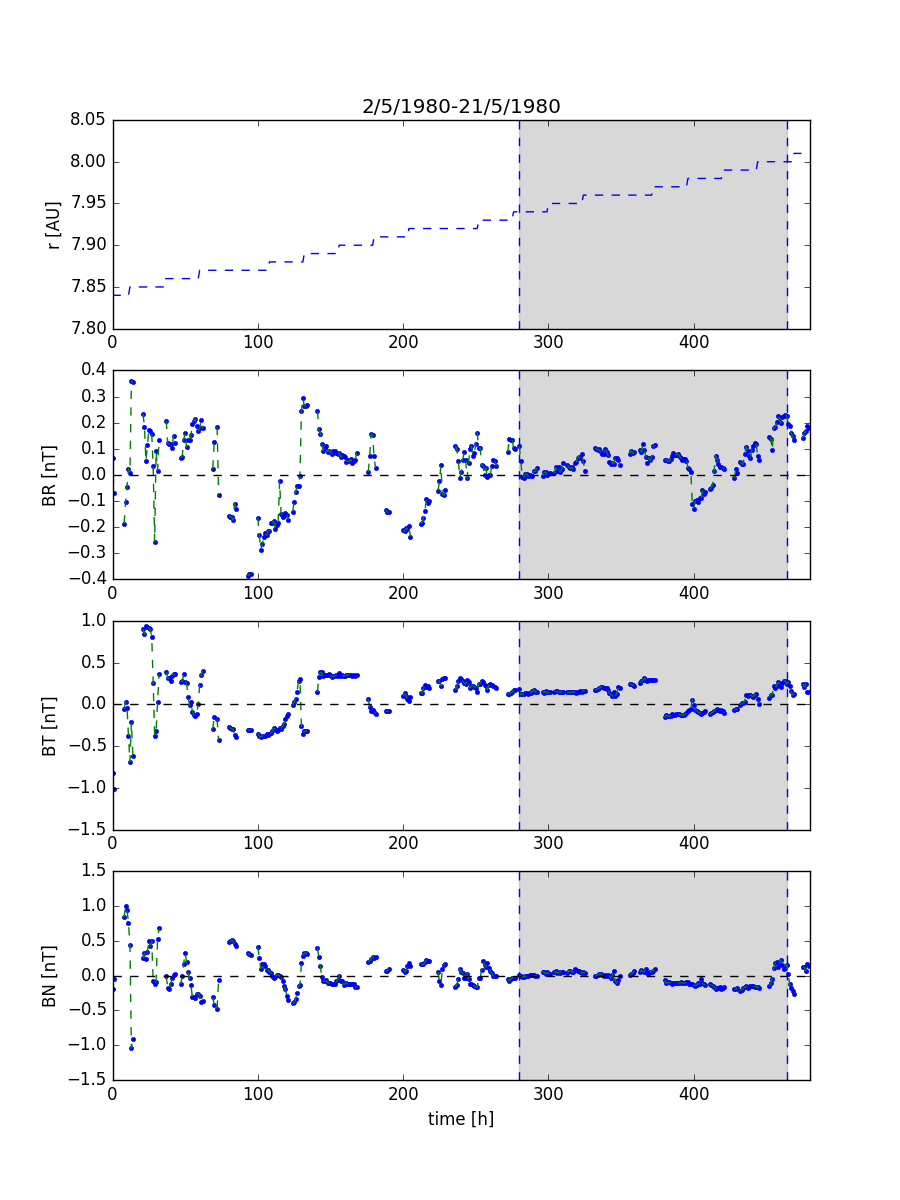}
  \includegraphics[trim=1cm 1.5cm 1.5cm 2cm, clip=true,width=0.495\linewidth]{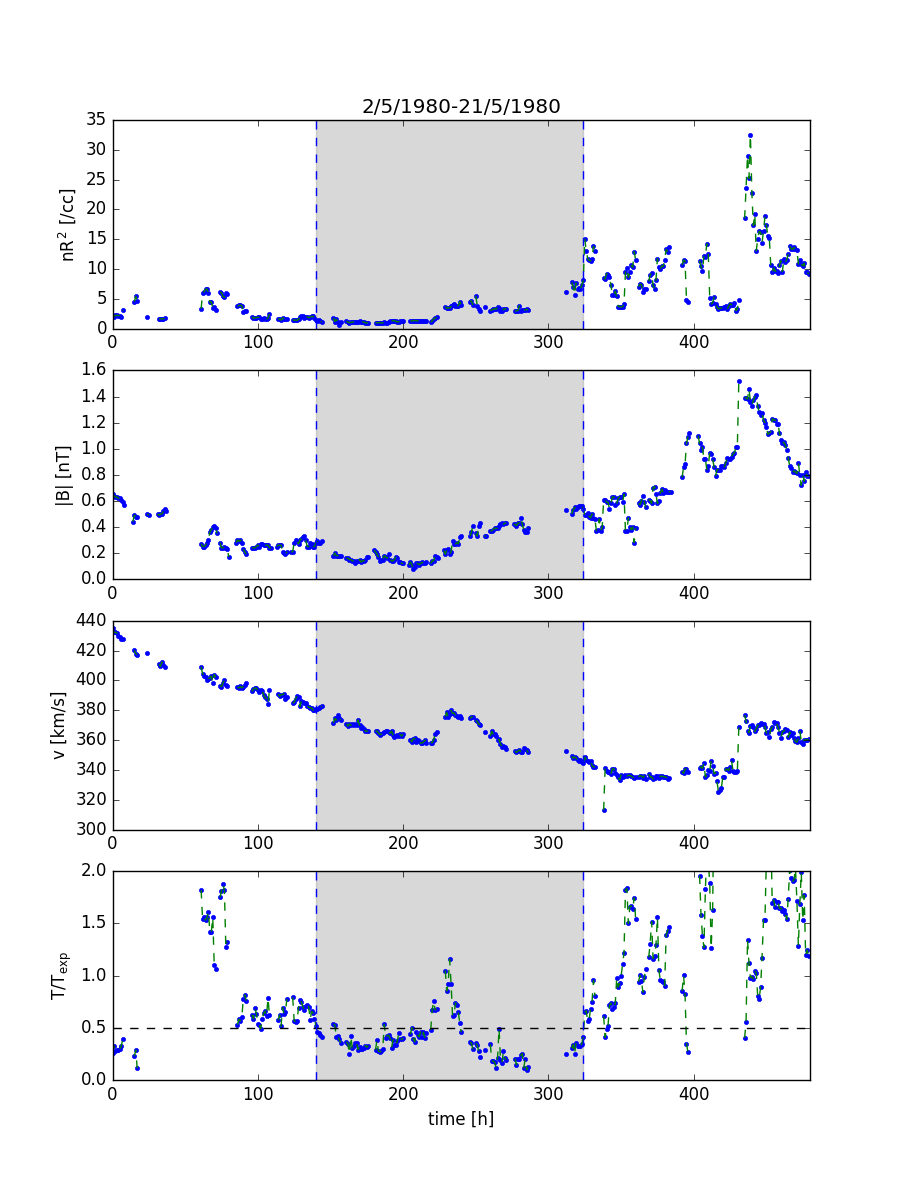}
  \includegraphics[trim=1cm 1.5cm 1.5cm 2cm, clip=true,width=0.495\linewidth]{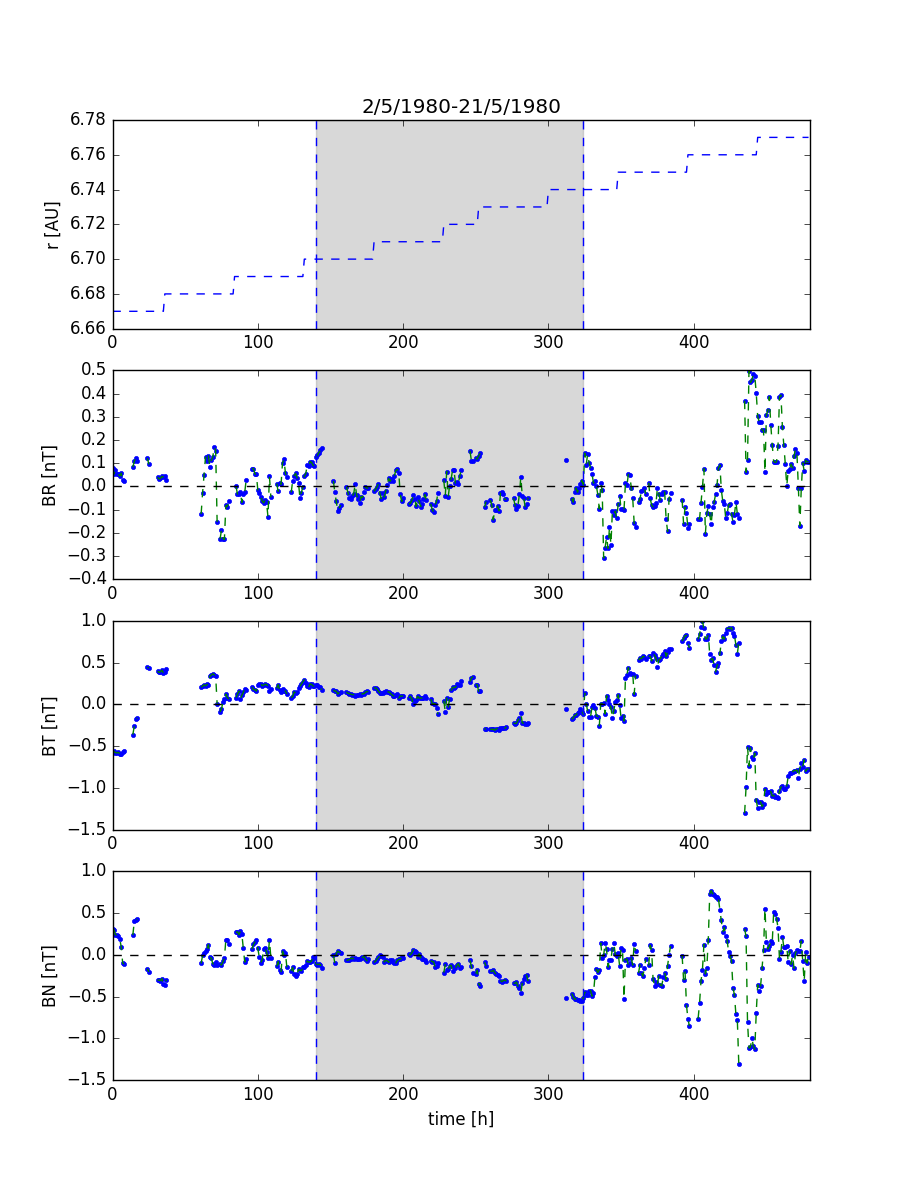} 
\caption{Voyager 1 (upper row) and 2 (lower row) time series for Event R.}\label{fig:eventR}
\end{figure*}

\begin{figure*}[!htb]
\centering
  \centering
  \includegraphics[trim=1cm 1.5cm 1.5cm 2cm, clip=true,width=0.495\linewidth]{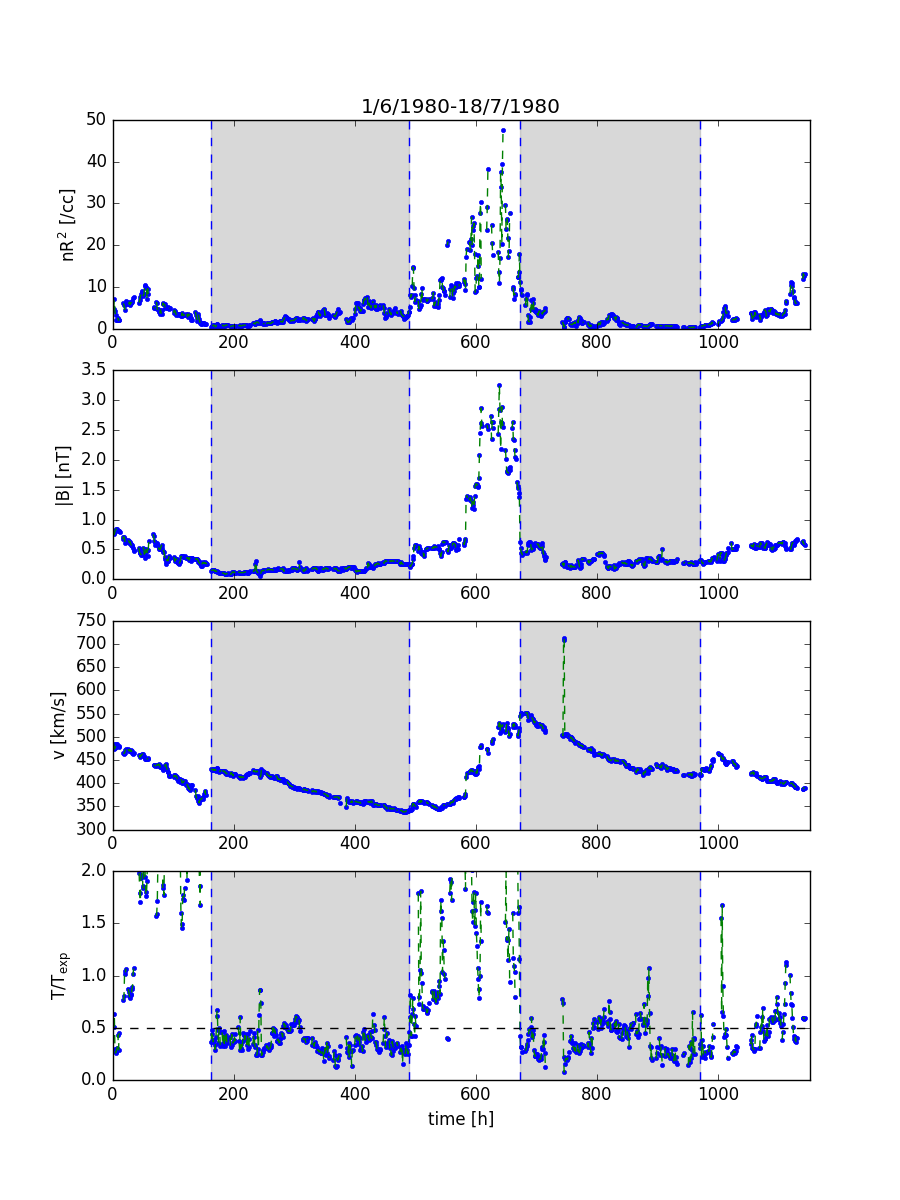} 
  \includegraphics[trim=1cm 1.5cm 1.5cm 2cm, clip=true,width=0.495\linewidth]{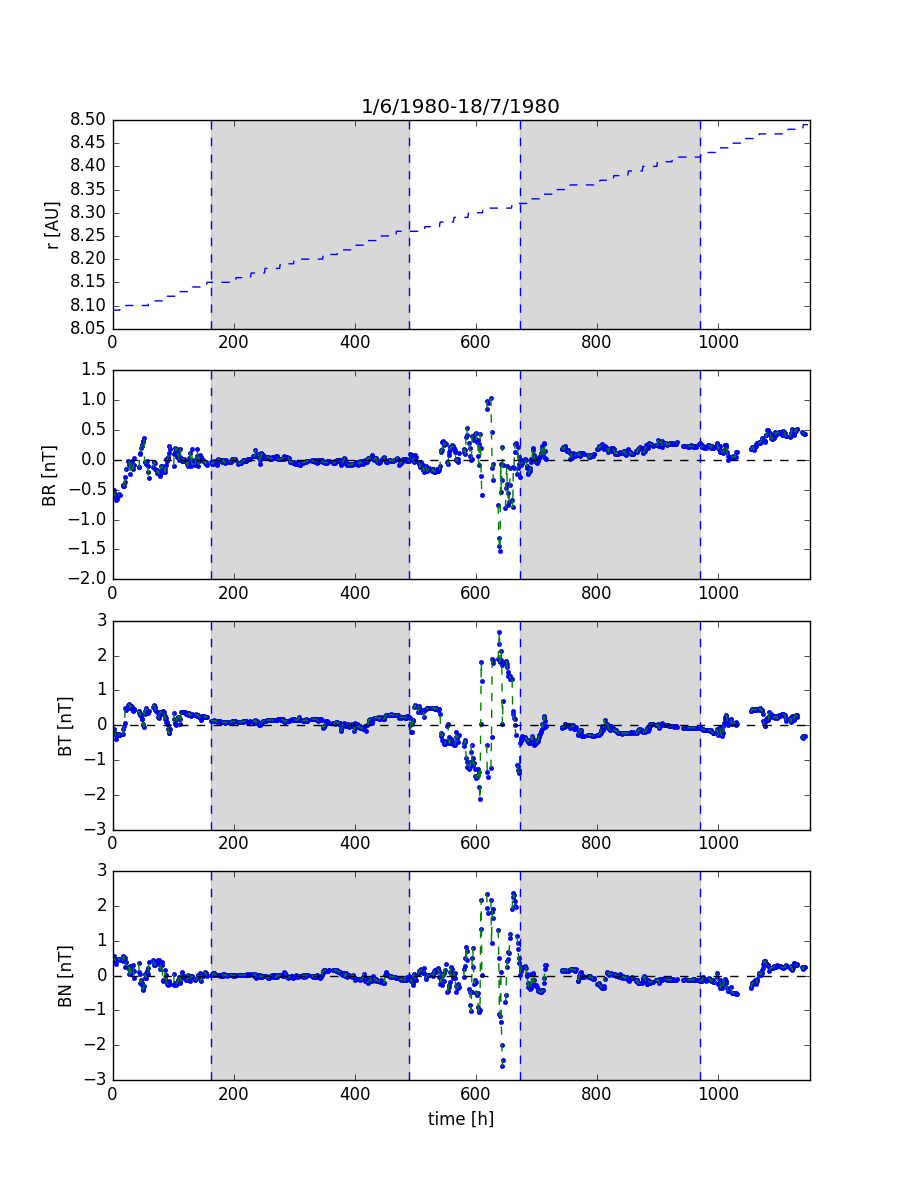}
  \includegraphics[trim=1cm 1.5cm 1.5cm 2cm, clip=true,width=0.495\linewidth]{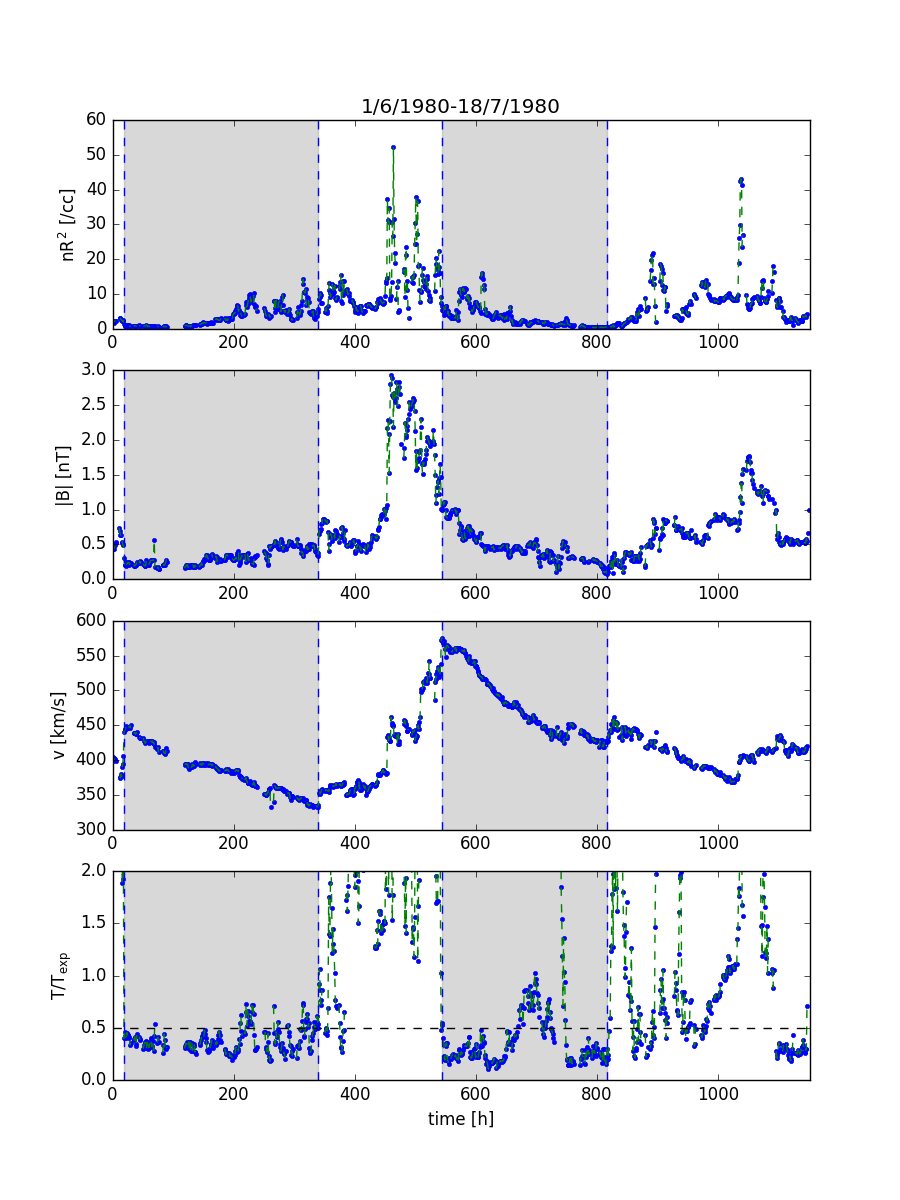}
  \includegraphics[trim=1cm 1.5cm 1.5cm 2cm, clip=true,width=0.495\linewidth]{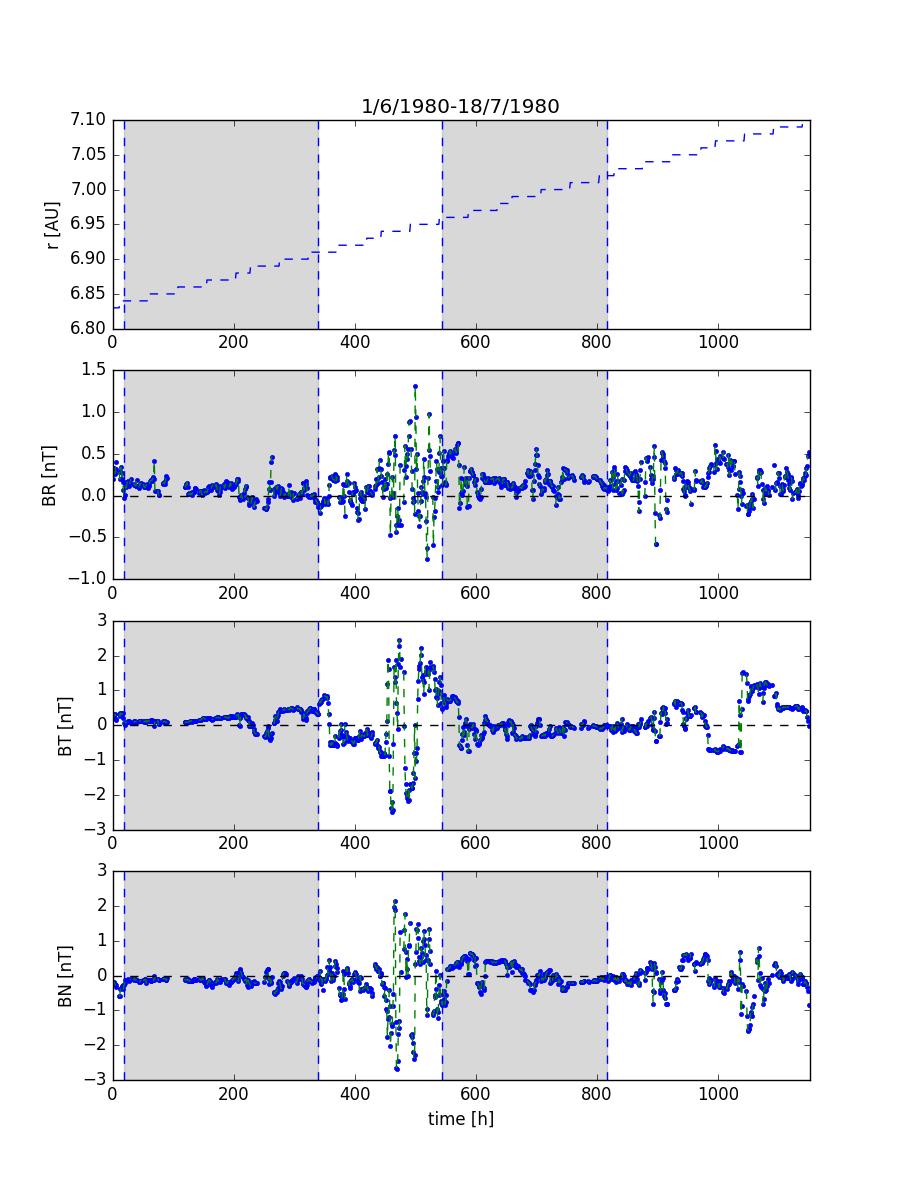} 
\caption{Voyager 1 (upper row) and 2 (lower row) time series for Events S and T.}\label{fig:eventST}
\end{figure*}

\begin{figure*}[!htb]
\centering
  \centering
  \includegraphics[trim=1cm 1.5cm 1.5cm 2cm, clip=true,width=0.495\linewidth]{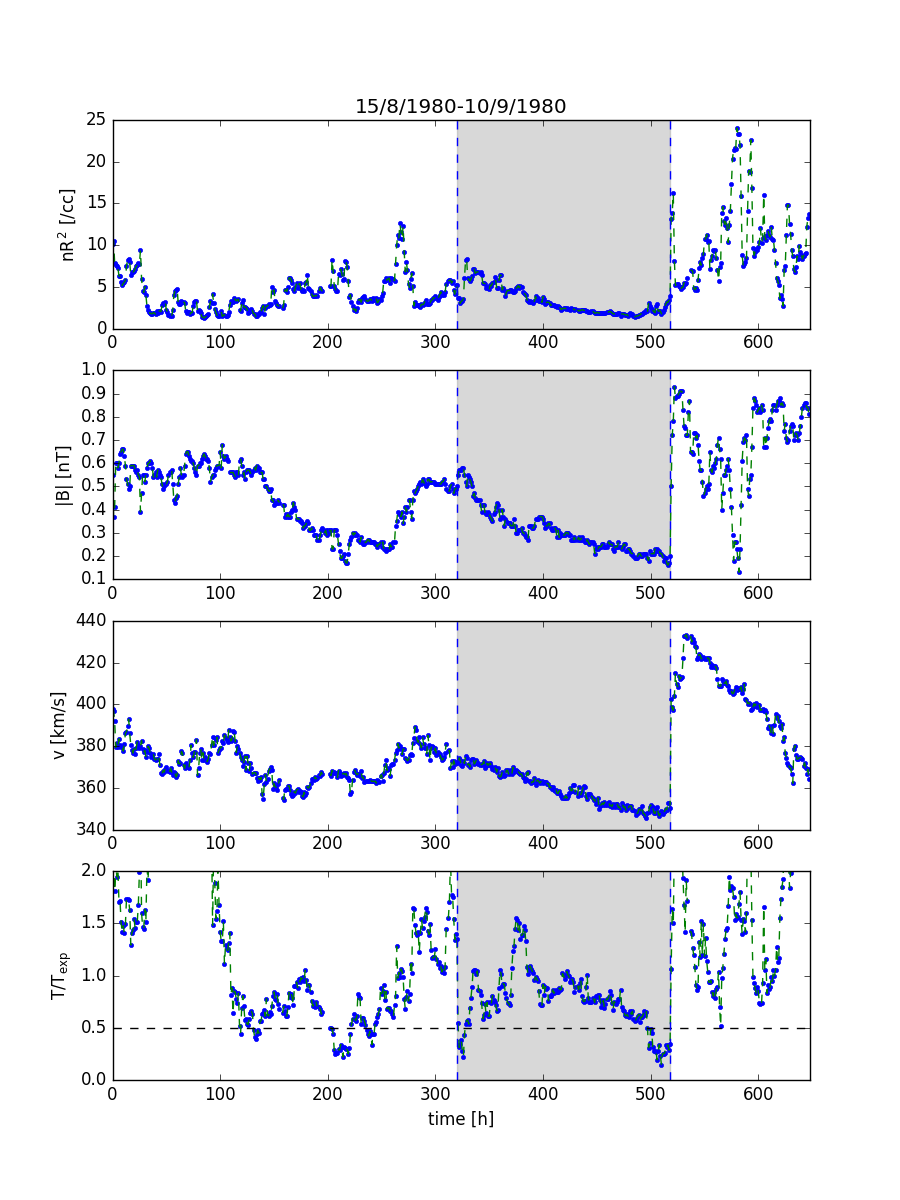} 
  \includegraphics[trim=1cm 1.5cm 1.5cm 2cm, clip=true,width=0.495\linewidth]{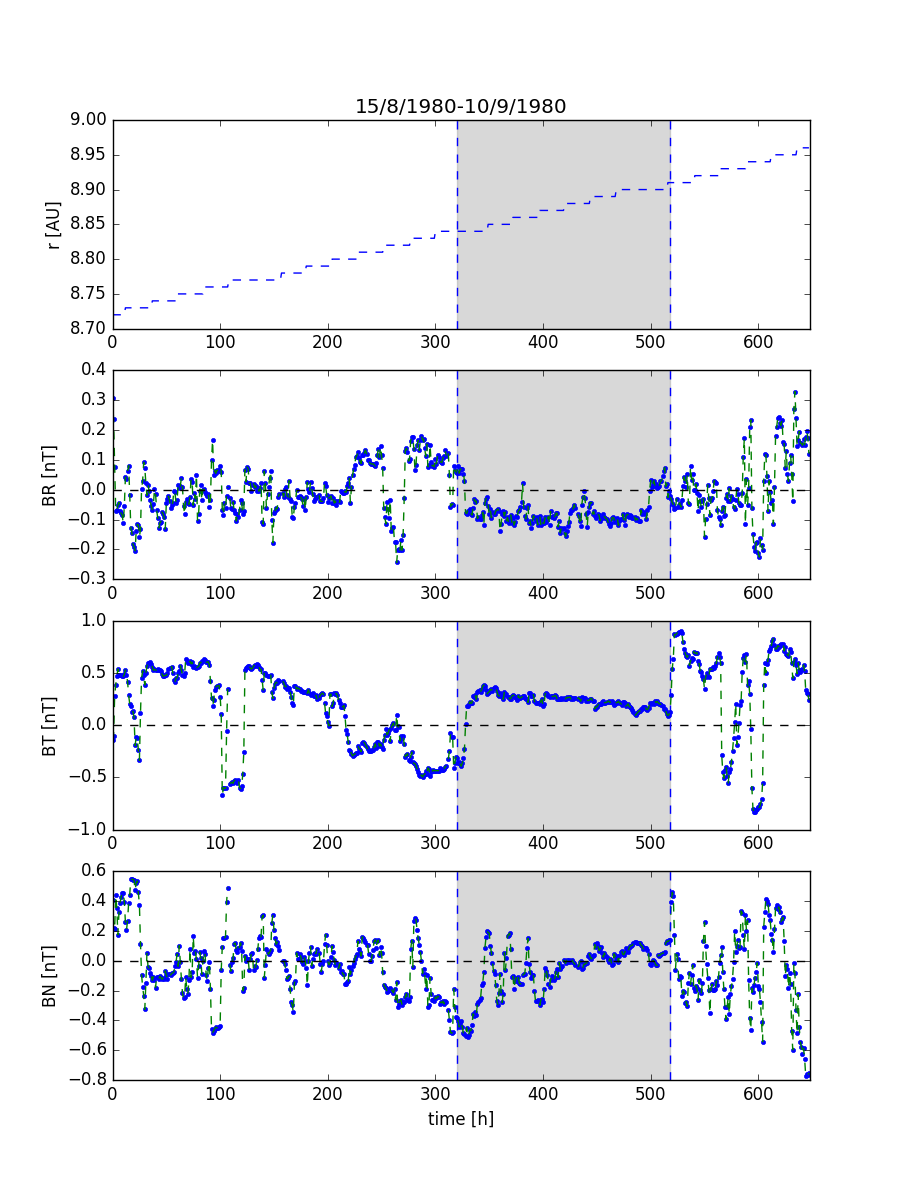}
  \includegraphics[trim=1cm 1.5cm 1.5cm 2cm, clip=true,width=0.495\linewidth]{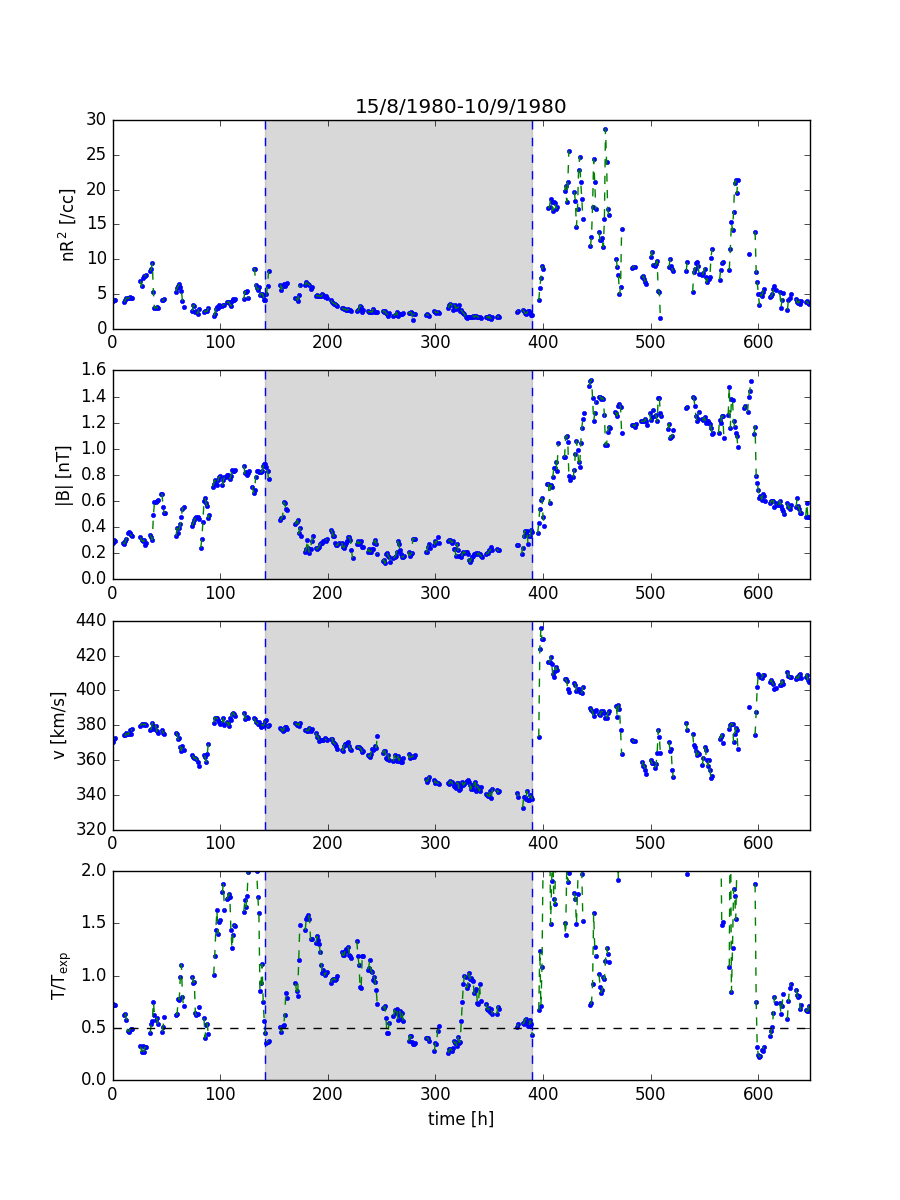}
  \includegraphics[trim=1cm 1.5cm 1.5cm 2cm, clip=true,width=0.495\linewidth]{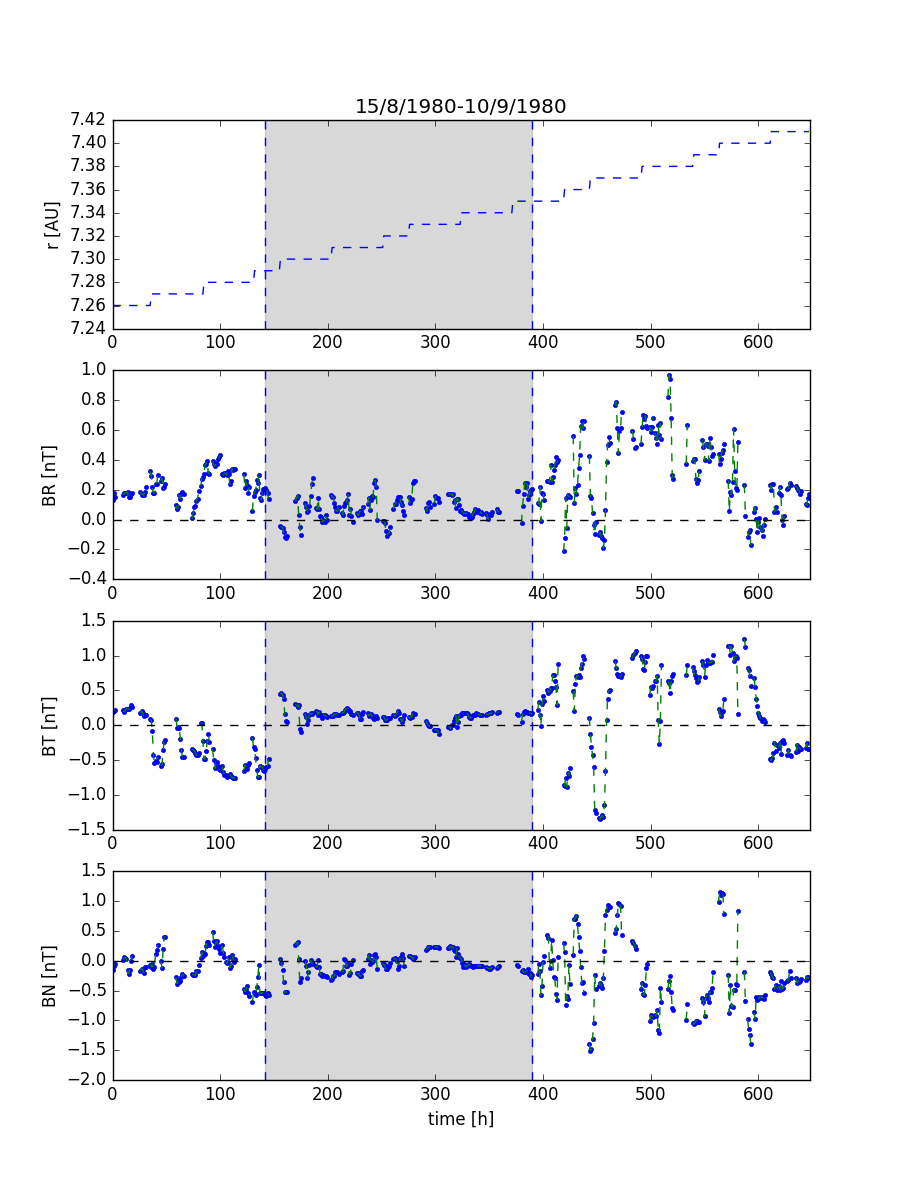} 
\caption{Voyager 1 (upper row) and 2 (lower row) time series for Event U.}
\end{figure*}

\cleardoublepage


\bibliographystyle{model5-names}
\biboptions{authoryear}
\bibliography{refs}

\end{document}